\documentclass[twocolumn]{aastex6}
\usepackage{natbib, chngcntr}
\usepackage{enumitem}
\renewcommand{\tabcolsep}{3.2pt}
\usepackage{mathtools}
\usepackage{color}

\begin{document}

\title{\texttt{HELIOS}: An Open-source, GPU-accelerated Radiative Transfer Code For Self-consistent Exoplanetary Atmospheres}
\author{Matej Malik\altaffilmark{1}, Luc Grosheintz\altaffilmark{1}, Jo\~{a}o M. Mendon\c{c}a\altaffilmark{1}, Simon L. Grimm\altaffilmark{1,2}, Baptiste Lavie\altaffilmark{1}, Daniel Kitzmann\altaffilmark{1}, Shang-Min Tsai\altaffilmark{1}, Adam Burrows\altaffilmark{3}, Laura Kreidberg\altaffilmark{4,5}, Megan Bedell\altaffilmark{4}, Jacob L. Bean\altaffilmark{4}, Kevin B. Stevenson\altaffilmark{4}, Kevin Heng\altaffilmark{1}}
\altaffiltext{1}{University of Bern, Center for Space and Habitability, Sidlerstrasse 5, CH-3012, Bern, Switzerland.  Emails: matej.malik@csh.unibe.ch, kevin.heng@csh.unibe.ch}
\altaffiltext{2}{University of Z\"{u}rich, Institute for Computational Science, Winterthurerstrasse 190, CH-8057 Z\"{u}rich, Switzerland}
\altaffiltext{3}{Department of Astrophysical Sciences, Princeton University Peyton Hall, Princeton, NJ 08544, U.S.A.}
\altaffiltext{4}{Department of Astronomy and Astrophysics, University of Chicago, 5640 S. Ellis Ave, Chicago, IL 60637, U.S.A.}
\altaffiltext{5}{Harvard-Smithsonian Center for Astrophysics, 60 Garden Street, Cambridge, MA 02138, U.S.A.}

\shorttitle{\texttt{HELIOS}: An Open-Source Radiative Transfer Code}
\shortauthors{M. Malik et al.}

\begin{abstract}
We present the open-source radiative transfer code named \texttt{HELIOS}, which is constructed for studying exoplanetary atmospheres.  In its initial version, the model atmospheres of \texttt{HELIOS} are one-dimensional and plane-parallel, and the equation of radiative transfer is solved in the two-stream approximation with non-isotropic scattering. A small set of the main infrared absorbers is employed, computed with the opacity calculator \texttt{HELIOS-K} and combined using a correlated-$k$ approximation. The molecular abundances originate from validated analytical formulae for equilibrium chemistry. We compare \texttt{HELIOS} with the work of Miller-Ricci \& Fortney using a model of GJ 1214b, and perform several tests, where we find: model atmospheres with single-temperature layers struggle to converge to radiative equilibrium; $k$-distribution tables constructed with $\gtrsim 0.01$ cm$^{-1}$ resolution in the opacity function ($ \lesssim 10^3$ points per wavenumber bin) may result in errors $\gtrsim 1$--10\% in the synthetic spectra; and a diffusivity factor of 2 approximates well the exact radiative transfer solution in the limit of pure absorption.  We construct ``null-hypothesis" models (chemical equilibrium, radiative equilibrium and solar element abundances) for 6 hot Jupiters. We find that the dayside emission spectra of HD 189733b and WASP-43b are consistent with the null hypothesis, while it consistently under-predicts the observed fluxes of WASP-8b, WASP-12b, WASP-14b and WASP-33b.  We demonstrate that our results are somewhat insensitive to the choice of stellar models (blackbody, Kurucz or \texttt{PHOENIX}) and metallicity, but are strongly affected by higher carbon-to-oxygen ratios.  The code is publicly available as part of the Exoclimes Simulation Platform (ESP; \url{exoclime.net}).
\end{abstract}


\section{Introduction}
\label{sec:intro}

The past few years have been marked by a slow, but steady, shift from the era of the detections of exoplanets to the new age of the characterization of their atmospheres.  Exoplanets transiting in front of their host stars allow for atmospheric features to be imprinted onto the total system light \citep{se00,br01,ch02}.  Secondary eclipses allow for photons from the exoplanetary atmosphere to be directly measured \citep{ch05,deming05}.  Extracting the spectroscopic signatures of these exoplanetary atmospheres is a challenging task, because they are typically many orders of magnitude fainter than the light from their host stars.  Interpreting these signatures requires a profound understanding of radiative transfer and atmospheric chemistry, in order to infer the thermal structure and atomic/molecular abundances of the atmosphere from the data.

Hot Jupiters are particularly accessible to atmospheric characterization via transits and eclipses.  They are hardly one-dimensional (1D) objects, but a reasonable first approach is to study them using 1D, plane-parallel model atmospheres \citep{su03,ba05,fo05,fo06,fo08,fo10,bu06,bu07,bu08}, which may be used to mimic the dayside- or nightside-integrated emission. The simplest model one may construct of a dayside emission spectrum (besides a Planck function) is a 1D model with an atmosphere in radiative and chemical equilibrium, if one neglects the effects of atmospheric dynamics and photochemistry.  Despite these simplifications, there are several non-trivial demands associated with such a model: it should be able to consider a rich variety of chemistries, metallicities, irradiation fluxes from the star and internal heat fluxes from the interior of the exoplanet.  It should be able to take, as an input, arbitrary combinations of molecules and their opacities.  The synthetic spectrum computed should be highly customizable, such that it may be readily compared to both photometric and spectroscopic data, often combined in a heterogeneous way across wavelength.  To explore such a broad range of parameter space, the numerical implementation of a model (short: ``code") needs to solve for radiative equilibrium very efficiently and also allow for numerical convergence to be checked in several different ways: number of model layers, spectral resolution of opacity function, number of wavelength bins used, etc. Such a code forms the basis of a flexible radiation package that one may couple to a chemical kinetics code or a three-dimensional general circulation model. The challenges of constructing a 1D radiative-convective model are also discussed in the review article by \cite{marley15}, where the ``convective" part stands for the additional consideration of convective stability, which marks the next step in sophistication of an atmospheric model.

In the current work, we present a customizable and built-from-scratch computer code named \texttt{HELIOS}\footnote{Named after the Greek god of the Sun.}, which  has or uses the following components.
\begin{itemize}

\item In this initial version, we use the analytical solutions of the radiative transfer equation in the two-stream approximation, as derived by \cite{he14}.  These solutions enable us to iteratively and self-consistently solve for the temperature-pressure profile of the atmosphere via iteration with its opacity function, which generally depends on temperature, pressure and wavelength.  The synthetic spectrum is obtained as a natural by-product of this self-consistent calculation.

\item For the opacity function of the atmosphere, we use our open-source and custom opacity calculator, \texttt{HELIOS-K}, which was previously published by \cite{gr15}.  The finest resolution we have used is $10^{-5}$ cm$^{-1}$ across the entire wavenumber range considered.  We then compute $k$-distribution tables from this finely-spaced grid of opacities across temperature, pressure and molecular species.

\item Throughout this work, we assume chemical equilibrium, which effectively means that the chemistry is described by only two parameters: the elemental abundances of carbon and oxygen.  Given the input values of these elemental abundances, we then use the validated analytical formulae of \cite{he16b} and \cite{he16c} to calculate the mixing ratios (abundances normalized to that of molecular hydrogen) of the various molecules.  We consider water (H$_2$O), carbon monoxide (CO), carbon dioxide (CO$_2$) and methane (CH$_4$).

\item We have built \texttt{HELIOS} to run on graphics processing units (GPUs) to maximize the computational throughput. A \texttt{HELIOS} calculation with 101 model layers and 300 wavelength bins takes only a few minutes to complete on a personal computer with a NVIDIA GeForce 750M GPU.\footnote{Note that these are fully converged and self-consistent models, which require iteration to solve for radiative equilibrium.} This level of efficiency allows us to effectively perform parameter studies.

\end{itemize}

In Section \ref{sec:meth}, we provide a detailed description of our methodology, including the equations and boundary conditions used, the numerical methods, the structure of our grid, the opacity calculations, the chemistry model, and the stellar models used.  In Section \ref{sec:res}, we subject \texttt{HELIOS} to various tests, use it to address several lingering ambiguities\footnote{We describe these issues as ``lingering", because studies in the published literature typically omit the details involved, which prevents us from directly comparing our results to them.} in the literature and also to examine 6 case studies of hot Jupiters.  In Section \ref{sec:sum}, we summarize our results, compare them to previous work and discuss opportunities for future work.

\section{Methodology}
\label{sec:meth}

\subsection{Radiative Transfer Scheme}
\label{sec:rad}

\subsubsection{Preamble}

Any scheme to represent the propagation of radiation through an atmosphere has to solve the radiative transfer equation \citep{ch60,mi70},
\begin{equation}
\mu \frac{\partial I_\lambda}{\partial \tau_\lambda} = I_{\lambda} - S_\lambda,
\label{eq:general}
\end{equation}
where $I_{\lambda}$ is the monochromatic and wavelength-dependent intensity, $\mu \equiv \cos\theta$ is the cosine of the incident angle ($\theta$) relative to the normal and $\tau_\lambda$ is the optical depth measured from the top of the atmosphere downwards. We denote the wavelength by $\lambda$. The crucial ``length" to adopt in radiative transfer is the optical depth.  (Only a non-vanishing $\Delta \tau_\lambda$ leads to a change in intensity $\Delta I_\lambda$.)  The source function $S_\lambda$ accounts for both radiation scattered into the line of sight and the thermal emission associated with each location in the medium. Equation (\ref{eq:general}) is generally difficult to solve, because it is a partial differential equation in $\tau_\lambda$ and $\mu$.

A commonly used simplification is to reduce equation (\ref{eq:general}) to an ordinary differential equation in $\tau_\lambda$ by integrating over the incoming ($-\pi/2 \le \theta \le 0$ or $-1 \le \mu \le 0$) and outgoing ($0 \le \theta \le \pi/2$ or $0 \le \mu \le 1$) hemispheres and assuming that the ratios of various moments of the intensity are constant and take on specific values. This is known as the ``two-stream approximation" \citep{me80}. One may then solve the ordinary differential equation analytically to obtain solutions for \textit{pairs} of model atmospheric layers \citep{he14}. The moments of the intensity are related by the so-called ``Eddington coefficients". Of particular interest to us is the first Eddington coefficient \citep{he14},
\begin{equation}
\epsilon = \frac{1}{{\cal D}},
\end{equation}
which is related to the ``diffusivity factor" ${\cal D}$.  In the next subsection, we show that ${\cal D}$ should take on a value between 1 and 2 depending on the thickness of the atmospheric layers.

In the current study, we use the two-stream solutions previously derived by \cite{he14}.  We note that these solutions allow for the inclusion of non-isotropic scattering via two functions: the single-scattering albedo ($\omega_0$) and the scattering asymmetry factor ($g_0$) \citep{go89,pi10}.  Pure absorption and scattering correspond to $\omega_0=0$ and $\omega_0=1$, respectively.  Forward, backward and isotropic scattering correspond to $g_0=1$, -1 and 0, respectively.  Our formulation allows for $\omega_0$ and $g_0$ to be specified as functions of wavelength/frequency/wavenumber, temperature and pressure.

Hereafter, the term ``flux" describes a wavelength-dependent quantity.\footnote{Accordingly, the units of the flux $F$ are $[F]=$ erg s$^{-1}$ cm$^{-3}$.} Integrating the flux over all wavelengths, one obtains the ``bolometric flux". We also neglect for readability the subscript $\lambda$ for $\tau$ and $B$.

\subsubsection{Exact solution of the radiative transfer equation in the pure absorption limit}

As previously shown by \cite{he14} (and references therein), the radiative transfer equation has an exact solution in the limit of pure absorption ($\omega_0=0$).  We use a staggered grid (see Section \ref{subsect:grid}), such that the two-stream solutions are applied to the \textit{interfaces} of a model layer.  We label the interfaces by ``1" and ``2" and our convention is to locate interace 2 above interface 1 in altitude.  If the layer has only one temperature throughout (i.e., it is isothermal), then the fluxes at the interfaces are given by
\begin{equation}
\label{eq:isodirect}
\begin{split}
F_{2,\uparrow} &=  \mathcal{T} F_{1,\uparrow} + \pi B_1 (1-\mathcal{T}) , \\
F_{1,\downarrow} &=  \mathcal{T} F_{2,\downarrow} + \pi B_1 (1-\mathcal{T}).
\end{split}
\end{equation} 
The $\uparrow$ and $\downarrow$ subscripts refer to the outgoing and incoming fluxes, respectively.  The blackbody intensity within this layer is given by $B_1$.

We can improve upon the isothermal-layer treatment by considering a (linear) temperature gradient within the layer \citep{toon89}.  If we instead Taylor-expand the Planck function in $\tau$ and retain only the constant and linear terms, we obtain
\begin{equation}
\label{eq:nonisodirect}
\begin{split}
F_{2,\uparrow} =& \mathcal{T} F_{1,\uparrow} + \pi B_1 (1 - \mathcal{T}) \\
+& \pi B^\prime \left\{ \frac{2}{3}\left[1-e^{-\Delta \tau} \right] - \Delta \tau \left(1-\frac{\mathcal{T}}{3}\right) \right\} , \\
F_{1,\downarrow} =& \mathcal{T} F_{2,\downarrow} + \pi B_2 (1  - \mathcal{T})  , \\
+& \pi B^\prime \left\{-\frac{2}{3}\left[1-e^{-\Delta \tau} \right] + \Delta \tau \left(1-\frac{\mathcal{T}}{3}\right) \right\}.
\end{split}
\end{equation}
following the derivation in \cite{he14}. The difference in optical depth between the layers is given by $\Delta \tau \equiv \tau_2 - \tau_1$.  The gradient of the Planck function is approximated by
\begin{equation}
\label{eq:planck}
B^\prime \approx \frac{B_2 - B_1}{\tau_2 - \tau_1},
\end{equation}
where $B_1$ and $B_2$ are now the Planck functions for the temperatures at the interfaces 1 and 2, respectively.

\begin{figure}
\begin{center}
\begin{minipage}[t]{0.48\textwidth}
\includegraphics[width=\textwidth]{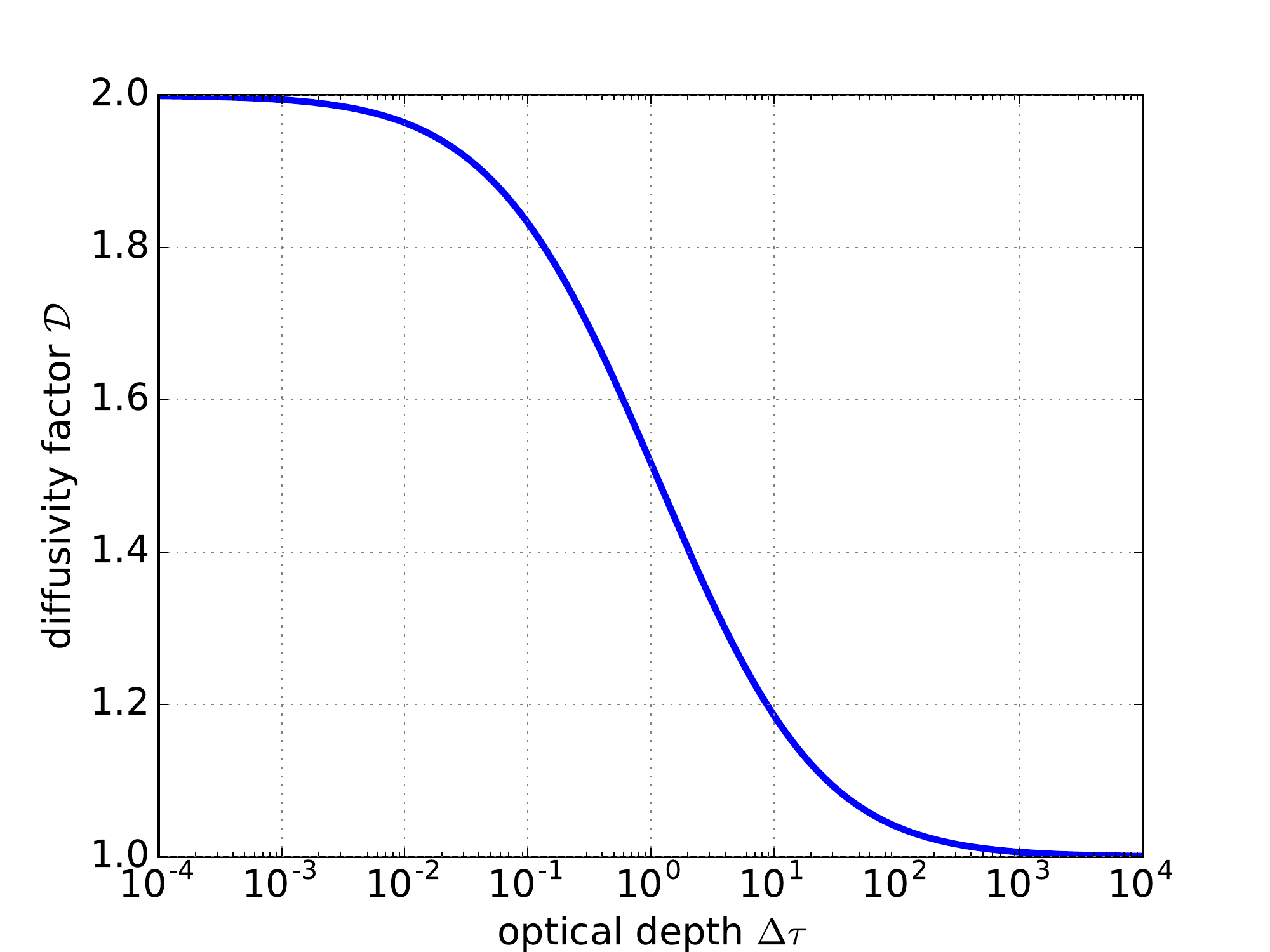}
\end{minipage}
\vspace{-0.3cm}
\caption{Diffusivity factor $\mathcal{D}$, as a function of the difference in optical depth $\Delta \tau$ across a layer, in the limit of pure absorption.}
\label{fig:diff}
\end{center}
\end{figure}

In both the isothermal and non-isothermal cases, the transmission function or transmissivity is
\begin{equation}
\begin{split}
\label{eq:direct_trans}
\mathcal{T} &= 2 \int^1_0 \mu \exp\left(-\frac{\Delta\tau}{\mu}\right) d\mu , \\
&= (1-\Delta \tau)\exp{\left(-\Delta \tau\right)}+\Delta \tau^2 \mathcal{E}_1(\Delta \tau) ,
\end{split}
\end{equation} 
where $\mathcal{E}_1$ is the exponential integral of the first order.  Unlike for the two-stream solutions, there is no need to specify $\mathcal{D}$ as an input, because it has an exact solution,
\begin{equation}
\label{eq:diff}
\mathcal{D} = - \frac{1}{\Delta \tau}\ln{\left[\left(1-\Delta \tau \right)\exp{\left(-\Delta \tau\right)}+\Delta \tau^2 \mathcal{E}_1\left(\Delta \tau \right) \right]} .
\end{equation}
For very thin layers ($\Delta \tau \ll 1$), ${\cal D}=2$ is an accurate approximation, but as the layer becomes optically thick the value of $\cal{D}$ approaches unity (Figure \ref{fig:diff}).  Operationally, since we pick our model grid to be equally spaced in the logarithm of pressure, it means that the value of $\Delta \tau$ is small near the top of the model atmosphere and gradually becomes large (and exceeds unity) at high pressures.  Within the context of the two-stream approximation, assuming ${\cal D}$ to be constant is equivalent to picking a representative or mean value, over the entire atmosphere, of the diffusivity factor.

As already pointed out by \cite{he14}, the analytical expression for ${\cal D}$ when scattering is present (equivalent to eq. \ref{eq:diff}) is unknown.

It is worth emphasizing that equations (\ref{eq:isodirect}) and (\ref{eq:nonisodirect}) are exact solutions and that the two-stream approximation is \textit{not} taken.  In Section \ref{subsect:diffuse}, we compare these exact solutions to the two-stream solutions to derive the value of $\mathcal{D}$.

\subsubsection{Different flavors of two-stream solutions}

We now rederive the two-stream solutions of \cite{he14} without setting ${\cal D}=2$, so as to facilitate comparisons with the exact solutions.  For all of the solutions presented in this subsection, the transmission function is
\begin{equation}
\label{eq:trans}
\mathcal{T} \equiv \exp{\left[- \mathcal{D} \sqrt{(1 - \omega_0 g_0)(1 - \omega_0)} \Delta \tau \right]}.
\end{equation}

The simplest two-stream solutions are derived in the limit of pure absorption and isothermal atmospheric layers,
\begin{equation}
\label{eq:isonoscat}
\begin{split}
F_{2,\uparrow} &=  \mathcal{T} F_{1,\uparrow} + 2\pi\epsilon B_1 (1-\mathcal{T}) , \\
F_{1,\downarrow} &=  \mathcal{T} F_{2,\downarrow} + 2\pi\epsilon B_1 (1-\mathcal{T}).
\end{split}
\end{equation} 
Without scattering ($\omega_0 = 0$), the coupling coefficients are $\zeta_+ = 1$ and $\zeta_- = 0$, and the transmission function simply becomes $ \mathcal{T} = \exp(-\mathcal{D}\Delta \tau)$.  If we increase the sophistication of the model by considering non-isothermal layers and pure absorption, we obtain
\begin{equation}
\begin{split}
F_{2,\uparrow} =& \mathcal{T} F_{1,\uparrow} + 2\pi\epsilon \left[ B_1 - \mathcal{T} B_2 + \epsilon B' (1 - \mathcal{T})  \right] , \\
F_{1,\downarrow} =& \mathcal{T} F_{2,\downarrow} + 2\pi\epsilon \left[ B_2 - \mathcal{T} B_1 - \epsilon B' (1 - \mathcal{T})  \right] .
\end{split}
\end{equation} 

For isothermal atmospheric layers with non-isotropic scattering being included, the two-stream solutions for the fluxes read
\begin{equation}
\label{eq:isoscat}
\begin{split}
F_{2,\uparrow} &= \frac{1}{\alpha} \left[\xi F_{1,\uparrow} - \beta F_{2,\downarrow} + 2\pi \epsilon B_1 (\beta - \upsilon) \right] , \\
F_{1,\downarrow} &= \frac{1}{\alpha} \left[\xi F_{2,\downarrow} - \beta F_{1,\uparrow} + 2\pi \epsilon B_1 (\beta - \upsilon) \right].
\end{split}
\end{equation} 
The coefficients $\alpha$, $\beta$, $\xi$, $\upsilon$ are defined as
\begin{equation}
\label{eq:cap}
\begin{split}
\alpha &\equiv \zeta^2_- \mathcal{T}^2 - \zeta^2_+ ,\\
\beta &\equiv \zeta_+ \zeta_- (1-\mathcal{T}^2) ,\\
\xi &\equiv (\zeta^2_- - \zeta^2_+) \mathcal{T} ,\\
\upsilon &\equiv (\zeta^2_- \mathcal{T} + \zeta^2_+)(1-\mathcal{T}) ,
\end{split}
\end{equation}
with the coupling coefficients being
\begin{equation}
\mathcal{\zeta_\pm} \equiv \frac{1}{2} \left[ 1 \pm \left(\frac{1 - \omega_0}{1 - \omega_0 g_0}\right)^{1/2} \right].
\end{equation}
In the limit of $\omega_0 = 1$, the equations in (\ref{eq:isoscat}) are replaced by
\begin{equation}
\label{eq:purescat}
\begin{split}
F_{2,\uparrow} &= F_{1,\uparrow} - \frac{\mathcal{D} (1 - g_0) \tau_0 (F_{1,\uparrow} - F_{2,\downarrow})}{2 + \mathcal{D}(1 - g_0) \tau_0} , \\
F_{1,\downarrow} &= F_{2,\downarrow} + \frac{\mathcal{D} (1 - g_0) \tau_0 (F_{1,\uparrow} - F_{2,\downarrow})}{2 + \mathcal{D}(1 - g_0) \tau_0} .
\end{split}
\end{equation} 
These solutions give the correct limits of a transparent or opaque atmosphere when $\omega_0=1$ \citep{he14}.  The general solutions stated before in equation (\ref{eq:isoscat}) do not reproduce this limit.

Our most sophisticated two-stream solutions include non-isotropic scattering and non-isothermal model atmospheric layers,
\begin{equation}
\label{eq:nonisoscat}
\begin{split}
F_{2,\uparrow} =& \frac{1}{\alpha} \left\{\xi F_{1,\uparrow} - \beta F_{2,\downarrow} +  2\pi \epsilon \left[ B_1 (\alpha + \beta) - B_2 \xi \right.{} \right.{} \\
&+ \left.{} \left.{} \frac{\epsilon}{1+\omega_0 g_0} B' (\alpha - \xi - \beta)  \right] \right\}, \\
F_{1,\downarrow} =& \frac{1}{\alpha} \left\{\xi F_{2,\downarrow} - \beta F_{1,\uparrow} + 2\pi \epsilon \left[ B_2 (\alpha + \beta) - B_1 \xi \right.{} \right.{} \\
&+ \left.{} \left.{}  \frac{\epsilon}{1+\omega_0 g_0} B' (\xi - \alpha + \beta)  \right] \right\}.
\end{split}
\end{equation} 
Note that in the non-isothermal approach a single constant gradient of $B'$ is assumed within a layer. Thus $B_1$ and $B_2$ are placed at the interfaces.\footnote{In practice, in the numerical implementation of the equations one layer has to be divided into two sublayers (see Sect. \ref{subsect:grid}).} The coefficients $\alpha$, $\beta$ and $\xi$, as well as the coupling coefficients $\zeta_\pm$, retain the same functional forms as in the case of having isothermal layers.

Generally, we find that the non-isothermal solutions attain more rapid numerical convergence (to radiative equilibrium). In principle, if a large enough number of isothermal layers is used, the isothermal and non-isothermal calculations should agree.
 
\subsubsection{Rayleigh scattering}

To include the effects of Rayleigh scattering by molecules, we use the cross section \citep{sn05},
\begin{equation}
\sigma_{\rm scat, \lambda} = \frac{24 \pi^3}{n_{\rm ref}^2 \lambda^4} \left(\frac{n_\lambda^2 - 1}{n_\lambda^2 + 2}\right)^2 K_\lambda,
\end{equation}
where $n_{\rm ref}$ is the number density at a reference temperature and pressure $n_\lambda$ is the wavelength-dependent refractive index and $K_\lambda$ is the King factor, which is a correction factor for polarization.

In the current study, we focus on Rayleigh scattering by hydrogen molecules, but our approach may be straightforwardly generalized to other molecules.  We ignore the contribution due to helium, which is less than 1\% compared to that of molecular hydrogen.  For H$_2$, we use $n_{\rm ref} = 2.68678\times10^{19}$cm$^{-3}$, $K = 1$ and
\begin{equation}
n_\lambda = 13.58\times10^{-5}\left(1+7.52\times10^{-11} \mbox{ cm}^2 ~\lambda^{-2} \right)+1.
\end{equation}
The influence of Rayleigh scattering enters via its inclusion, as $\sigma_{\rm scat,\lambda}/\bar{m}$, to the opacity of each model layer, where $\bar{m}$ is the mean molecular mass, and also via the single-scattering albedo $\omega_0$. The dashed line in Figure \ref{fig:opac} shows the opacity of Rayleigh scattering by H$_2$, which dominates in the optical but becomes subdominant, compared to molecular absorption, in the infrared due to its dropoff with $\lambda^{-4}$.

If the scattering dominates and $(1- \omega_0) < 10^{-6}$ in this layer and waveband, then we switch to the pure scattering solutions (eq. \ref{eq:purescat}).

\subsection{Numerical Method}
\label{sec:numerical}

\subsubsection{Model grid}
\label{subsect:grid}

For the isothermal treatment, a staggered grid is used with the layers being separated by interfaces.  There are $n$ layers and $n+1$ interfaces.  The grid is evenly spaced in height or the logarithm of pressure, which serves as the vertical coordinate.  The thickness of the $i$-th layer is given by 
\begin{equation}
\Delta z_i = \frac{k_{\rm B} T_i}{\bar{m}g}\ln \left(\frac{P_{i,{\rm inter}}}{P_{i+1,{\rm inter}}}\right),
\end{equation}
with $k_{\rm B}$ being the Boltzmann constant, $g$ the surface gravity.  For hydrogen-dominated atmospheres, we set $\bar{m}=2.4 m_p$ with $m_p$ being the mass of the proton. The pressures at the interfaces are represented by $P_{i,{\rm inter}}$ and $P_{i+1,{\rm inter}}$.  The preceding expression is obtained from integrating the equation of hydrostatic balance over a model layer and assuming isothermality and the equation of state for an ideal gas.

The contribution to the optical depth\footnote{To be pedantic, the optical depth is a coordinate.  It is the difference in optical depth that is needed for radiative transfer.  The analogy is to distance versus displacement.} by the $i$-th layer is
\begin{equation}
\Delta\tau_{i} = \Delta m_{{\rm col},i} \kappa_i = \frac{P_{i,{\rm inter}} - P_{i+1,{\rm inter}}}{g} \kappa_i,
\label{eq:layertau}
\end{equation}
where $\kappa_i$ is the opacity and $\Delta m_{{\rm col},i}$ is the difference in column mass, which can be further written in terms of pressure and surface gravity.

\begin{figure}
\begin{center}
\begin{minipage}[t]{0.4\textwidth}
\includegraphics[width=\textwidth]{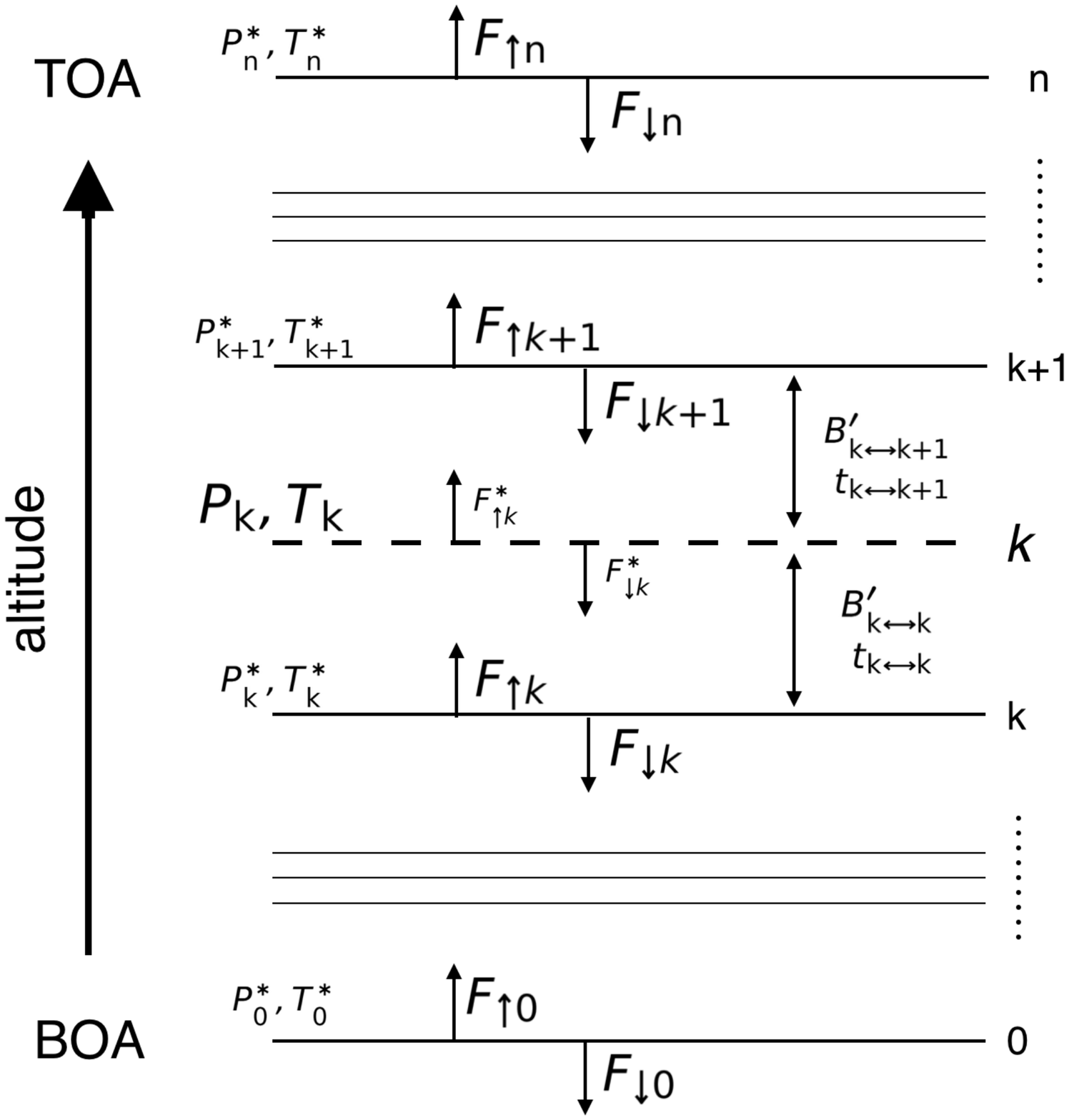}
\end{minipage}
\vspace{-1.3cm}
\caption{Staggered grid used for models with non-isothermal layers.  The boundary conditions are applied at the top (stellar irradiation) and bottom (internal heat flux) of the model atmosphere, which are also the $n$-th and 0th interfaces, respectively.  The pressure and temperature are located at the center of each layer, while the fluxes tranversing a layer are computed at the layer interfaces.  We further divide each layer into two sublayers during the iteration for radiative equilibrium (see text for details).  In the schematic, we have focused on the $k$-th layer and the various quantities associated with its center and interfaces.  Quantities marked with an asterisk are temporarily used in the computation, but not stored as the final output.  The layers are evenly spaced in the logarithm of pressure.}
\label{fig:grid}
\end{center}
\end{figure}

For the non-isothermal grid, we require a more sophisticated grid layout, which is shown in Figure \ref{fig:grid}.  Each layer has a temperature and pressure, located at its center.  To compute the fluxes, we need to interpolate across the temperature and pressure grids to obtain their values at the interfaces.  A key quantity to compute is the Planck function $B$, which relates the temperature to the thermal emission of a layer. If one constructs the grid using a single gradient $B{^\prime}$ of the Planck function over the whole layer, one is essentially decoupling the radiative transfer process from the temperature at the center of the layer. We solve this problem by splitting each layer into two sublayers, leading to two $B{^\prime}$ values within a layer. The fluxes are propagated first from the lower interface to the layer center, then from the layer center to the upper interface (and vice versa), similar to the approach taken in e.g. \cite{mendonca2015}. In this manner, both the layer centers and interfaces are involved in the iteration for radiative equilibrium.

Finally, in the non-isothermal grid, a numerical caveat arises in the upper atmosphere. There, the optical depth difference $\Delta\tau_{i}$ of a layer $i$ is tiny (due to the very small pressure) and thus the denominator of eq. (\ref{eq:planck}) vanishes, which in turn leads to numerical issues for $B^\prime$ in eq. (\ref{eq:nonisoscat}). To prevent this, we keep the sub-layered grid of the non-isothermal approach, but switch in each sublayer from the non-isothermal (eq. \ref{eq:nonisoscat}) to the isothermal prescription (eq. \ref{eq:isoscat}) whenever $\Delta\tau_{i} < 10^{-4}$ occurs.

\subsubsection{Boundary conditions}
\label{sec:bound}

At the top of the atmosphere (TOA), which is also the $n$-th interface of the model atmosphere, the flux is given by
\begin{equation}
F_{n, \downarrow} = f\left(\frac{R_\star}{a}\right)^2 \pi B_\star ,
\end{equation}
where $R_\star$ is the stellar radius, $a$ is the orbital distance of the planet and $B_\star$ is the stellar blackbody function.  This represents the heating from the incident stellar flux.  Most of the quantities in the preceding expression are astronomical observables (or quantities that may be inferred from the observations).  It is possible to replace $B_\star$ by a more sophisticated model of the stellar spectrum (see Section \ref{sec:star}).

The quantity $f$ is a parameter that describes the redistribution of heat from the dayside to the nightside of a tidally-locked hot Jupiter, which is dictated by an interplay between atmospheric dynamics and radiative cooling.  In principle, its value may be inferred from infrared phase curves. Theoretically, it is bounded between $f=1/4$ (full redistribution) and $f=1$ (no redistribution). Since we are using our 1D, plane-parallel model to describe the dayside emission spectra of hot Jupiters, the value of $f$ is a proxy for the dayside integrated absorption and re-emission of radiation. In the current study, we adopt $f=2/3$ following the arguments in e.g. \cite{bu08} and \cite{sb10}.

At the bottom of the model atmosphere (BOA), we have included the option to specify an internal radiative heat flux ($\pi B_{\rm intern}$), such that 
\begin{equation}
\int \pi B_{\rm intern} ~d\lambda= \sigma_{\rm SB} T_{\rm intern}^4, 
\end{equation}
where $\lambda$ is the wavelength, $\sigma_{\rm SB}$ is the Stefan-Boltzmann constant, $T_{\rm intern}$ is the internal temperature, $B_{\rm intern} \equiv B(T_{\rm intern})$ and $B$ is the Planck function. The internal heat flux reflects the thermal heating due to gravitational contraction.  The BOA is also the 0-th interface.  It is important to note that any form of atmospheric heating is associated with the \textit{net flux} (the difference between the outgoing and incoming fluxes) \citep{he14},
\begin{equation}
\pi B_{\rm intern} = F_{0, \uparrow} - F_{0, \downarrow}.
\end{equation}
In our current study, we set $T_{\rm intern}=0$ K in the absence of such constraints on hot Jupiters.

\subsubsection{Iterating for radiative equilibrium}
\label{sec:iter}

Within each model layer of the atmosphere, its temperature and pressure determine its absorption and scattering properties, given by molecular abundances and opacities, which in turn determines the transmission function and fluxes.  However, as flux enters and exits the layer, the temperature changes, which in turn changes the opacity.  Clearly, this is an iterative process.  It turns out that one is iterating for radiative equilibrium, which is a statement of \textit{local} energy conservation \citep{he14}.  Local energy conservation implies global energy conservation, but not vice versa \citep{he16b}.

We integrate the fluxes ($F_\uparrow$ and $F_\downarrow$) over the entire spectral range to obtain the bolometric fluxes ($\mathcal{F}_\uparrow$ and $\mathcal{F}_\downarrow$), which in turn allows us to construct the bolometric net flux ($\mathcal{F}_- \equiv \mathcal{F}_\uparrow - \mathcal{F}_\downarrow$).  For the $i$-th layer, the divergence\footnote{In 1D, the divergence is simply the vertical gradient.} of the bolometric net flux becomes
\begin{equation}
\frac{ \Delta \mathcal{F}_{i,-}}{\Delta z_i} = \frac{(\mathcal{F}_{i+1,\uparrow} - \mathcal{F}_{i+1,\downarrow}) - (\mathcal{F}_{i,\uparrow} - \mathcal{F}_{i,\downarrow})}{\Delta z_i} .
\end{equation}
Between successive timesteps, the change in temperature of the $i$-th layer then becomes \citep{he14}
\begin{equation}
\label{eq:step}
\Delta T_i = \frac{1}{\rho_i c_{\rm p}}\frac{ \Delta \mathcal{F}_{i,-}}{\Delta z_i} \Delta t_i , 
\end{equation}
where $\rho_i$ is the local density and $\Delta t_i$ is the numerical timestep.  The specific heat capacity of an ideal gas at constant pressure is \citep{pi10}
\begin{equation}
c_{\rm p} = \frac{2+n_{\rm dof}}{2\bar{m}}k_{\rm B} ,
\end{equation}
where we set the number of degrees of freedom of the gas to be $n_{\rm dof} = 5$, as is valid for a diatomic molecule (ignoring the vibrational modes) like molecular hydrogen, the main component of gas planet atmospheres. This simplification does not hold should the atmospheric composition change, e.g. by dissociation of molecular hydrogen at very high temperatures. Since in our model the only occurence of $c_{\rm p}$ is in the timestepping algorithm this flaw is for our cause of only minor concern, but would render e.g. the calculation of the entropy inaccurate.
 
In practice, we start with an arbitrary temperature profile with $\Delta \mathcal{F}_{i,-} \neq 0$.  We perform the iteration described until $\Delta \mathcal{F}_{i,-}$ vanishes for each layer, which is our numerical condition for radiative equilibrium (c.f. eq. \ref{eq:condition}).  Physically, each atmospheric layer emits the same amount of energy which it receives.

When scattering is present, the flux solutions become coupled.   Each array of outgoing or incoming fluxes cannot be populated independently of the other.  This is solved iteratively by starting with the coupled dependencies as zero and populating the flux arrays multiple times in each temperature step. We include 4 additional scattering iterations in the full radiative transfer calculation as the flux values are still known from the previous timestep and 80 scattering iterations for pure post-processing (propagating only once through the atmosphere) purposes.

\subsubsection{Numerical timestepping}

For the numerical timestepping, there are two options in \texttt{HELIOS}.  The first option uses a fixed and uniform timestep ($\Delta t$) for every model layer.  Typically, we choose $10^2 \lesssim \Delta t \lesssim 10^4$ s.  The challenge is that $\Delta T \propto \rho^{-1}$ and $\rho$ may vary by several orders of magnitude across our model atmosphere.  With a uniform timestep, the upper layers of the atmosphere attain convergence much more rapidly than the lower atmosphere.  Thus, this approach is plausible and technically correct, but infeasible.

A more efficient approach is to implement an adaptive timestepping scheme that uses a different timestep for each model layer.  Specifically, the timestep in the $i$-th layer is related to the radiative timescale ($t_{i,\rm rad}$), 
\begin{equation}
\Delta t_i = f_{i,\rm pre} ~t_{i,\rm rad} ,
\end{equation}
where $f_{i,\rm pre}$ is a pre-factor to adjust to the optimal value of $\Delta t_i$. The radiative timescale is approximated by
\begin{equation}
t_{i,\rm rad} \approx \frac{c_{\rm p} P_i}{\sigma_{\rm SB} g T_i^3},
\end{equation}
where the temperature and pressure of the $i$-th layer is given by $T_i$ and $P_i$, respectively.  With this improved timestepping scheme, the timestep becomes larger as one goes deeper into the model atmosphere.  The evolution of the model does not strictly correspond to a physical evolution, but is rather a convenient way of reaching a numerical steady state.  

To further optimize the efficiency of \texttt{HELIOS}, we also allow the timestep to vary in time as the model approaches radiative equilibrium.  Specifically, the algorithm checks in each layer whether the temperature has oscillated for the most recent 6 successive timesteps.  We find oscillations in temperature to be a robust and practical indicator of having adopted too large a timestep.  If oscillations are detected, the timestep is reduced by 33\%.  By contrast, if no oscillations are detected (i.e., the change in temperature is monotonic), then the timestep is increased by 10\%.

We note that the purpose of the pre-factor ($f_{i, {\rm pre}}$) is to dampen sudden spikes in $\Delta \mathcal{F}_{i,-}$.  For practical purposes, it takes the form of
\begin{equation}
f_{i, {\rm pre}} = \frac{10^5}{\left[|\Delta \mathcal{F}_{i,-}|/\left({\rm erg \; s}^{-1} {\rm cm}^{-2}\right)\right]^{0.9} },
\end{equation}
which leads the temperature iteration step $\Delta T_i$ to depend only on $\Delta \mathcal{F}_{i,-}^{0.1}$, which guarantees the correct direction of the evolution but substantially smoothes irregularities, making the iteration process significantly more stable.

Finally, we need a condition to judge if radiative equilibrium has been established.  Usually, one would assume a criterion demanding the rate of temperature change to be below a certain threshold, $\Delta T/\Delta t < \delta_{\rm limit}$, and evaluate whether this is satisfied in every layer. However, if $\Delta t$ is variable and not representing a physical time, then the utility of this approach becomes suspect.  Instead of setting a threshold on the consequence of radiative equilibrium (changes in temperature), we set one on its cause (a vanishing bolometric flux divergence).  We use the dimensionless convergence criterion,
\begin{equation}
\label{eq:condition}
\frac{\Delta \mathcal{F_-}}{\sigma_{\rm SB} T^4} < 10^{-7},
\end{equation}
where the change in bolometric net flux is normalized by the thermal emission associated with each layer.  In practice, this criterion results in changes in temperature of less than 4 K at the BOA and less than 1 K in the photospheric regions, which impacts the emission spectrum by less than 0.5\%.

\subsection{Calculating Opacities and Transmission Functions}
\label{sec:opac}

\begin{figure}
\begin{center}
\begin{minipage}[t]{0.48\textwidth}
\includegraphics[width=\textwidth]{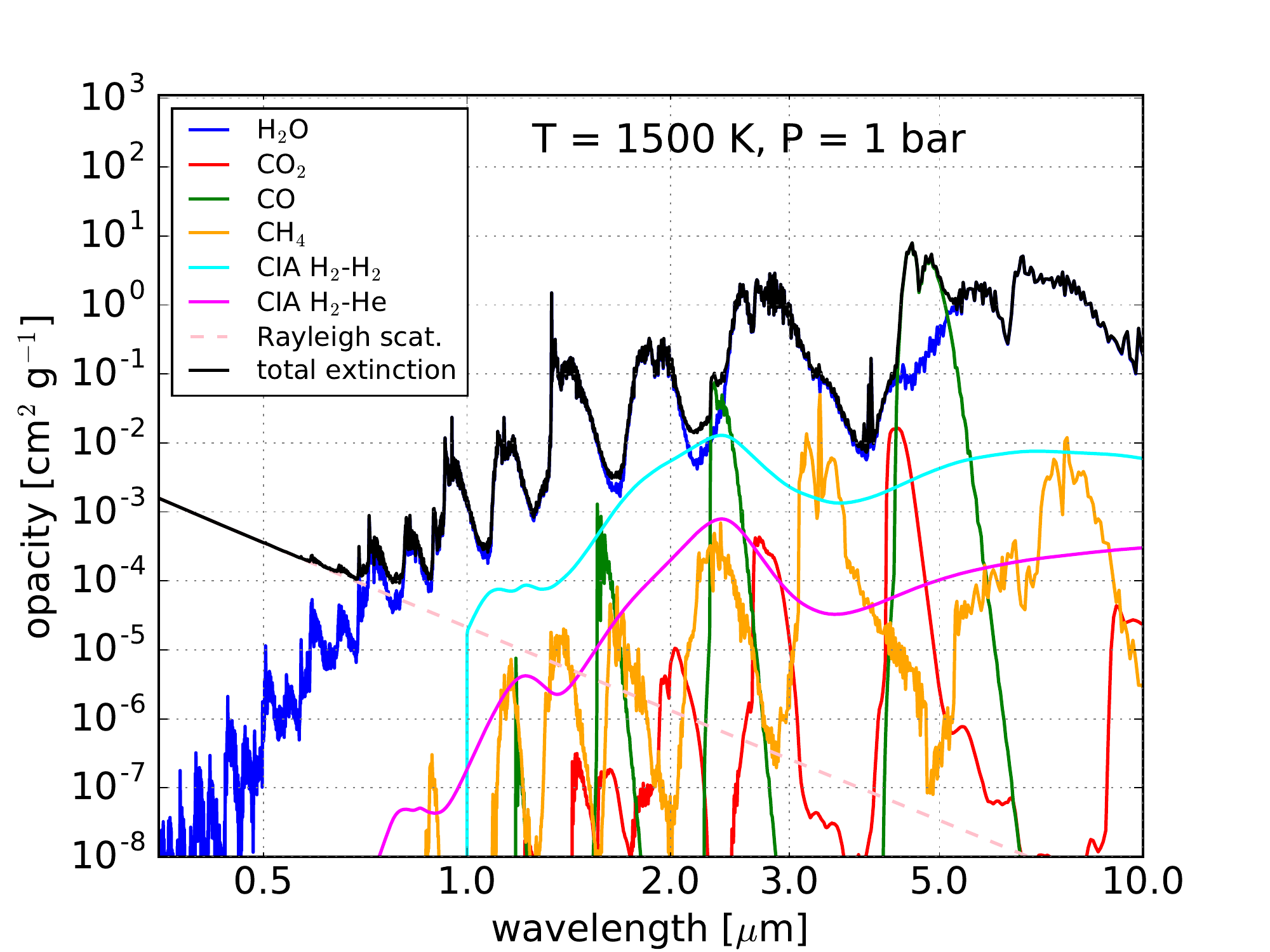}
\end{minipage}
\vspace{-0.3cm}
\caption{Opacities, as functions of wavelength, for all of the opacity sources used in the current study, computed using \texttt{HELIOS-K} \citep{gr15}.  For illustration, we set $T=1500$ K and $P=1$ bar.  Each opacity is weighted by its mass mixing ratio.  We include only Rayleigh scattering by molecular hydrogen, but CIA associated with both H$_2$-H$_2$ and H$_2$-He pairs.}
\label{fig:opac}
\end{center}
\end{figure}

Our method for computing the opacities (cross sections per unit mass) of molecules has previously been elucidated in \cite{gr15}, who published an opacity calculator named \texttt{HELIOS-K} that is part of the \texttt{HELIOS} radiation package.  As such, we do not repeat the detailed explanations of \cite{gr15} and instead highlight only the salient points.  We include the opacities associated with the four main infrared absorbers: H$_2$O, CO$_2$, CO and CH$_4$.  We also include the opacities associated with the collision-induced absorption (CIA) of H$_2$-H$_2$ and H$_2$-He pairs.  Table \ref{tab:opac} states the spectroscopic line lists used to compute our opacities, while Figure \ref{fig:opac} displays the final weighted opacities used in the code at one temperature and pressure.\footnote{The reader should be aware that, in this first version of \texttt{HELIOS}, we omit greenhouse gases like NH$_3$, HCN, C$_2$H$_2$ and the alkali metals Na and K, which may have an impact on the atmospheric structure.  H and H$^-$ absorption may also be important at high temperatures.  Nevertheless, our starting set of four molecules is sufficient for us to build up the first version of a radiative transfer code, and we intend to augment this set in the future.}

The first step involves calculating the opacity function (cross section per unit mass as a function of wavelength, temperature and pressure), which includes all of the molecules previously mentioned, at a given spectral resolution.  If the spectral resolution is too coarse, then spectral lines may be missed or omitted, which leads to an under-estimation of the true opacity.  To avoid this pitfall, we use a resolution of $10^{-5}$ cm$^{-1}$.  Since the wavenumber range goes up to $\sim 10^4$ cm$^{-1}$, this means that we are sampling the opacity function at $\sim 10^9$ points, which approaches a true line-by-line calculation.

The shape of each spectral line is described by a Voigt profile.  A major uncertainty associated with this approach, which remains an unsolved physics problem, is that the far line wings of the Voigt profile over-estimate or under-estimate the true opacity contribution depending on the molecule (see \citealt{gr15} for a discussion).  The common practice is to truncate each Voigt profile at some fixed spectral width.  For example, \cite{sh07} use a line-wing cutoff of $\mbox{min}(25P/1\mbox{ atm}, 100)$ cm$^{-1}$.  We use a cutoff of 100 cm$^{-1}$ except for water, where we instead use 25 cm$^{-1}$.  We emphasize that the correct functional form of these far line wings is unknown.

To speed up our calculations, we wish to avoid having to deal with integrating over $\sim 10^9$ points in the opacity function to obtain the transmissivities.  Instead, we employ the $k$-distribution method to calculate the transmission function within each wavelength bin,
\begin{equation}
\mathcal{T} = \int^1_0 \psi ~dy,
\label{eq:trans_gen}
\end{equation}
where the integrand, which is given by $\psi \equiv \exp{(-\mathcal{D} \Delta \tau)}$, is a function of a new variable ($y$) that is bounded between 0 and 1.  We refer the reader to \cite{gr15} for a detailed explanation of the $k$-distribution method and instead focus on our method for numerically evaluating the preceding integral, which we solve by applying the Gauss-Legendre quadrature rule,
\begin{equation}
\int^1_0 \psi ~dy = \frac{1}{2} \sum^{20}_{g=1} w_g ~\psi\left(\frac{1 + y_g}{2} \right),
\end{equation}
where $y_g$ is $g$-th root of the 20-th order Legendre polynomial $P_{20}$. The corresponding Gaussian weight $w_g$ is \citep{abramowitz72}
\begin{equation}
w_g = \frac{2}{[1-y_g^2]P_{20}^\prime[y_g]^2},
\end{equation}
with $P_{20}^\prime$ being the derivative of $P_{20}$.  We find that using a 20th order Gaussian quadrature rule is sufficient by comparing our calculations to direct integration using Simpson's rule (not shown).

The obvious advantage of using Gaussian quadrature over direct integration is the enhanced computational efficiency.  In \texttt{HELIOS}, we propagate the fluxes through the model atmosphere for each of the 20 Gaussian points and perform the Gaussian quadrature sum at the end of the propagation to obtain the flux associated with a wavelength bin.  Since the fluxes follow inhomogeneous paths across pairs of layers (i.e., the temperatures and pressures are not constant along these paths) and we also add the $k$-distribution functions of the various molecules, we have to invoke the correlated-$k$ approximation \textit{twice} \citep{gr15}.  

Computing the flux through each Gaussian point is equivalent to expressing the transmission function through layer $i$ and waveband $l$ by

\begin{equation}
\mathcal{T}_{i,l} = \sum^{20}_{g = 1} w_g e^{-\kappa_{i,l,g} \Delta m_{{\rm col},i}} ,
\label{eq:trans_sum}
\end{equation}
which is nothing else than a discrete form of equation (\ref{eq:trans_gen}) applied to our model. The $g$-th $k$-coefficient in waveband $l$ is written as 

\begin{equation}
\kappa_{i,l,g} = \sum^6_{j=1} \mathcal{X}_j\left(T_i, P_i \right) ~\kappa_{j,l,g}\left(T_i, P_i \right),
\label{eq:kappa_gauss}
\end{equation}
where $T_i$ and $P_i$ are the temperature and pressure at the center of the $i$-th layer in the isothermal layer grid and also at the interfaces in the non-isothermal layer grid where we have sublayers. In the latter case, we calculate the opacity in the center and at the interface and take their average to obtain the value in the connecting sublayer. The mixing ratios and opacities are generally functions of temperature and pressure.  At this point, we have to distinguish between the mixing ratios by volume ($X_j$) versus the mixing ratios by mass ($\mathcal{X}_j$). The chemistry formulae (see Sect. \ref{sec:chem}) are constructed to compute $X_j$.  However, to construct $\kappa_i$ we need
\begin{equation}
\label{eq:mixing}
\mathcal{X}_j = \frac{X_j m_j}{\bar{m}},
\end{equation}
where $m_j$ is the mass of the $j$-th molecule.

In equations (\ref{eq:kappa_gauss}) and (\ref{eq:mixing}), the indices $j=1,2,3,4$ refer to the 4 molecules being included in the current study: CO, CO$_2$, H$_2$O and CH$_4$.  For these molecules, $X_j$ is computed using the chemistry model.  The indices $j=5$ and $j=6$ refer to the CIA opacities associated with H$_2$-H$_2$ and H$_2$-He, respectively.  For these, we use $X_5=1$ and $X_6=0.1$ to approximately reflect cosmic abundance.  We use $m_5 = 2 m_p$ and $m_6 = 4 m_p$.

By using equation (\ref{eq:kappa_gauss}), we inherently assume the spectral lines of the various molecules to be {\it perfectly correlated}.  In general, there are three limits: perfectly correlated, randomly overlapping (perfectly uncorrelated) and disjoint lines (see \citealt{pi10} for  a review).  Real spectral lines behave in a way that is intermediate between these limits.  \cite{lacis91} and, more recently, \cite{amundsen16} have implemented a randomly overlapping method for combining the opacities of the different molecules, which is computationally more expensive as it involves multiple summations.  As the spectral resolution increases (and the bin size decreases), these approaches should converge to the same answer.  The true accuracy of these approaches remains unquantified in the hot atmosphere regime and needs to be tested by a true line-by-line calculation, where each of the $\gtrsim 10^9$ line shapes is numerically resolved.  This is the subject of future work and is beyond the scope of the present paper.

In \texttt{HELIOS}, the $k$-coefficients are read in from a four-dimensional, pre-computed table in temperature ($100 \le T \le 2900$ K, $\Delta T = 200$ K), pressure ($10^{-6} \le P \le 10^3$ bar, $\Delta \log_{10} P$ = 0.5)\footnote{If the layer pressure or temperature exceeds the range of the values in the table, the opacity is simply taken to correspond to the closest pre-tabulated value.} and wavelength  ($0.33 \le \lambda \le 10^5$ $\mu$m), with the bins subdivided by 20 Gaussian points.  The opacities are used at the constructed wavelength (and Gaussian point) values, but are linearly interpolated across $T$ and $\log{P}$.

Finally, we note that we use 300 wavelength bins (equally spaced in wavenumber) when running \texttt{HELIOS} to solve for radiative equilibrium.  Upon obtaining the converged temperature-pressure profile, we then use it to compute synthetic spectra in 3000 wavelength bins as a post-processing step.  We find that this approach produces essentially identical results to performing the entire calculation using 3000 wavelength bins (not shown).

\begin{table*}
	\caption{Opacity Sources used in this work.}
	\label{tab:opac}
	\vspace{-0.4cm}
\begin{center}
\bgroup
\def\arraystretch{1.5}
  \begin{tabular}{| l | l |}
    \hline
Name & Source \\ \hline
H$_2$O &  HITEMP database\footnote{\label{f:hitran}hitran.org/hitemp/} \citep{ro10}  \\
CO$_2$ &  HITEMP database\textsuperscript{\ref{f:hitran}} \\
CO &   HITEMP database\textsuperscript{\ref{f:hitran}} \\
CH$_4$ &  HITRAN database\footnote{www.cfa.harvard.edu/hitran/} \citep{ro13} \\
CIA  &  HITRAN CIA database \citep{ri12} \\
Rayleigh scattering & \citealt{sn05} \\
    \hline
  \end{tabular}
	\egroup
	\end{center}
\end{table*}

\subsection{Chemistry Model}
\label{sec:chem}

\begin{figure}
\begin{center}
\begin{minipage}[t]{0.48\textwidth}
\includegraphics[width=\textwidth]{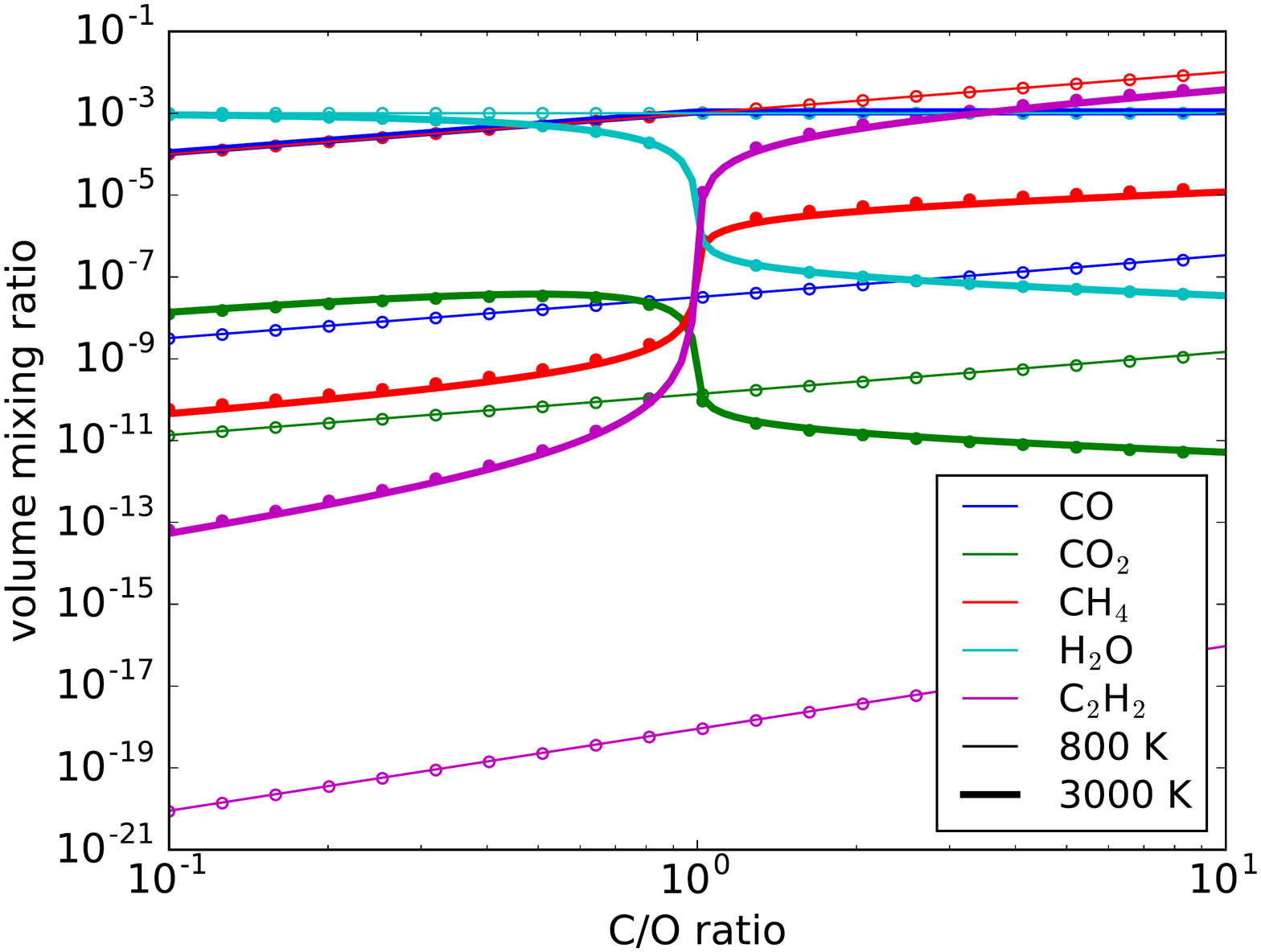}
\end{minipage}
\vspace{-0.3cm}
\caption{Validation of our analytical chemistry model (\citealt{he16b}; circles) by calculations using the Gibbs free energy minimization code, \texttt{TEA} (\citealt{bl15}; solid curves).  For illustration, we have computed the volume mixing ratios as functions of C/O and examined $P=1$ bar and $T=800$ and 3000 K.}
\label{fig:tea}
\end{center}
\end{figure}

Given the elemental abundances of carbon ($n_{\rm C}$) and oxygen ($n_{\rm O}$), we would like to compute the mixing ratios (number densities normalized by that of molecular hydrogen) of the 4 molecules used in our model as functions of temperature and pressure.  This requires a chemistry model. To this end, we use the analytical calculations of \cite{he16b}.  Specifically, \cite{he16a} laid out the theoretical formalism, which led to the formulae in equations (12), (20) and (21) in \cite{he16b} that we are using. \cite{he16c} demonstrated that these formulae are accurate compared to a Gibbs free energy minimization code, even when nitrogen is added to the system.  We explicitly demonstrate the agreement between equations (12), (20) and (21) of \cite{he16b} and the calculations from the \texttt{TEA} code of \cite{bl15} in Figure \ref{fig:tea}.  Since we do not study atmospheres with C/O $>1$, we ignore C$_2$H$_2$.

We define ``solar element abundance" to be $n_{\rm C} = 2.5 \times 10^{-4}$ and $n_{\rm O} =5 \times 10^{-4}$, such that ${\rm C/O} \equiv n_{\rm C}/n_{\rm O} = 0.5$.  In this study, we keep the value of $n_{\rm O}$ fixed and vary $n_{\rm C}$ when we vary C/O.  For example, a model with ${\rm C/O} = 0.1$ has $n_{\rm C} = 5 \times 10^{-5}$ and $n_{\rm O} = 5 \times 10^{-4}$.

Following the convention of the astronomers, we refer to the ``metallicity" as the set of values of the elemental abundances that have atomic numbers larger than that of helium.  In our model, these are $n_{\rm C}$ and $n_{\rm O}$.  These numbers are simply decreased or increased by a constant factor when the metallicity is varied.  For example, a model with $3\times$ solar metallicity has $n_{\rm C} = 7.5 \times 10^{-4}$ and $n_{\rm O} = 1.5 \times 10^{-3}$, but still retains ${\rm C/O} =0.5$.

\begin{figure}
\begin{center}
\begin{minipage}[t]{0.48\textwidth}
\includegraphics[width=\textwidth]{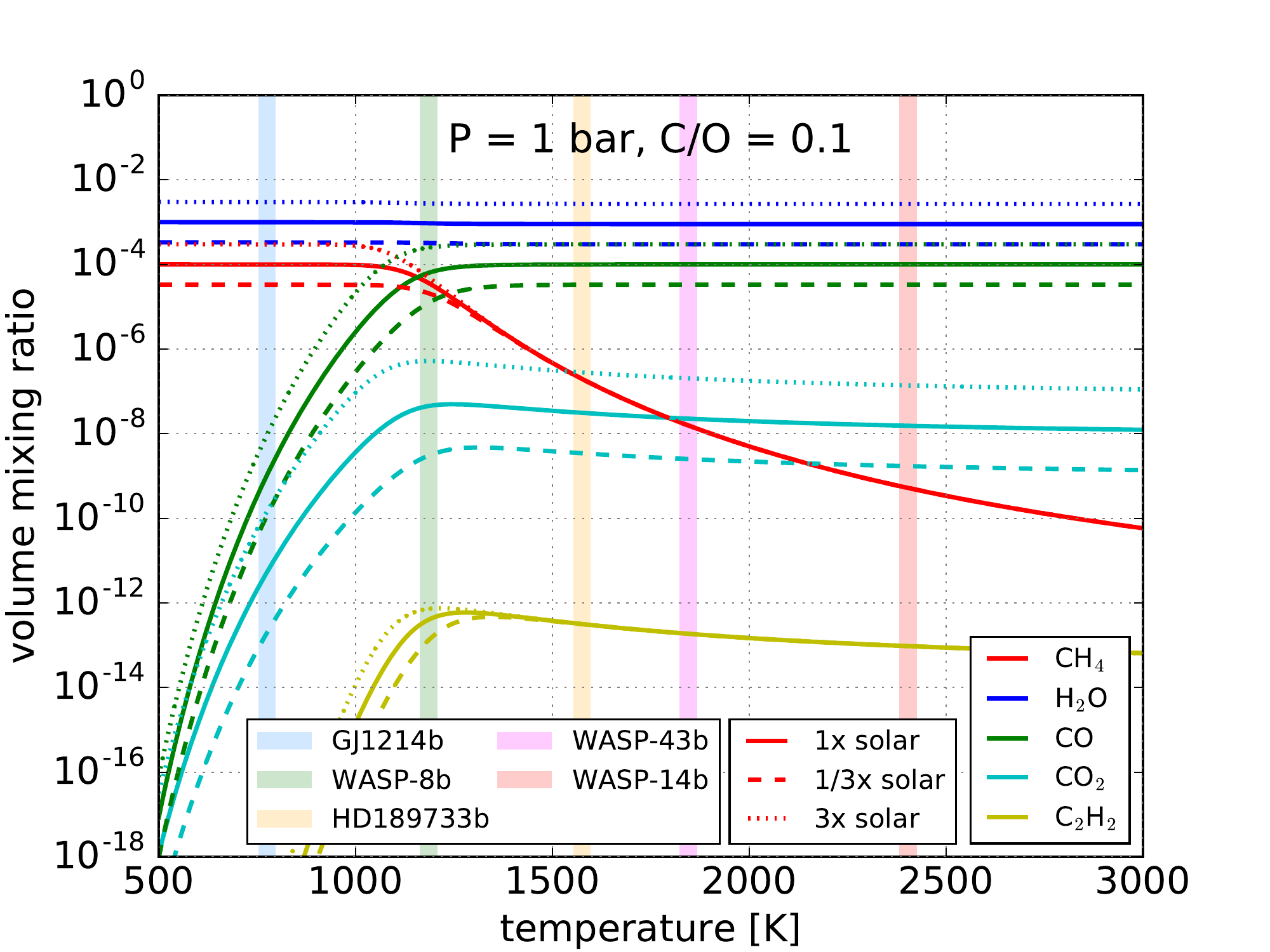}
\end{minipage}
\vspace{-0.1cm}
\begin{minipage}[t]{0.48\textwidth}
\includegraphics[width=\textwidth]{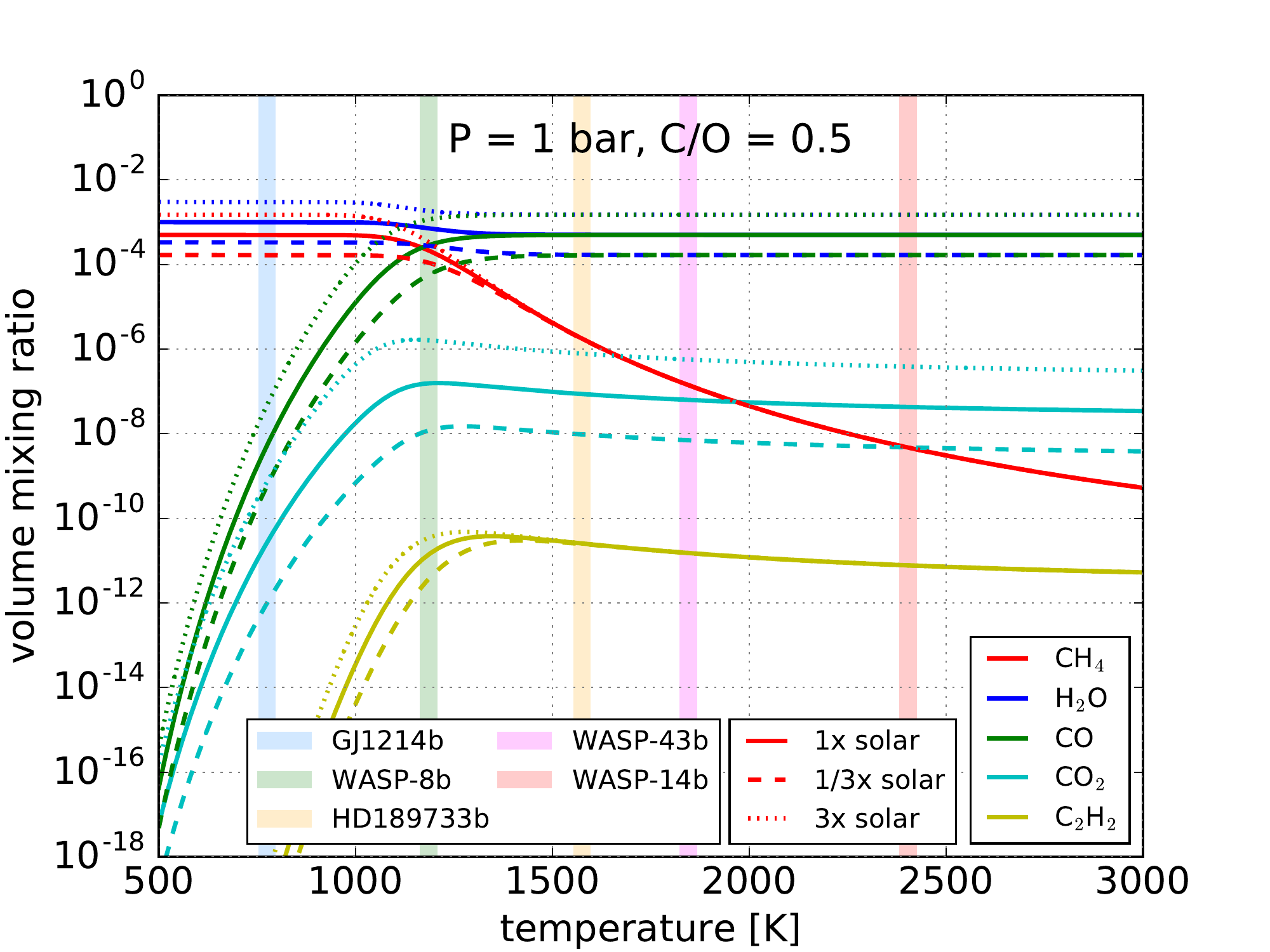}
\end{minipage}
\vspace{-0.1cm}
\begin{minipage}[t]{0.48\textwidth}
\includegraphics[width=\textwidth]{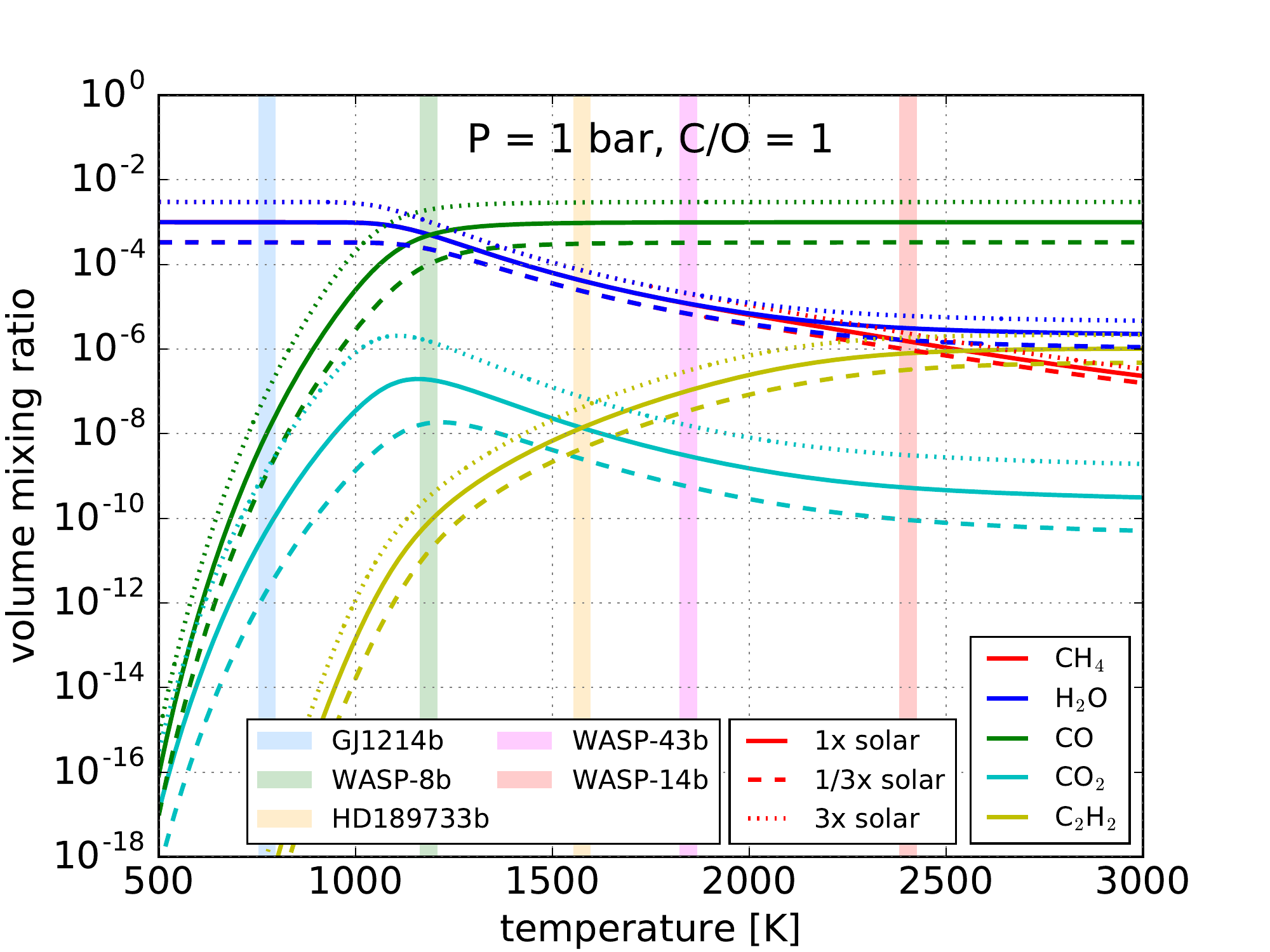}
\end{minipage}
\vspace{-0.1cm}
\caption{Elucidating the temperature dependence of the volume mixing ratios of the molecules used in the current study.  For illustration, we set $P=1$ bar and explore C/O$=0.1$ (top panel), C/O$=0.5$ (middle panel) and C/O$=1$ (bottom panel).  Within each panel, we explore the effects of varying the metallicity by $1/3\times$ and $3\times$ the solar value.}
\label{fig:abundances}
\end{center}
\end{figure}

Figure \ref{fig:abundances} shows examples of our calculations of the molecular mixing ratios as functions of temperature, C/O and metallicity.  To develop some intuition for the relative abundances of molecules present in our model atmospheres, we have included shaded columns indicating the dayside-averaged temperatures of 5 of the 7 exoplanets being studied in the current paper.\footnote{Two planets are hotter than 3000 K and not visible in Fig. \ref{fig:abundances}. As we have tabulated Gibbs free energies only up to 3000 K, we assume the chemistry to be that at 3000 K if the temperatures exceed 3000 K.}

\subsection{Stellar Models}
\label{sec:star}

\begin{figure*}
\begin{center}
\begin{minipage}[t]{0.48\textwidth}
\includegraphics[width=\textwidth]{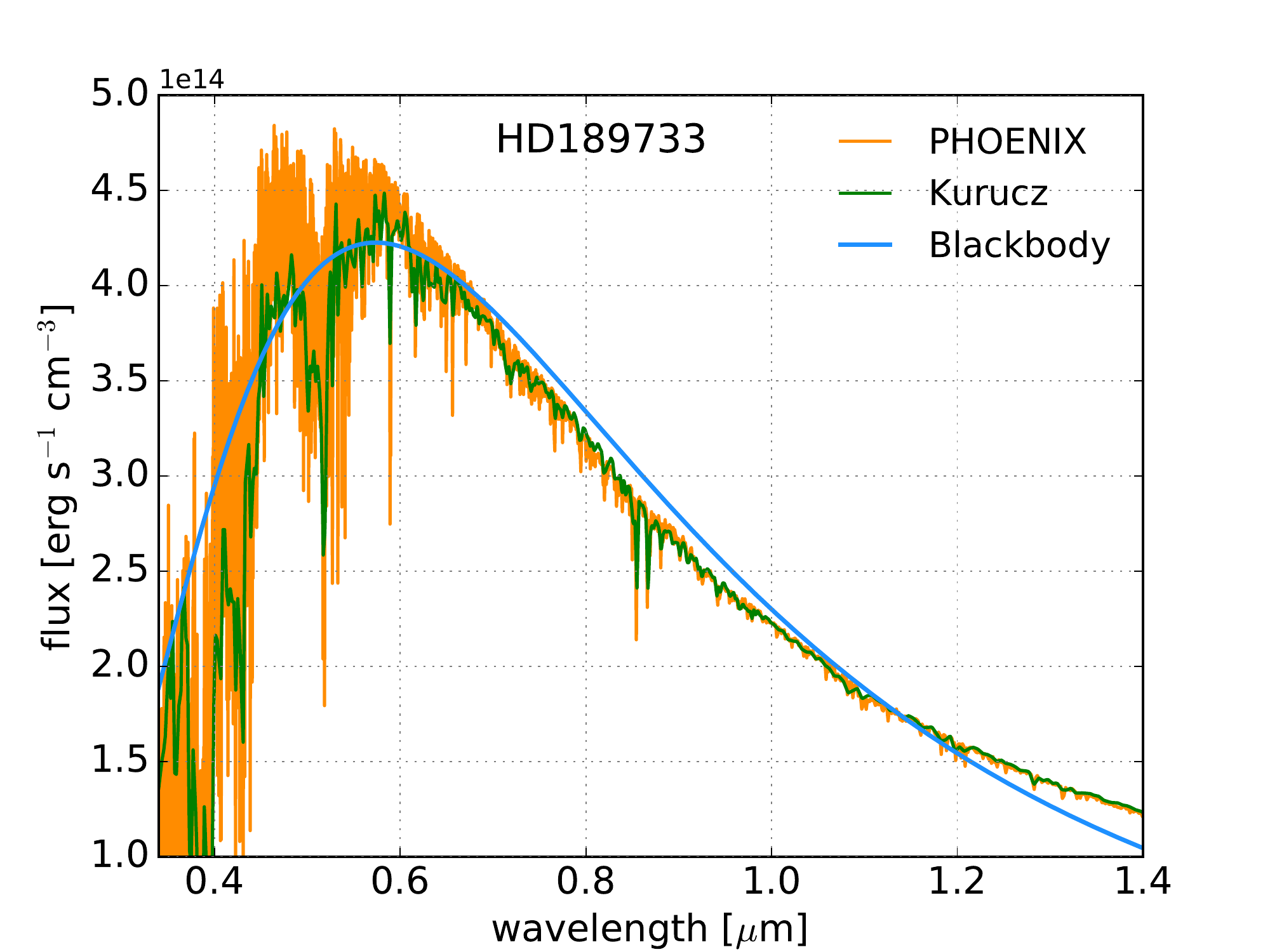}
\end{minipage}
\hfill
\begin{minipage}[t]{0.48\textwidth}
\includegraphics[width=\textwidth]{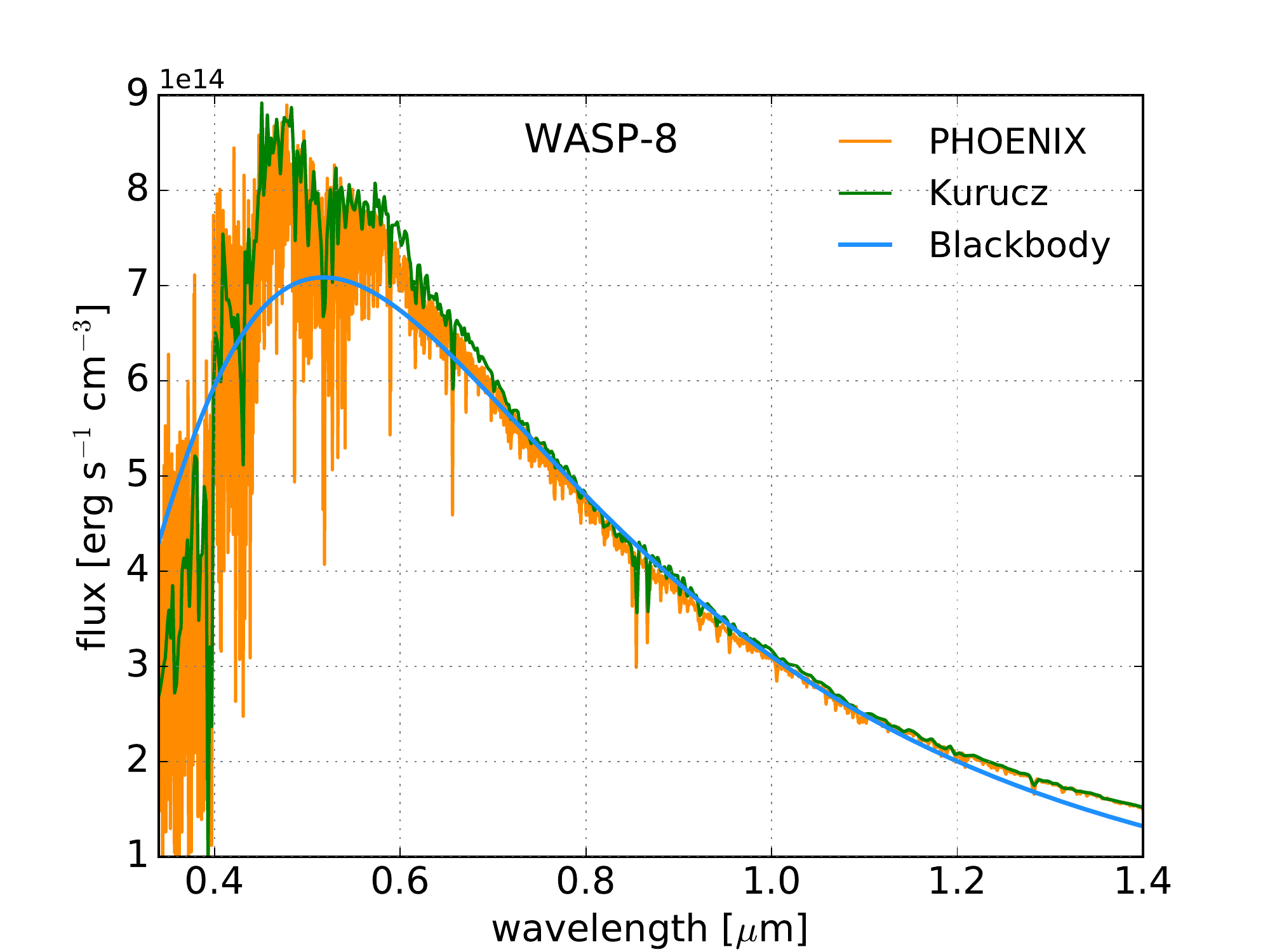}
\end{minipage}
\begin{minipage}[t]{0.48\textwidth}
\includegraphics[width=\textwidth]{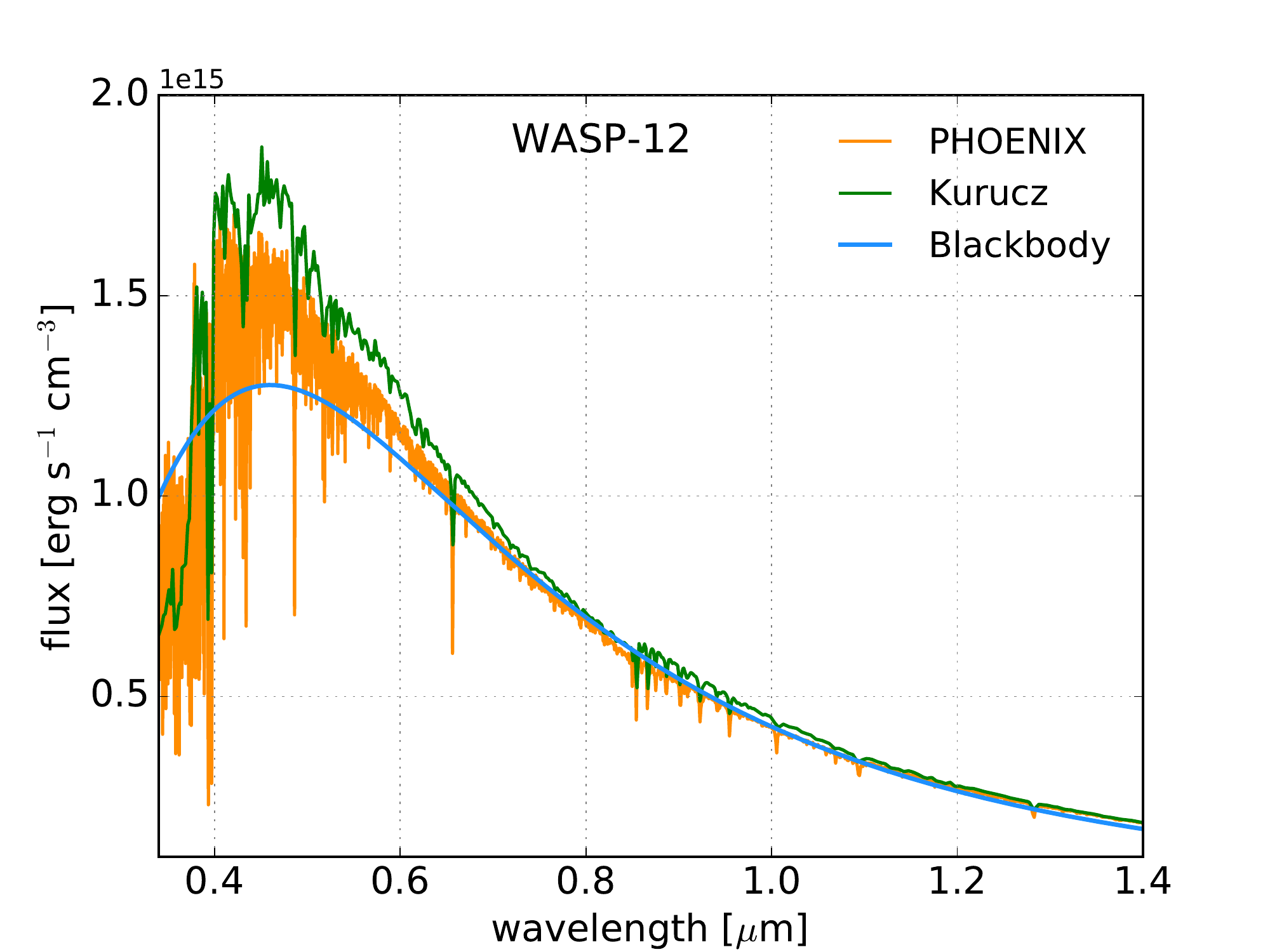}
\end{minipage}
\hfill
\begin{minipage}[t]{0.48\textwidth}
\includegraphics[width=\textwidth]{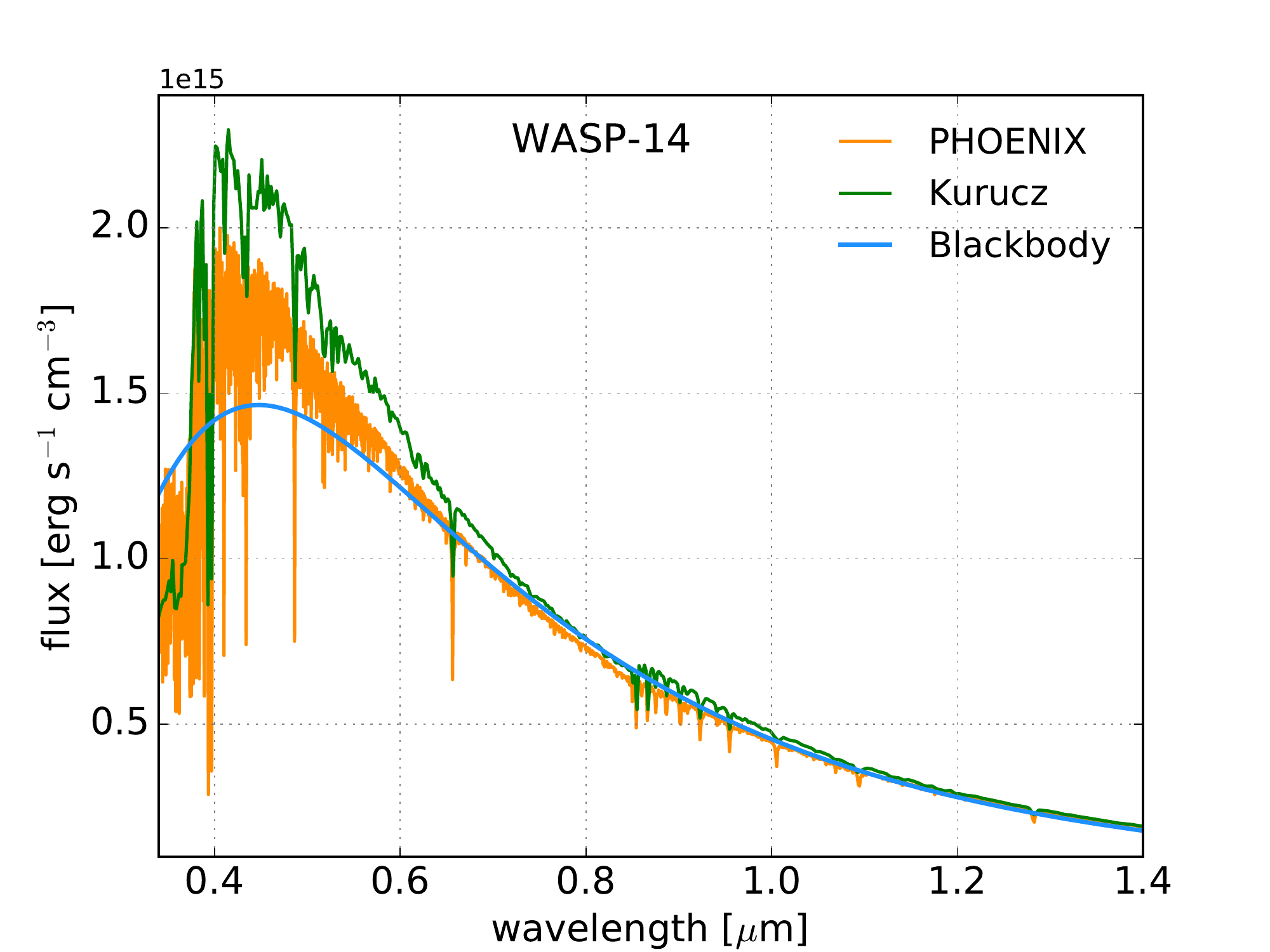}
\end{minipage}
\begin{minipage}[t]{0.48\textwidth}
\includegraphics[width=\textwidth]{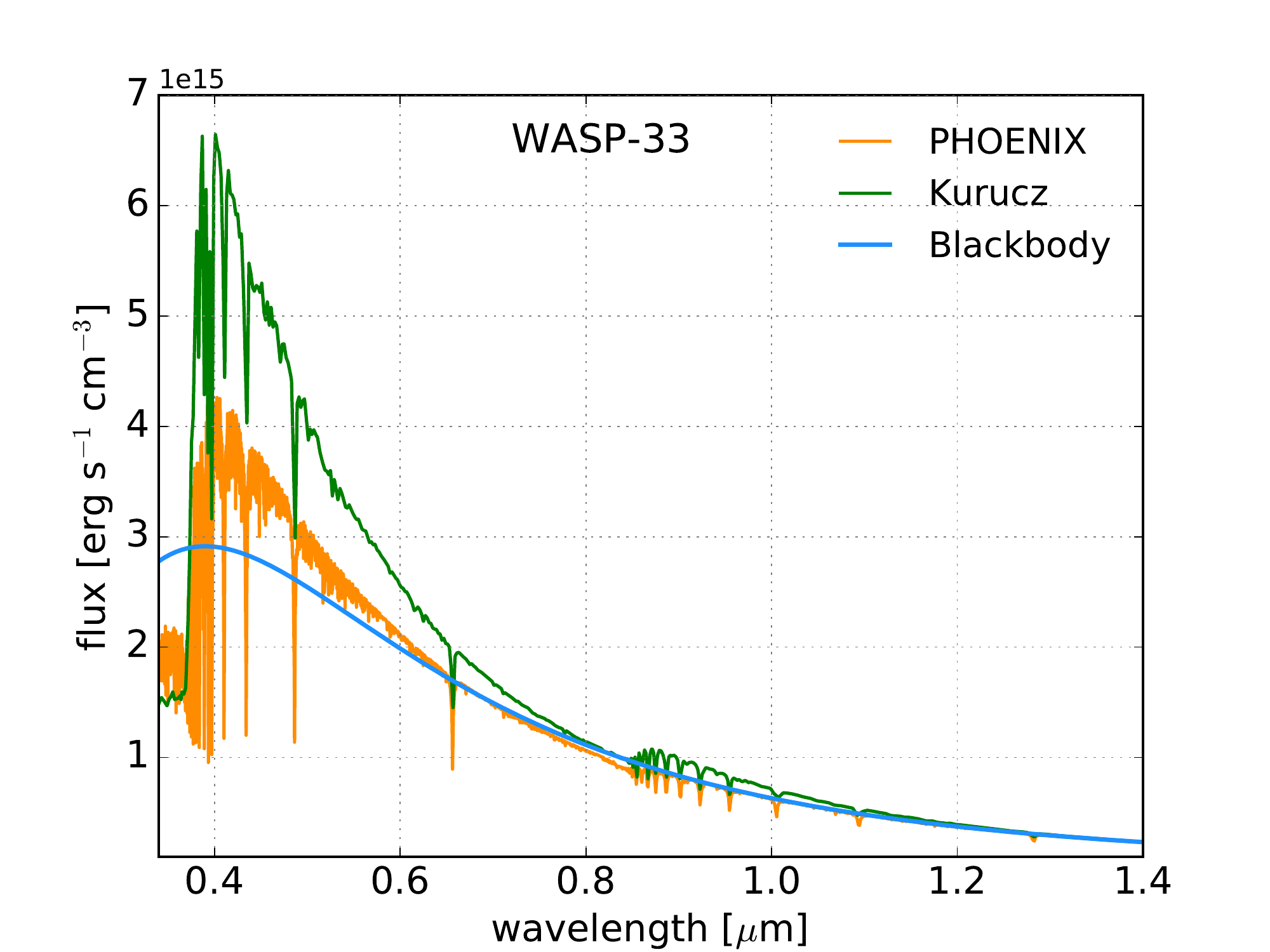}
\end{minipage}
\hfill
\begin{minipage}[t]{0.48\textwidth}
\includegraphics[width=\textwidth]{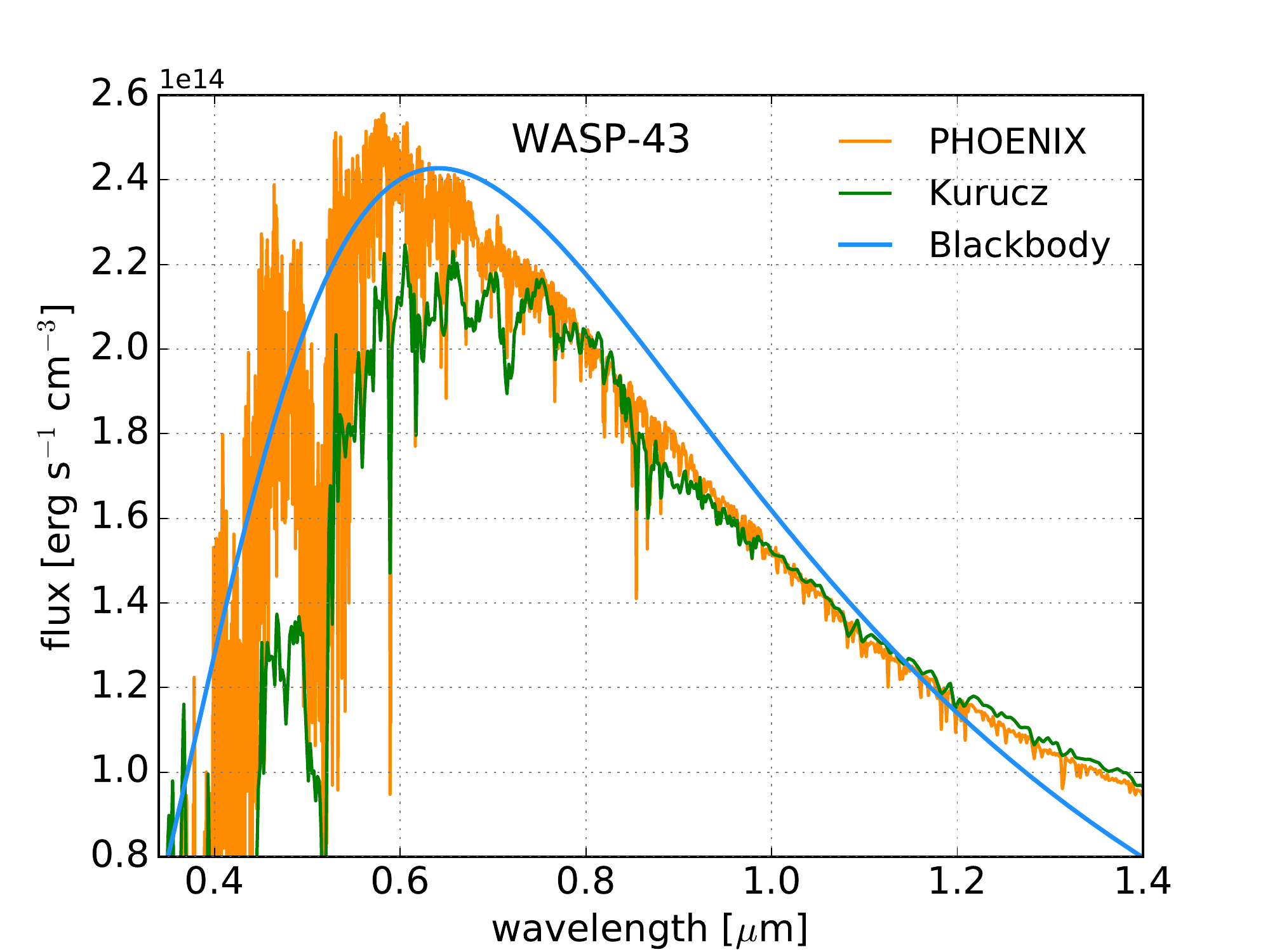}
\end{minipage}
\vspace{0.0cm}
\caption{Comparison of the \texttt{PHOENIX} and Kurucz stellar models with the stellar blackbody function for the 6 hot Jupiters examined in the current study.  Each stellar model was customized according to the specified stellar effective temperature, surface gravity and metallicity, as stated in Table \ref{tab:para}.}
\label{fig:star}
\end{center}
\end{figure*}

For any atmosphere of the exoplanet irradiated by the host star, one needs a description of the incident stellar flux.  The simplest approach is to adopt a Planck function, where the only input is the effective temperature of the stellar photosphere ($T_\star$).  The next level of sophistication requires the use of models such as \texttt{MARCS}, \texttt{PHOENIX} or Kurucz (\texttt{ATLAS}) that predict the photospheric emission from a star.  Specifically in this work, we use the latter two: \texttt{PHOENIX} \citep{al95,hu13} and Kurucz models \citep{ku79,mu04,mu05}.\footnote{The \texttt{PHOENIX} spectra are downloaded directly from their online library at ftp://phoenix.astro.physik.uni-goettingen.de/HiResFITS/ and interpolated in stellar temperature $T_{\star}$, surface gravity $g_\star$, and metallicity to fit the stellar parameters shown in Table \ref{tab:para}. The Kurucz spectra are interpolated in $T_{\star}$ and $g_\star$.} For completeness, Figure \ref{fig:star} shows the stellar spectra we used to model our sample of 6 hot Jupiters in Section \ref{sec:bench}.  The choice of stellar model has two primary effects.  First, since the secondary emission spectrum is the ratio of the exoplanet's to the star's flux, features in the stellar spectrum are imprinted onto it.  Second, differences in the stellar spectrum cause changes in the way the model atmosphere is being heated, which ultimately affects the temperature-pressure profile and synthetic spectrum.  As both the \texttt{PHOENIX} and Kurucz stellar models do not extend across the entire wavelength range included in our calculations (0.33 $\mu$m to 10 cm), we patch them using a Planck function.

\subsection{Numerical Implementation}
\label{sec:num}

The computationally intensive parts of \texttt{HELIOS} are written in \texttt{CUDA C++} \citep{cuda}, a proprietary language extension of \texttt{C++} for general purpose computations on suitable NVIDIA GPUs. Due to the GPU's main purpose of providing a fast 2D graphical image where each pixel needs to be updated simultaneously, their architecture is designed to maximize the throughput of parallel calculations and memory bandwidth.  A radiative transfer problem is naturally amenable to parallelization as the flux propagation through the atmosphere can be computed for each wavelength separately if we assume coherent scattering (i.e., no change in the wavelength of the radiation). We also parallelize the interpolation of the pre-computed $k$-distribution tables to determine the correct layer values. For further speed-up, the code offers the possibility to tabulate the Planck and the transmission functions at the model's wavelength values and a grid in $T$ ($\Delta T = 10$ K) and opacity ($ \Delta \log_{10} \kappa$ = 0.1), respectively. These grid resolutions are found to be sufficient for a converged behaviour of the model (not shown here). 

With this implementation the temperature iteration, the procedure needs typically the following time: with a NVIDIA Geforce 750M, the atmospheric temperatures converge within 2 to 15 minutes; with a NVIDIA K20 GPU, this takes between 0.5 to 4 minutes. These times have been found for a typical atmospheric set-up with 101 layers and 300 wavelength bins, including a separate iteration for scattering during each numerical timestepping. Without scattering, the convergence times are usually a factor of 2 smaller. Once the converged temperature-pressure profile is found, the calculation of the emission spectrum with very high spectral resolution (3000 wavelength bins), as a post-processing step, takes less than 30 seconds.  In our experience, we have found it to be sufficient to run \texttt{HELIOS} on a personal computer with a NVIDIA GPU (i.e., a GPU cluster is unnecessary).

\section{Results}
\label{sec:res}

We first subject \texttt{HELIOS} to a battery of tests.  We then use it to address several lingering issues in the literature concerning the radiative transfer of exoplanetary atmospheres.  Finally, we present 1D, benchmark calculations for the emission spectra of 6 hot Jupiters (HD 189733b, WASP-8b, WASP-12b, WASP-14b, WASP-33b and WASP-43b) that serve as ``null hypothesis" models.

By default, we use 300 wavelength bins and 101 non-isothermal layers in our calculations to solve for radiative equilibrium.  The emission spectra are computed using 3000 bins and isothermal layers as a post-processing step.  These bins are evenly distributed in wavenumber and cover a range of $0.1 \le \lambda^{-1} \le 30000$ cm$^{-1}$, which corresponds to 0.33 $\mu$m $\le \lambda \le 10$ cm. The layer pressures at the TOA and BOA  are set at 1 $\mu$bar and 1 kbar, respectively.  Stellar heating is represented by a Planck function.  The diffusivity factor is set to $\mathcal{D}=2$ and the redistribution efficiency factor is set to $f=2/3$.  Isotropic scattering ($\omega_0 \ne 0$, $g_0=0$) and equilibrium chemistry with solar abundances are assumed.  Unless otherwise stated, our fiducial model adopts these default parameter values.

\subsection{Tests}

To check \texttt{HELIOS} for consistency of the implementation, we focus on the case study of the super Earth GJ 1214b.  The parameter values used are listed in Table \ref{tab:para}.

\begin{table*}
	\caption{Planetary and stellar parameters used in this study.}
	\label{tab:para}
	\vspace{-0.4cm}
\begin{center}
\bgroup
\def\arraystretch{1.5}
  \begin{tabular}{| l | l | l | l | l | l | l | l |}
    \hline
Object & GJ 1214b\footnote{\cite{bouchy05}, \cite{an13}, \cite{ha13}} & HD 189733b\footnote{\cite{so10}, \cite{de13}, \cite{bo15}} & WASP-8b\footnote{\cite{qu10}} & WASP-12b\footnote{\cite{he09}, \cite{ch11}} & WASP-14b\footnote{\cite{jo09}} & WASP-33b\footnote{\cite{co10}, \cite{ko13}, \cite{le15}}  & WASP-43b\footnote{\cite{gi12}} \\ \hline
mean molecular mass ${\bar m}$ [$m_{\rm p}$] & \multicolumn{7}{c|}{2.4\footnote{Our choice value for a hydrogen dominated atmosphere.}} \\ \hline
surface gravity $g$ [cm s$^{-2}$] & 768 & 1950 & 5510\footnote{This value has been obtained from Newton's law of gravity assuming a spherical shape of the planet and neglecting rotation.} & 1164 & 10233 & 2884 & 4699   \\
orbital separation $a$ [AU] & 0.01411 & 0.03142 & 0.0801 & 0.02293 & 0.036 & 0.0259 & 0.0152  \\
effective temp. $T_{\rm eff}$\footnote{Assuming day-side heat redistribution using a factor $f = 2/3$.} [K]& 775 (660\footnote{\label{f:x}{This value is used for the model comparison with \cite{mi10}.}}) & 1575 & 1185 & 3241 & 2403 & 3494 & 1845 \\
planet. radius $R_{\rm pl}$ [$R_{\rm Jup}$] & 0.2479 & 1.216 & 1.038 & 1.776 & 1.281 & 1.679 & 1.036 \\
stell. temp. $T_{\star}$ [K] & 3252 (3026\textsuperscript{\ref{f:x}}) & 5050 & 5600 & 6300 & 6475 & 7430 & 4520 \\
stell. radius $R_{\star}$ [$R_{\odot}$]& 0.211 & 0.805 & 0.945 & 1.595 & 1.306 & 1.509 & 0.667 \\
stell. s. grav., $\log$ $g_{\star}$ [cm s$^{-2}$]& 5.04 & 4.53 & 4.5 & 4.16 & 4.07 & 4.3 & 4.645 \\
stell. metallicity [F/H] & 0.13 & 0.0 & 0.2 & 0.2 & 0.0 & 0.0 & 0.0 \\
\hline
\end{tabular}
	\egroup
	\end{center}
	\vspace{-0.1cm}
\end{table*}

\subsubsection{Comparison to GJ 1214b model of Miller-Ricci \& Fortney}

We test \texttt{HELIOS} against the results of \cite{mi10} for the planet GJ 1214b, who used the code originally developed by \cite{mc89} and \cite{ma99} for the atmospheres of Solar System planets. It was later adapted to exoplanetary atmospheres by \cite{fo05}. They utilize a radiative transfer technique based on \cite{toon89}, which is a multi-stream approach with a simplified two-stream solution for the scattering, further explained in \cite{cahoy10}, and add a convection model  for unstable atmospheric layers. Furthermore, \cite{mi10} use the opacities associated with H$_2$O, CO$_2$, CO, CH$_4$ and NH$_3$ \citep{fr08}, as well as  the CIA opacities associated with H$_2$-H$_2$, H$_2$-He, H$_2$-CH$_4$ and CO$_2$-CO$_2$. Their chemistry model is taken from \cite{lo02,lo06} and they include a treatment of Rayleigh scattering by molecular hydrogen. Still, we choose to compare \texttt{HELIOS} with the results of \cite{mi10}, because the employed radiative transfer technique and also the list of absorbers, together with the Rayleigh scattering, are similar to ours.

As a reference, we take their solar-abundance model that has a dayside-averaged temperature of 660 K (see the red, dashed curve in their Figure 1). To permit any reasonable comparison, we use the same astronomical parameters as \cite{mi10}. For instance, we set the stellar temperature to 3026 K and tune the redistribution parameter $f$ (in this test only) so that the dayside-effective temperature attains 660 K like in their set-up. Furthermore, to mimic their use of a stellar spectrum for GJ 1214 from \cite{hauschildt99} we also employ a PHOENIX stellar spectrum (from the updated online database) for the same stellar parameters, extrapolated by a blackbody fit to cover the whole wavelength range.

\begin{figure*}
\begin{center}
\begin{minipage}[t]{0.48\textwidth}
\includegraphics[width=\textwidth]{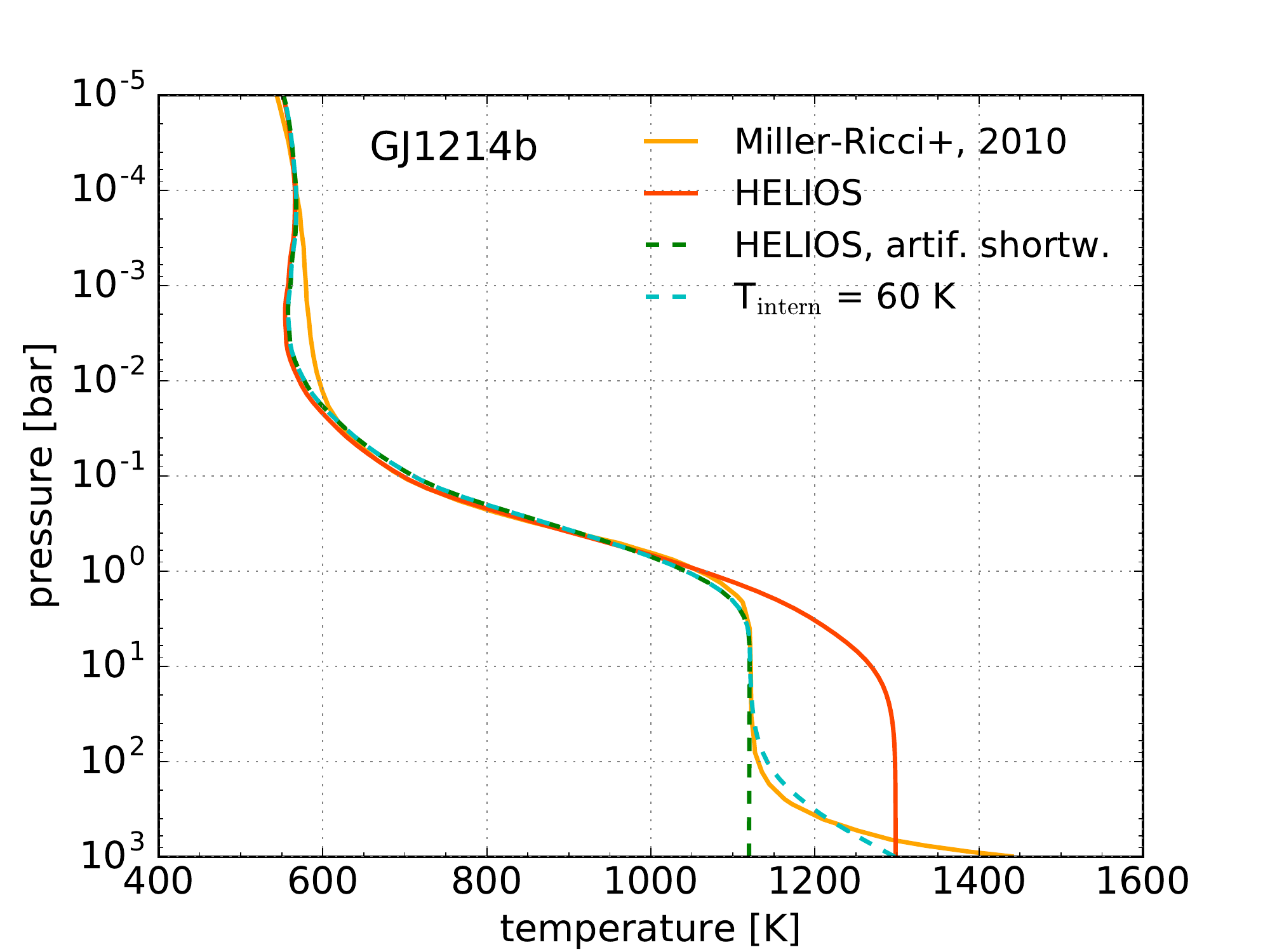}
\end{minipage}
\hfill
\begin{minipage}[t]{0.48\textwidth}
\includegraphics[width=\textwidth]{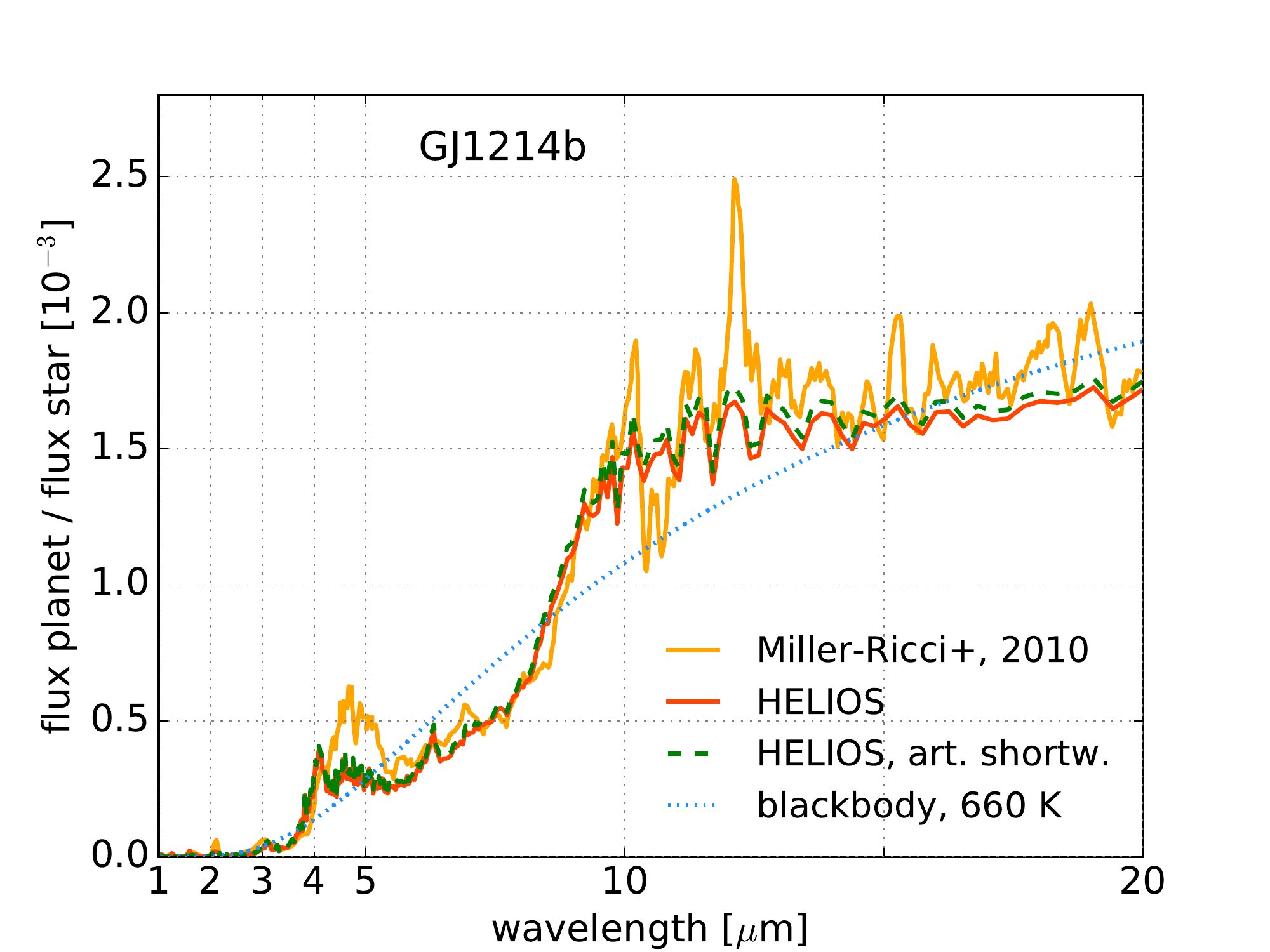}
\end{minipage}
\vspace{-0.0cm}
\caption{Comparison with the atmospheric model of GJ 1214b from \cite{mi10}.  The left panel shows the day-side temperature-pressure profile at $T_{\rm eff} = 660$ K. The temperatures in the infrared photosphere ($\sim 10^{-2} - 1$ bar) match very well. We also reproduce the deep atmosphere temperatures when an artificial opacity of $6 \times 10^{-4}$ cm$^2$g$^{-1}$ is inserted into the visible wavelengths (dashed curve). We can mimic the convective tail by adding internal heat flux; here shown for $T_{\rm intern} = 60$ K (cyan). The right panel depicts the corresponding planetary emission for three models of the left panel, together with a blackbody emission at the same effective temperature for comparison. The spectrum of \cite{mi10} shows more pronounced features, but overall has the same magnitude. The \texttt{HELIOS} runs are similar as the main temperature difference lies below the emitting photosphere.}
\label{fig:mrcomp}
\end{center}
\end{figure*}

In Figure \ref{fig:mrcomp}, left panel, we show the temperature-pressure profiles for GJ 1214b, by \cite{mi10}, and as computed with \texttt{HELIOS}. There is excellent agreement around $P=10^{-2} - 1$ bar---essentially, the calculations produce infrared photospheric temperatures that coincide.  At $P > 1$ bar, the \texttt{HELIOS} temperature-pressure profile is about 200 K hotter.  We suspect that this discrepancy is due to our simpler treatment of the opacities, as we only consider 4 molecules.  This leads to greater transparency particularly in the visible wavelengths of our model atmosphere, which in turn produces more heating in the deep atmosphere. To support this hypothesis, we have successfully reproduced the deep atmospheric structure of \cite{mi10}  by artificially introducing an opacity of $6 \times 10^{-4}$ cm$^2$ g$^{-1}$ to the shortwave below 1 $\mu$m (see green dashed curve in Fig. \ref{fig:mrcomp}). Since our model does not have any convective treatment, we cannot reproduce the adiabat in their model at the bottom boundary. However, by introducing an internal heat flux, we can somewhat mimic their deep temperatures (shown for $T_{\rm intern}=60$ K).

In the right panel of Figure \ref{fig:mrcomp} we show the ratio of the planetary and the stellar emission for \cite{mi10}'s model and ours. The spectra are of the same magnitude and show similar trends. Their results show a larger variation in intensity across wavelength, particularly enabling emission from deeper, and thus hotter, atmospheric regions. This could be a consequence of several factors: differences in employed molecular line lists, combination of the opacities, chemistry model or the stellar spectrum. Considering all those components it is not surprising that the individual spectral features do not match perfectly and we conclude that \texttt{HELIOS} is still rather consistent with the results of \cite{mi10}. 
For completion, we show both the spectra of our fiducial set-up and the one with an added artificial shortwave opacity. As expected, those are very similar because around the emitting photosphere the models only differ slightly.

\subsubsection{Trends associated with scattering}

As a further consistency check of \texttt{HELIOS}, we examine calculations with idealized descriptions of scattering and check if the trends match our physical intuition.  

For illustration, we set $\omega_0=0.5$ across all wavelengths.  We then examine models with $g_0=-0.5, 0$ and 0.5, which are also constant across all wavelengths.  We emphasize that the two-stream solutions used in \texttt{HELIOS}, which are taken from \cite{he14}, are generally able to take $\omega_0$ and $g_0$ as input \textit{functions} (rather than just scalars/numbers).  

\begin{figure*}
\begin{center}
\begin{minipage}[t]{0.48\textwidth}
\includegraphics[width=\textwidth]{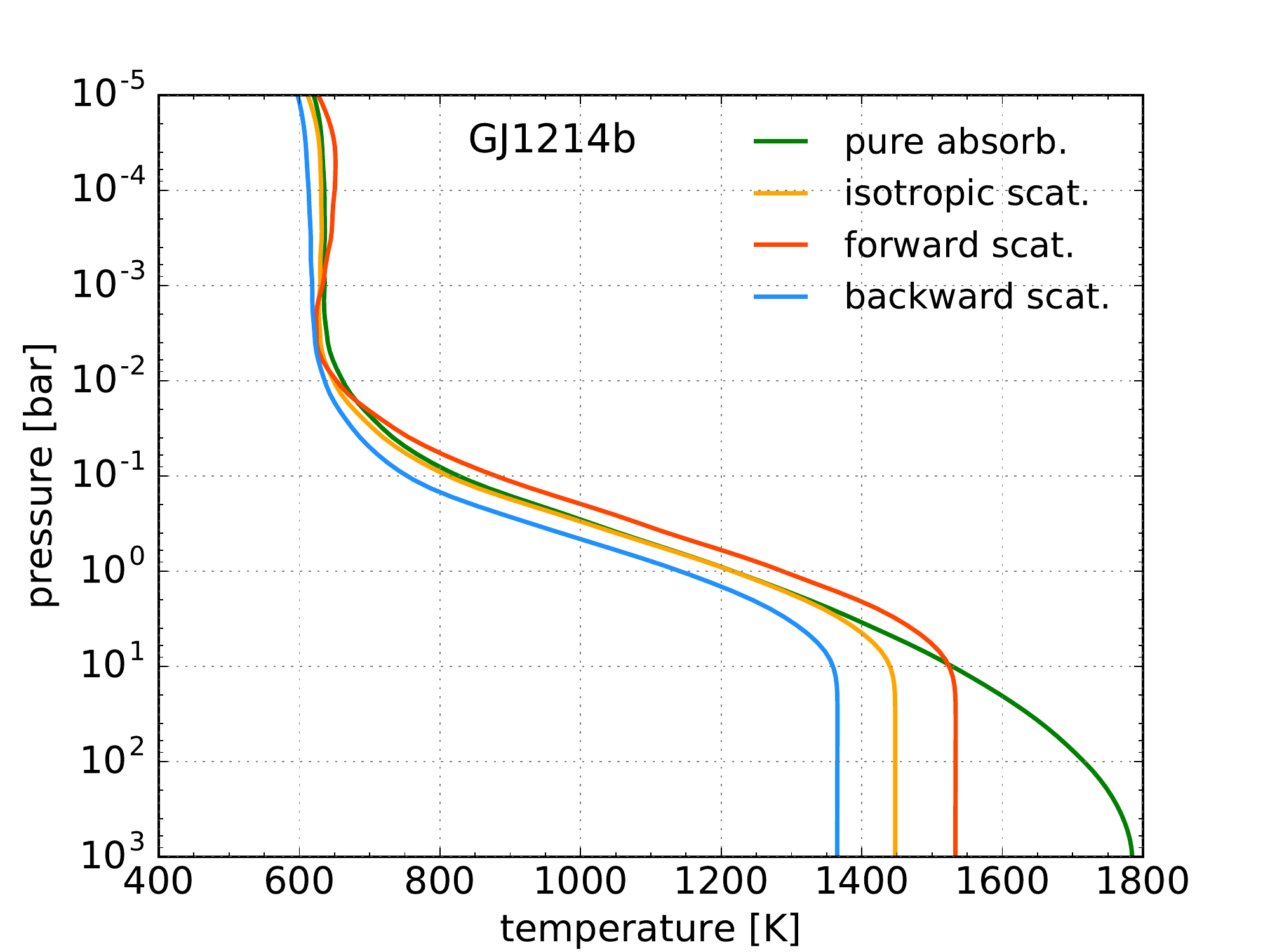}
\end{minipage} 
\hfill
\begin{minipage}[t]{0.48\textwidth}
\includegraphics[width=\textwidth]{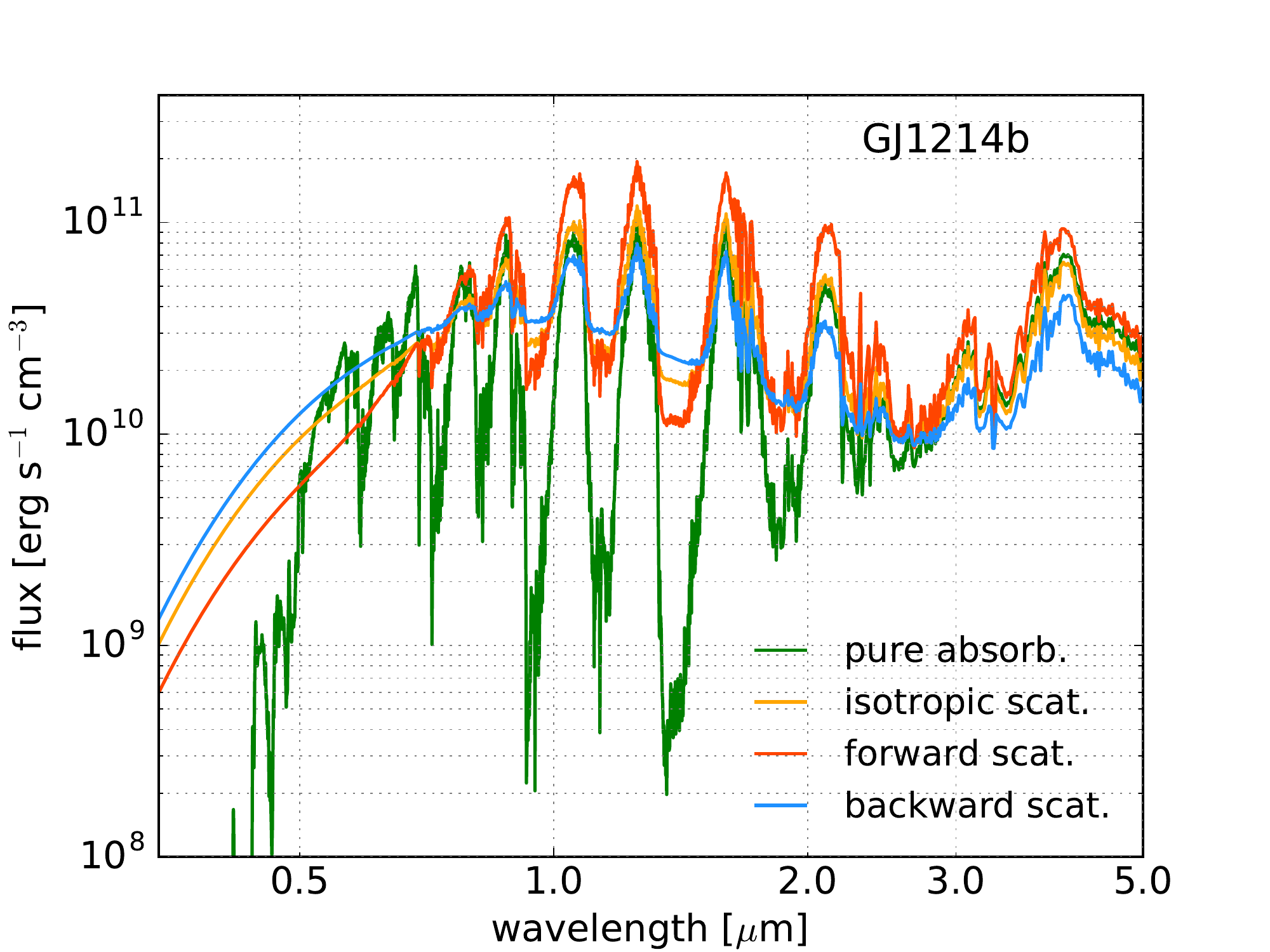}
\end{minipage}
\vspace{0.0cm}
\caption{Consistency check of \texttt{HELIOS} by examining the temperature-pressure profiles (left panel) and synthetic spectra (right panel) in the idealized limits of scattering: $g_0=0$ (isotropic scattering), $g_0=0.5$ (forward scattering) and $g_0=-0.5$ (backward scattering).  For illustration, we set $\omega_0=0.5$.  Both the single-scattering albedo ($\omega_0$) and scattering asymmetry factor ($g_0$) are assumed to be constant across wavelength, but we note that the two-stream solutions we implemented allow for them to generally be specified as functions of wavelength, temperature and pressure.  The qualitative trends associated with the temperature-pressure profiles and synthetic spectra are consistent with physical expectations (see text).}
\label{fig:scat}
\end{center}
\end{figure*}

Figure \ref{fig:scat} shows the fiducial pure absorption model compared against the 3 models with idealized descriptions of isotropic, backward and forward scattering. For $g_0=-0.5$ and 0, scattering generally shifts the absorption profile of starlight upwards (towards lower pressures), which cools the model atmosphere.  As the scattering shifts from being isotropic to being backward, the deep atmosphere becomes cooler. We also observe a trend of the reflected light at $\lesssim 1$ $\mu$m being the strongest for backward scattering, but of the thermal emission at $\gtrsim 1$ $\mu$m being the strongest for forward scattering, which is expected. 

Scattering also has the general effect of muting the spectral features in the synthetic spectra.  It effectively raises the level of the infrared continuum.  This effect is stronger as the scattering becomes more backward-dominated (Figure \ref{fig:scat}).  Such an effect mimicks the presence of aerosols or condensates.  Overall, these expected trends provide a ``proof-of-concept" validation of \texttt{HELIOS}.

\subsubsection{Isothermal versus non-isothermal layers}
\label{sec:reslayer}

An essential ingredient of 1D models of atmospheres in radiative equilibrium is the number of layers used in the computation.  We perform a series of convergence tests by considering different numbers of layers and employing isothermal versus non-isothermal layer models.  We again use the parameters of GJ 1214b as an illustration.  

\begin{figure*}
\begin{center}
\begin{minipage}[t]{0.48\textwidth}
\includegraphics[width=\textwidth]{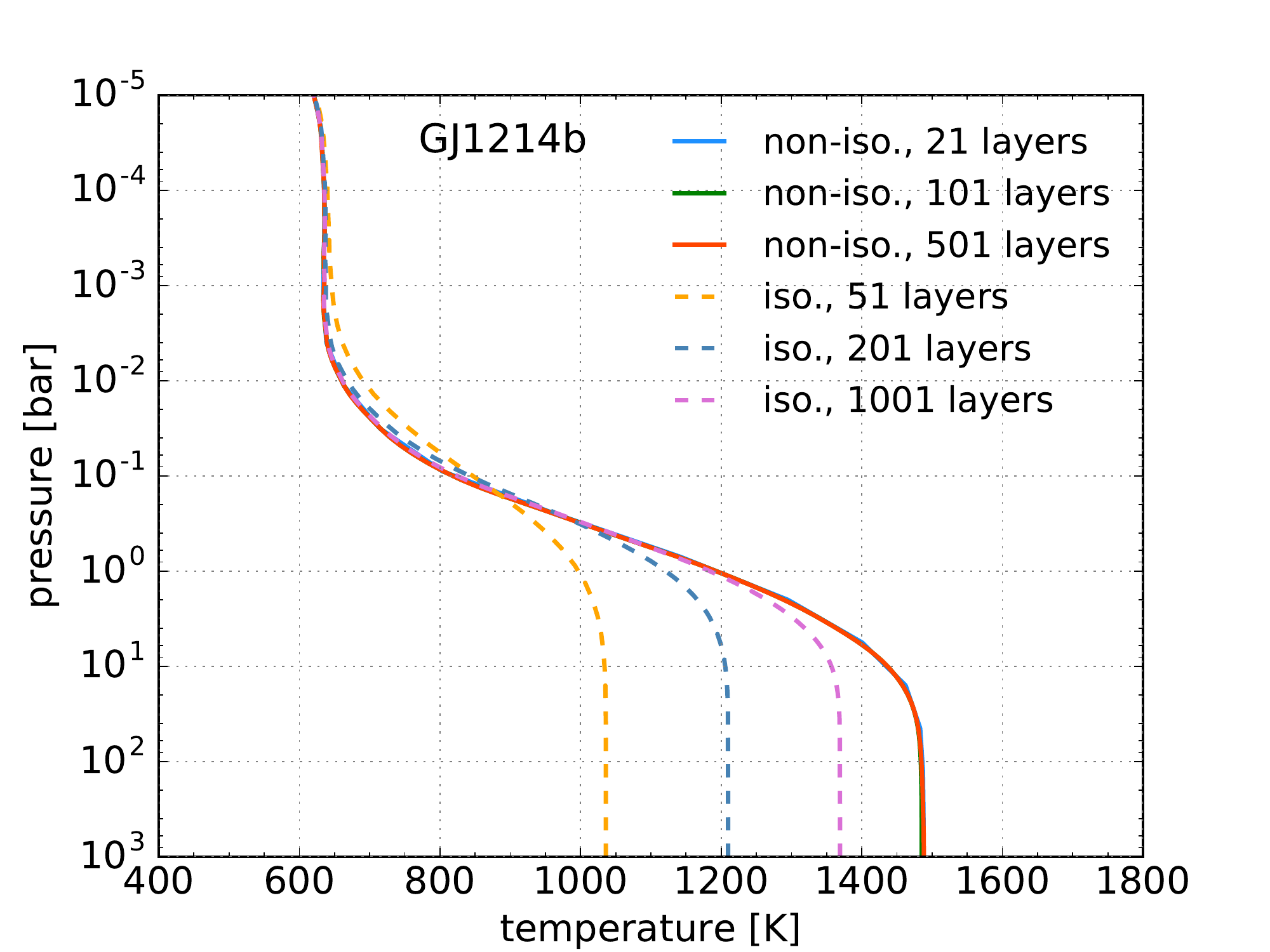}
\end{minipage}
\hfill
\begin{minipage}[t]{0.48\textwidth}
\includegraphics[width=\textwidth]{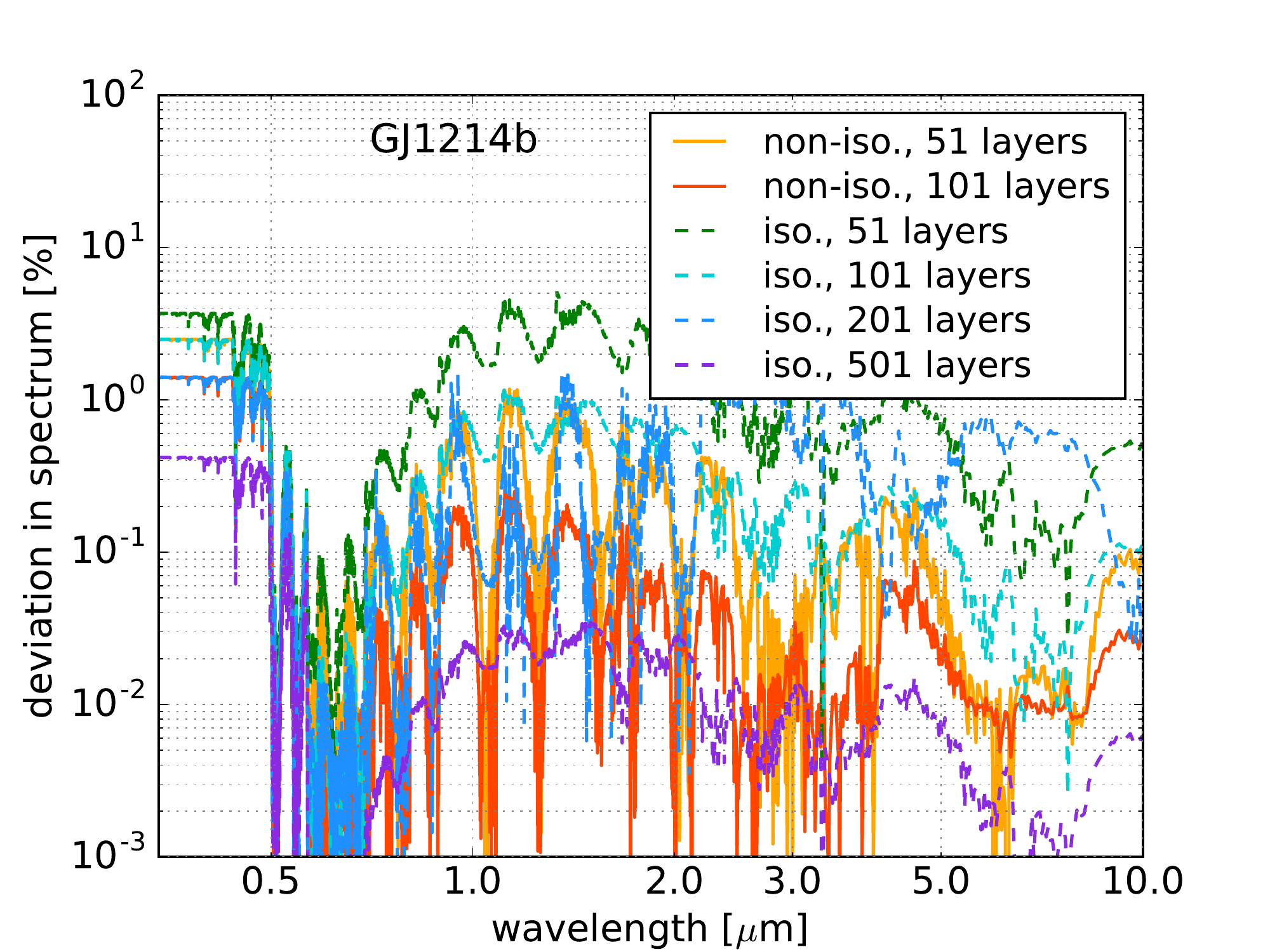}
\end{minipage}
\vspace{-0.1cm}
\caption{Resolution tests to determine the minimum number of isothermal versus non-isothermal layers needed for numerical convergence.  The left panel shows various temperature-pressure profiles computed using 51, 201 and 1001 isothermal layers versus 21, 101 and 501 non-isothermal layers, demonstrating that the use of isothermal layers is not an efficient approach.  The right panel shows the deviation or error in the synthetic spectrum, as a function of wavelength, using the model with 501 non-isothermal layers as a reference.}
\label{fig:restest}
\end{center}
\end{figure*}

Figure \ref{fig:restest} shows the temperature-pressure profiles associated with models having 51, 201 and 1001 isothermal layers, and also those with 21, 101, 501 non-isothermal layers.  First, we note that the temperature-pressure profiles of the models with non-isothermal layers coincide (with differences of less than 3 K), implying that 21 non-isothermal layers is sufficient to attain convergence.  By contrast, even with 1001 layers, no convergence is seen for the models with isothermal layers.  These results illustrate the superiority of using non-isothermal layers.  We recover the same behavior even when a different case study (e.g., WASP-12b) is considered (not shown).

Next, we compute the synthetic spectrum of the model with 501 non-isothermal layers and use it as a reference.  We then consider models with 51 and 101 non-isothermal layers, as well as models with 51, 101, 201 and 501 isothermal layers.  For each model, we compute the deviation in the synthetic spectrum, from the reference model, as a function of wavelength.  Figure \ref{fig:restest} shows that, as expected, the deviation decreases as the resolution increases.  Only the model with 51 isothermal layers produces deviations that exceed 1\% in the flux.  The model with 101 isothermal layers produces deviations that are typically less than 1\%.  Since models with isothermal layers are faster to compute, this motivates us to adopt a model with 101 isothermal layers for our post-processing step.  In other words, we use non-isothermal layers to iterate for radiative equilibrium.  Upon attaining radiative equilibrium, we post-process the converged temperature-pressure profile, using a model with 101 isothermal layers, to produce synthetic spectra.

\subsubsection{Obtaining convergence for the $k$-distribution tables}

Another essential ingredient of 1D models of atmospheres in radiative equilibrium is the spectral resolution used in constructing the opacity function, which is then used to construct the $k$-distribution tables.  We wish to investigate the errors associated with using different spectral resolutions.  The reference case is taken to be a model with a spectral resolution of 10$^{-5}$ cm$^{-1}$.  We examine models with resolutions of $10^{-1}$, $10^{-2}$, $10^{-3}$ and $10^{-4}$ cm$^{-1}$ and compare the errors in the synthetic spectra, after we have iterated for radiative equilibrium, as a function of wavelength, relative to the reference.  As we are using 3000 wavelength bins, these sampling resolutions correspond to $10^2$, $10^3$, $10^4$ and $10^5$ points per bin, respectively.

\begin{figure*}
\begin{center}
\begin{minipage}[t]{0.48\textwidth}
\includegraphics[width=\textwidth]{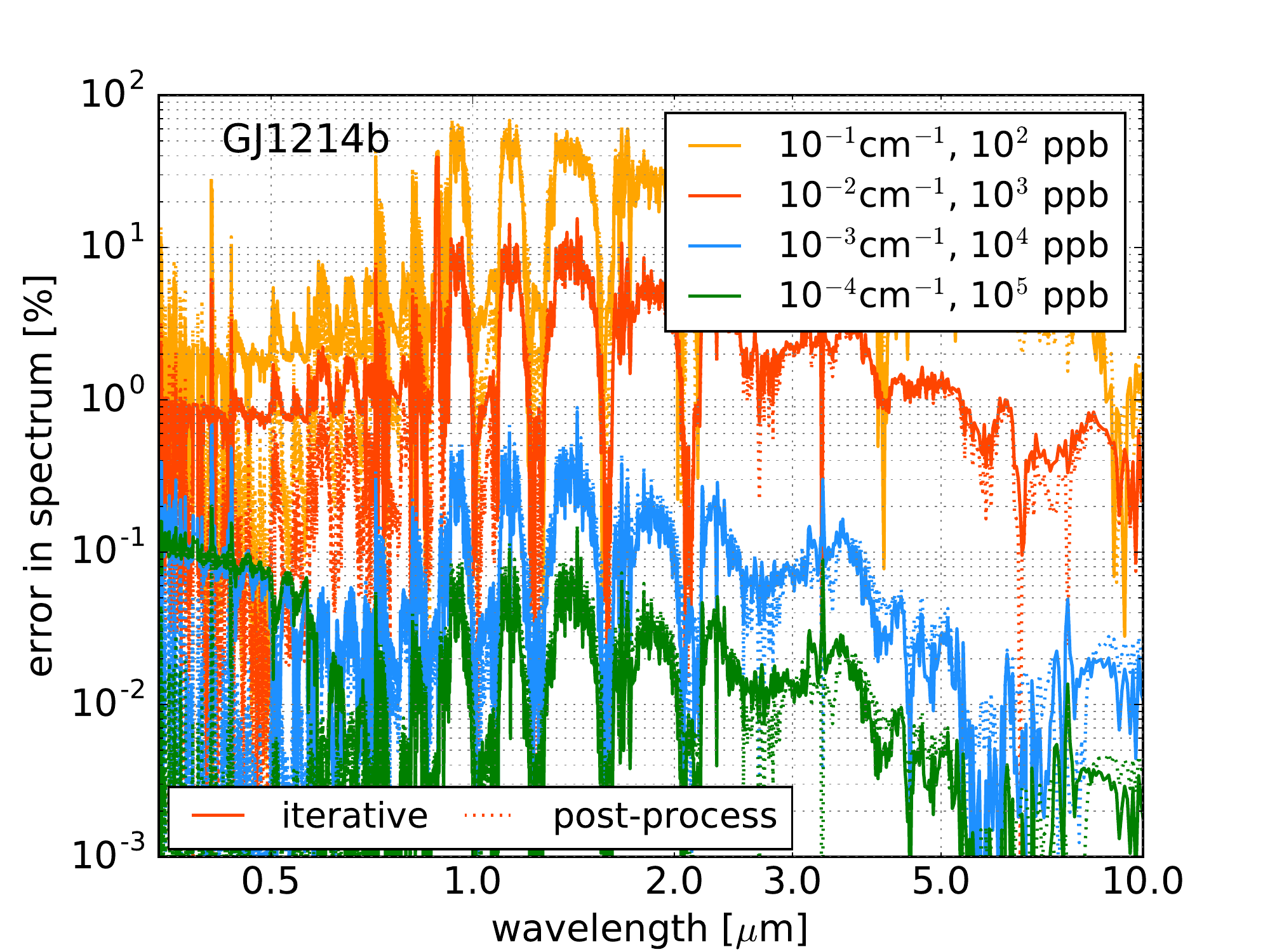}
\end{minipage}
\hfill
\begin{minipage}[t]{0.48\textwidth}
\includegraphics[width=\textwidth]{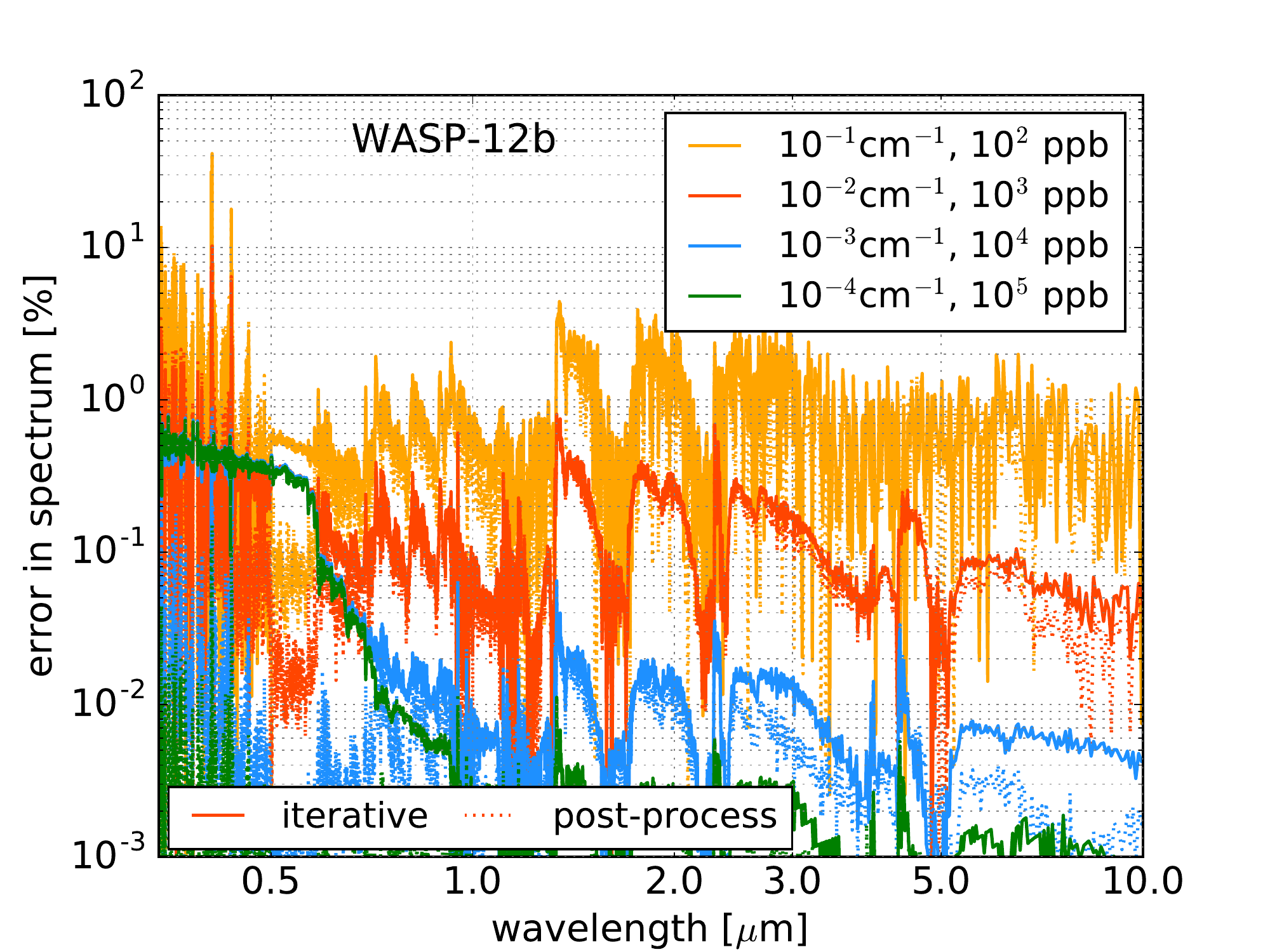}
\end{minipage}
\vspace{-0.1cm}
\caption{Elucidating the errors, in the synthetic spectra, associated with using different spectral resolutions to construct the $k$-distribution tables.  The reference case uses a spectral resolution of $10^{-5}$ cm$^{-1}$.  The label ``ppb" refers to the number of points per bin.  All of the synthetic spectra were computed for model atmospheres in chemical and radiative equilibrium using the correlated-$k$ approximation. We either run the whole radiative transfer iterative process (solid) or solely post-process the $T$-$P$ profile of the reference case (dotted).  For illustration, we examine models of cool (GJ 1214b; left panel) and hot (WASP-12b; right panel) exoplanetary atmospheres.}
\label{fig:resopac}
\end{center}
\end{figure*}

Figure \ref{fig:resopac} shows our results for the case studies of GJ 1214b and WASP-12b, which were illustrated to span the range of temperatures for the currently characterizable exoplanetary atmospheres.  We find the expected trend that the error decreases as the resolution increases from $10^{-1}$ cm$^{-1}$ to $10^{-4}$ cm$^{-1}$.  Using a spectral resolution of only $10^{-1}$ cm$^{-1}$ ($10^{-2}$ cm$^{-1}$) results in errors that are $> 10\%$ ($\sim 1\%-10\%$) in the near-infrared flux.  To reduce the error to $\sim 1$\%, we find a minimum resolution of $10^{-3}$ cm$^{-1}$ to be required in our model. This value might change if one is using opacity sampling. We also show the error in the spectra produced by purely post-processing the temperature profile of the reference case, which demonstrates that the errors are not merely associated with iterating for radiative equilibrium.

\subsubsection{Using the correct value of the diffusivity factor}
\label{subsect:diffuse}

As discussed previously, one may obtain an exact solution of the radiative transfer equation, without invoking the two-stream approximation, only in the limit of pure absorption.  This solution may be compared to two-stream calculations with different assumed values of the diffusivity factor. 

\cite{amundsen14} have recently advocated for the use of ${\cal D}=1.66$ from comparing their two-stream calculations to a different set of calculations computed using the discrete-ordinates radiative transfer method.  \cite{armstrong69} also advocate for ${\cal D}=1.66$ based on radiative transfer calculations of water in the atmosphere of Earth.  However, the correct value for ${\cal D}$ should depend on the vertical resolution of the model (c.f. Fig. \ref{fig:diff}), which motivates us to perform our own comparisons.

\begin{figure*}
\begin{center}
\begin{minipage}[t]{0.48\textwidth}
\includegraphics[width=\textwidth]{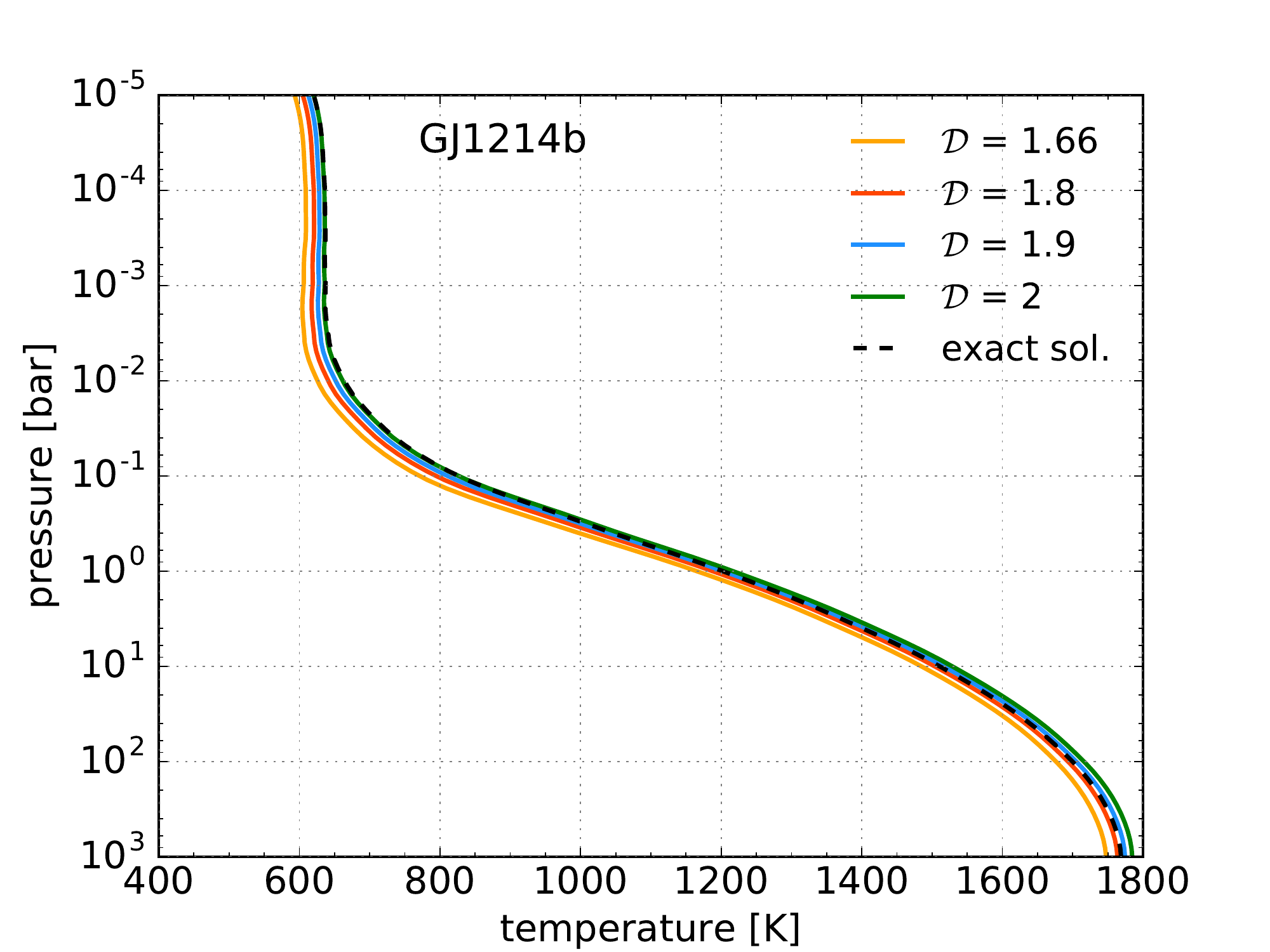}
\end{minipage}
\hfill
\begin{minipage}[t]{0.48\textwidth}
\includegraphics[width=\textwidth]{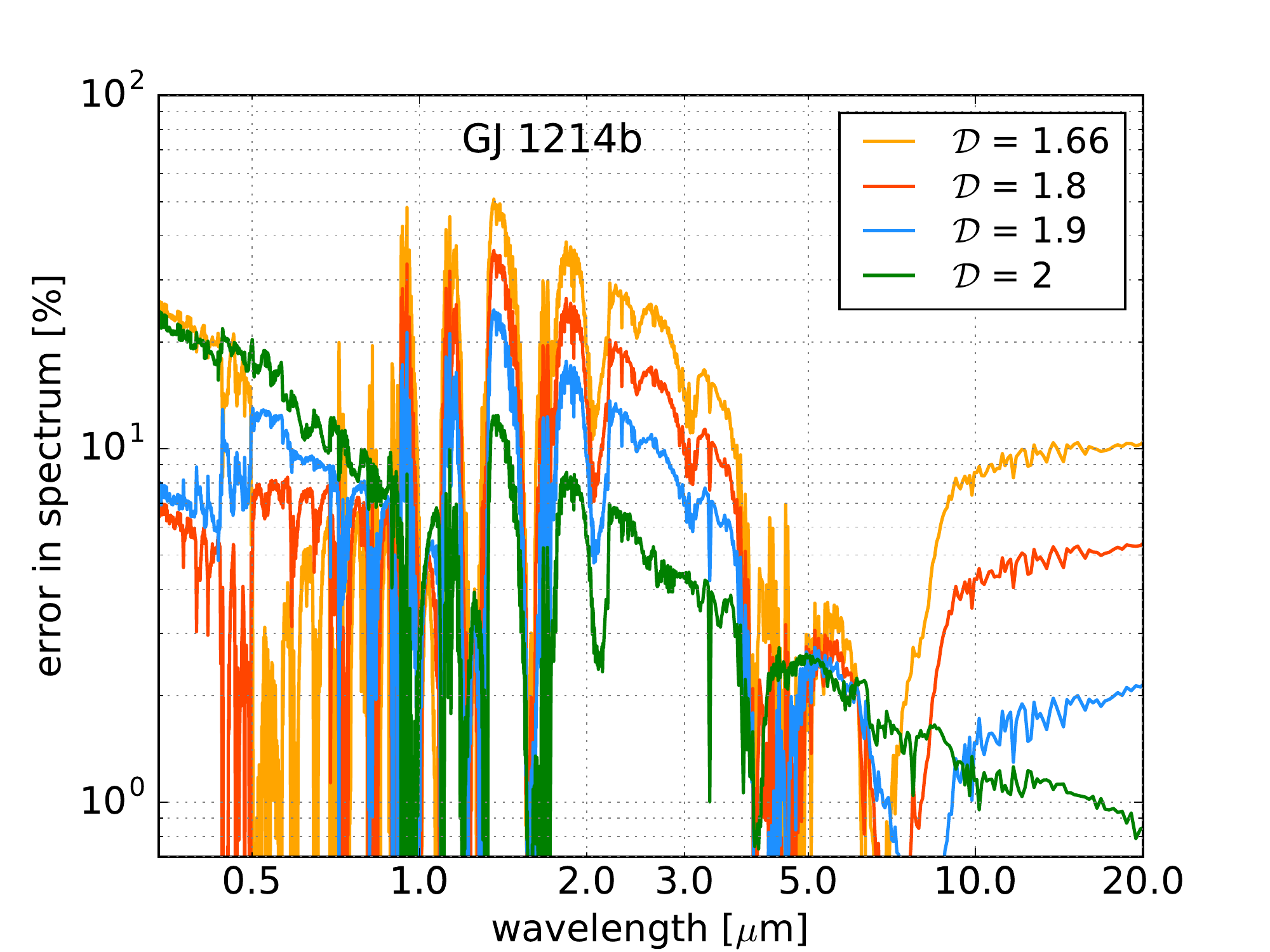}
\end{minipage}
\begin{minipage}[t]{0.48\textwidth}
\includegraphics[width=\textwidth]{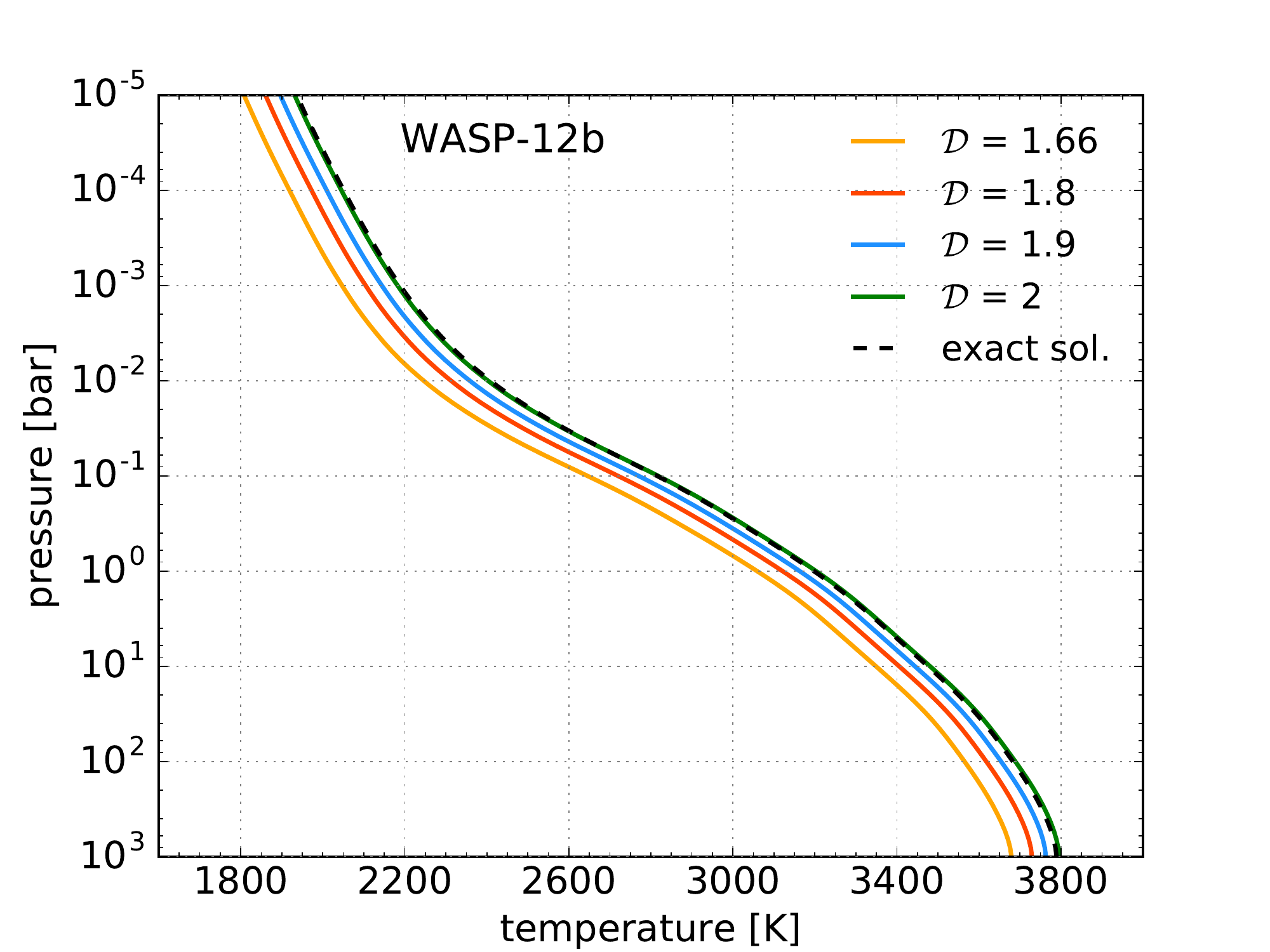}
\end{minipage}
\hfill
\begin{minipage}[t]{0.48\textwidth}
\includegraphics[width=\textwidth]{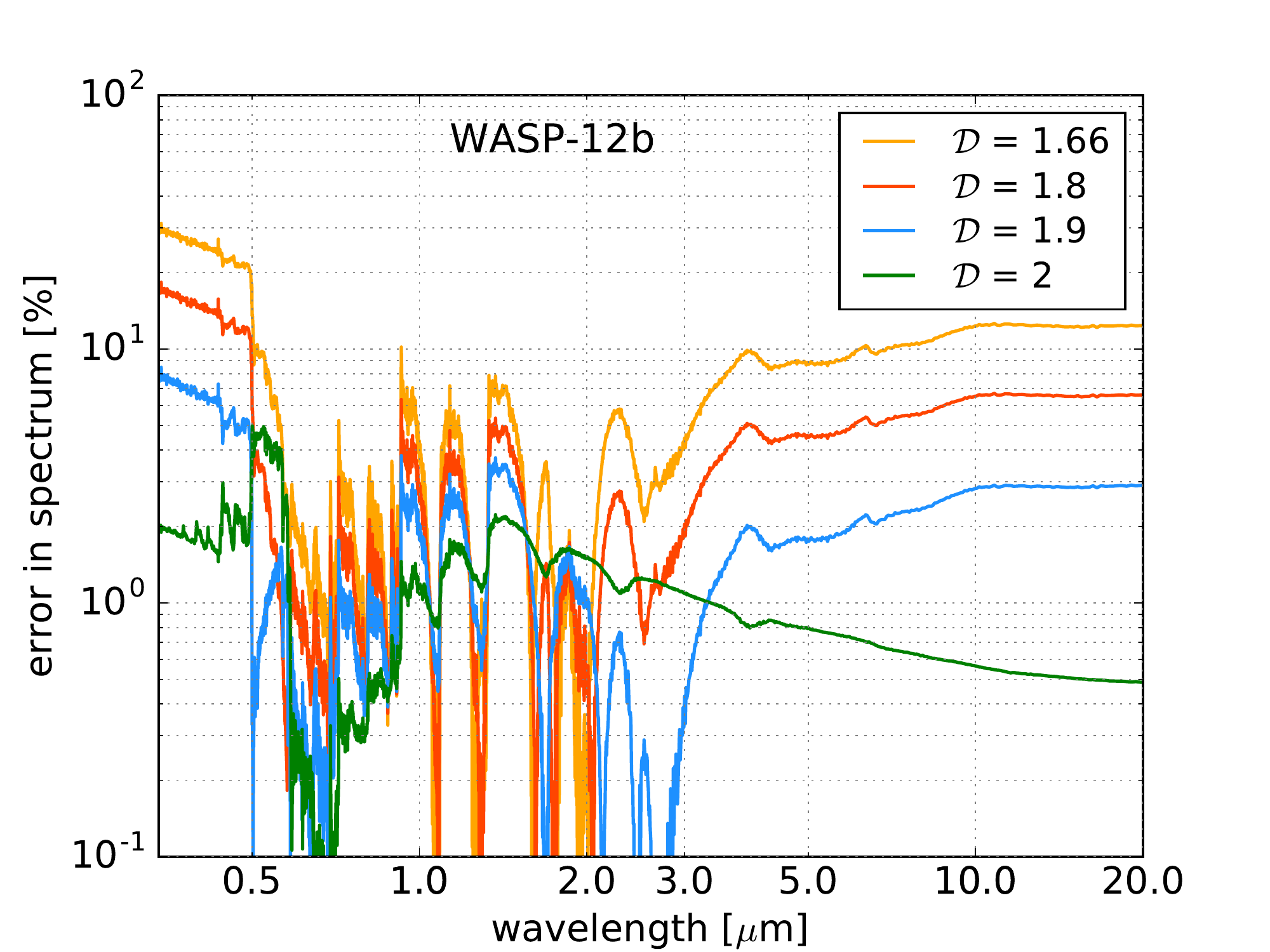}
\end{minipage}
\vspace{-0.1cm}
\caption{Determination of the diffusivity factor ($\mathcal{D}$) by comparing two-stream and exact solution in the limit of pure absorption.  For illustration, we study the warm super Earth GJ 1214b (top panels) and the hot Jupiter WASP-12b (bottom panels).  We show the temperature profiles on the left and the error in the resulting synthetic spectrum compared to the exact solution on the right. A diffusivity factor of $\mathcal{D}=2$ appears to produce the best match to the exact solutions, following closely the temperature-pressure profile of the exact solution and leading on average to the smallest error in the spectrum.}
\label{fig:difffit}
\end{center}
\end{figure*}

Figure \ref{fig:difffit} displays the computed temperature-pressure profiles and the error in the resulting synthetic spectrum for GJ 1214b for ${\cal D}=1.66, 1.8, 1.9$ and 2 compared to the exact solution. Regarding temperature, the ${\cal D}=1.9$ and 2 models produce the best match to the exact solution. However, ${\cal D}=2$ leads, on average, to the smallest error in the spectrum. We also consider the same set of calculations for a hotter exoplanet, WASP-12b.  In this case, ${\cal D}=2$ clearly produces the best match to the exact solution in terms of the temperature as well as the spectrum. In general, the error in the spectrum is smaller than for the cooler planet. It is not unsurprising, that ${\cal D}=2$ provides the most accurate results, because we expect the diffusivity factor to approach a value of 2 when the vertical resolution of the model is sufficient (see Figure \ref{fig:diff}), i.e. the difference in optical depth between the layers is small, at least in the photospheric regions of the atmosphere.

\subsection{Testing the Null Hypothesis and Variations on a Theme: Benchmark 1D Models for Hot Jupiters}
\label{sec:bench}

Despite heroic efforts to obtain data for exoplanetary atmospheres, exoplanets are spatially unresolved point sources---and will probably remain so for the foreseeable future---although phase curves and eclipse maps provide some spatial information.  As a first approach, theorists have resorted to interpreting the spectra of exoplanetary atmospheres using simple models: 1D, plane-parallel, just as we have constructed.  There is a precedent of using 1D models to interpret spectra (see Introduction for references).  As a null hypothesis, we make the following assumptions: chemical equilibrium, radiative equilibrium (which we solve for using \texttt{HELIOS}) and solar abundances.  This would be the second simplest model after a blackbody emission spectrum \citep{hansen14}.  Upon constructing the null hypothesis, we then examine variations in the metallicity and C/O.

\begin{table}
	\caption{Spectral data sources}
	\label{tab:data}
	\vspace{-0.4cm}
\begin{center}
\bgroup
\def\arraystretch{1.5}
  \begin{tabular}{| l | l |}
    \hline
Planet & Source \\ \hline
HD 189733b &  \cite{cr14}, \cite{to14}  \\
WASP-8b &  \cite{cu13} \\
WASP-12b &   \cite{st14} \\
WASP-14b &  \cite{bl13} \\
WASP-33b &  \cite{de12}, \cite{ha15} \\
WASP-43b & \cite{bl14}, \cite{kr14} \\
    \hline
  \end{tabular}
	\egroup
	\end{center}
\end{table}

We have chosen the sample of hot Jupiters to include in this analysis based on a literature search for planets with non-blackbody emission spectra. We have started from \cite{hansen14}, which catalogs all planets with secondary eclipse measurements in at least two bandpasses as of 2014. They found 7 planets that are poorly fit by a blackbody model. We also searched for any more recent non-blackbody results. To select the most precise, reliable measurements from our search, we consider space-based data only. We have also stipulated that the data were reduced with state-of-the-art techniques.  Specifically, we only consider Spitzer results that used sophisticated models of the intrapixel sensitivity such as BLISS mapping or pixel-level decorrelation \citep{stevenson12, deming15}. This approach has been demonstrated to be the best practice in Spitzer data analysis \citep{ingalls16}.  This search has resulted in the selection of six planets: HD 189733b, WASP-8b, WASP-12b, WASP-14b, WASP-33b, and WASP-43b. Their model parameter values and spectral data sources are shown in Tables \ref{tab:para} and \ref{tab:data}, respectively.

\begin{figure*}
\begin{center}
\begin{minipage}[t]{0.48\textwidth}
\includegraphics[width=\textwidth]{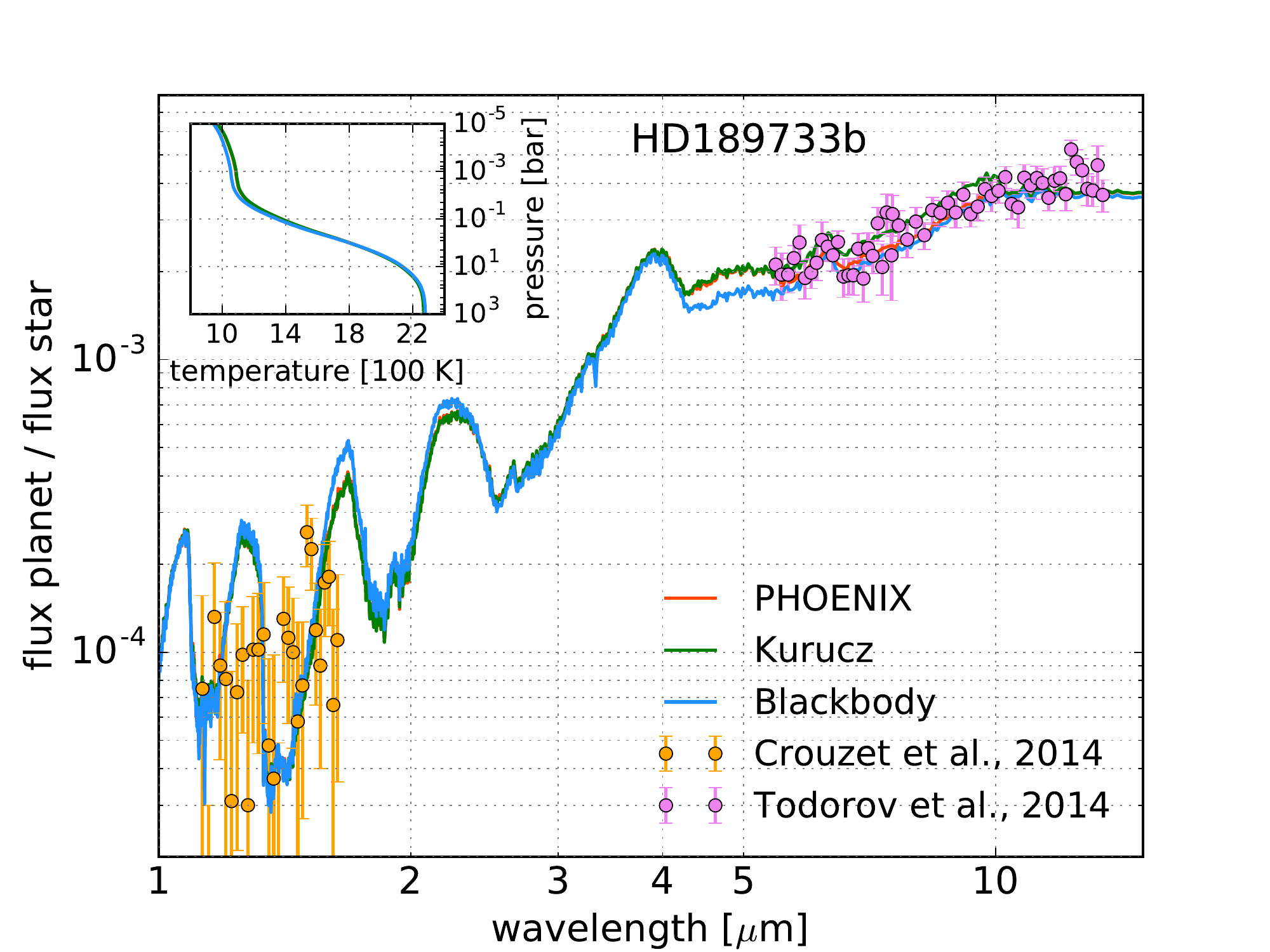}
\end{minipage}
\hfill
\begin{minipage}[t]{0.48\textwidth}
\includegraphics[width=\textwidth]{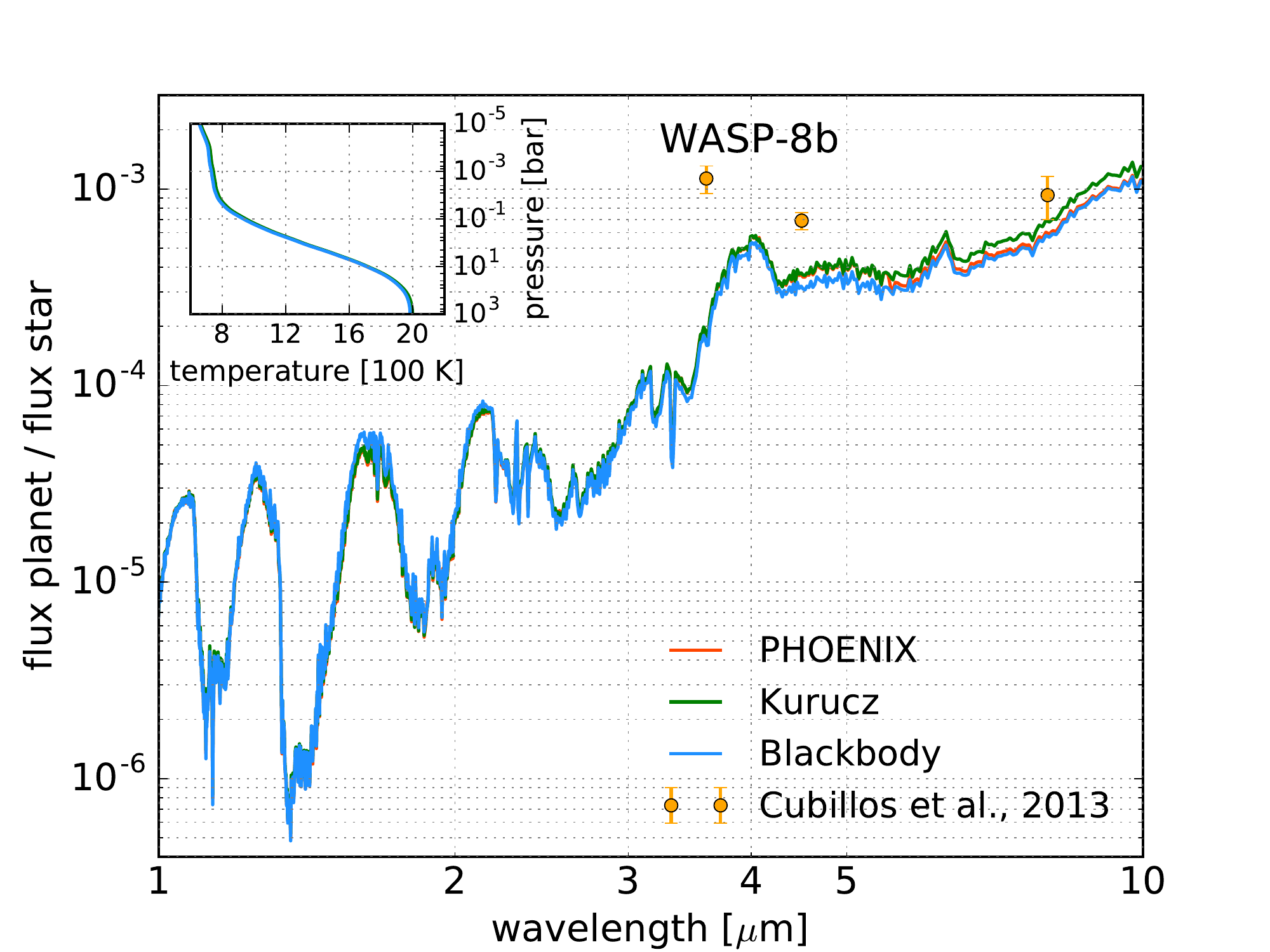}
\end{minipage}
\begin{minipage}[t]{0.48\textwidth}
\includegraphics[width=\textwidth]{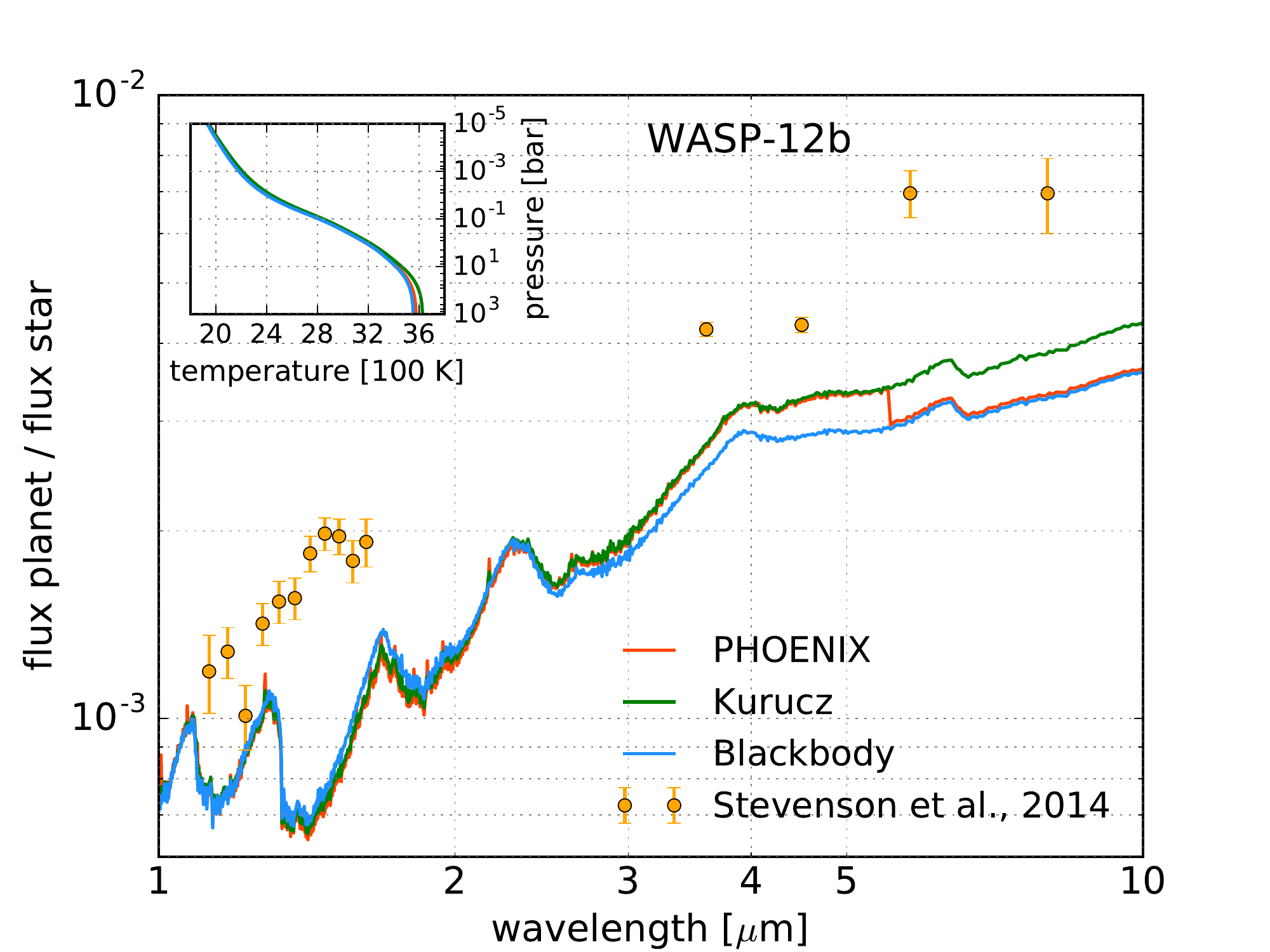}
\end{minipage}
\hfill
\begin{minipage}[t]{0.48\textwidth}
\includegraphics[width=\textwidth]{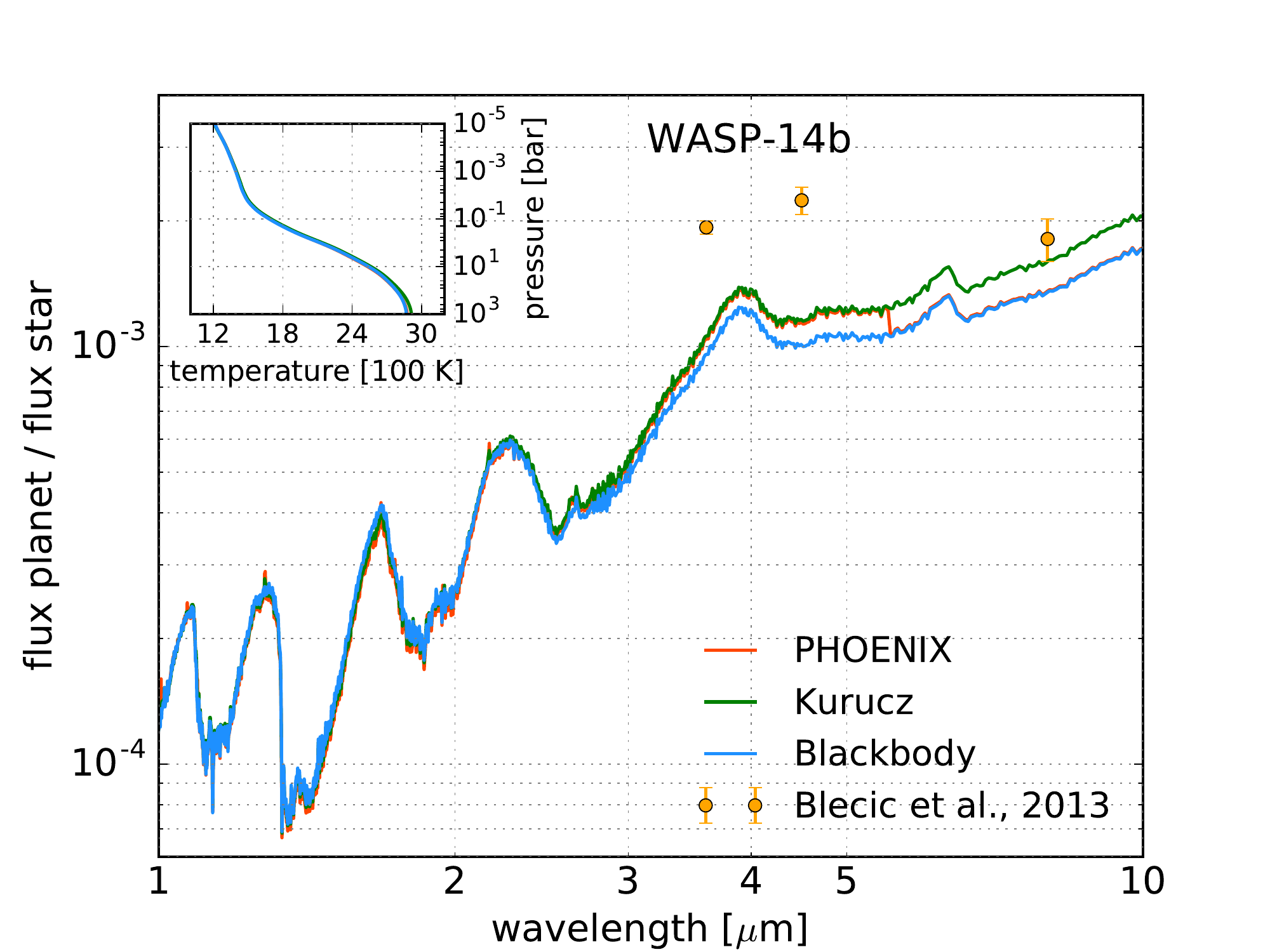}
\end{minipage}
\begin{minipage}[t]{0.48\textwidth}
\includegraphics[width=\textwidth]{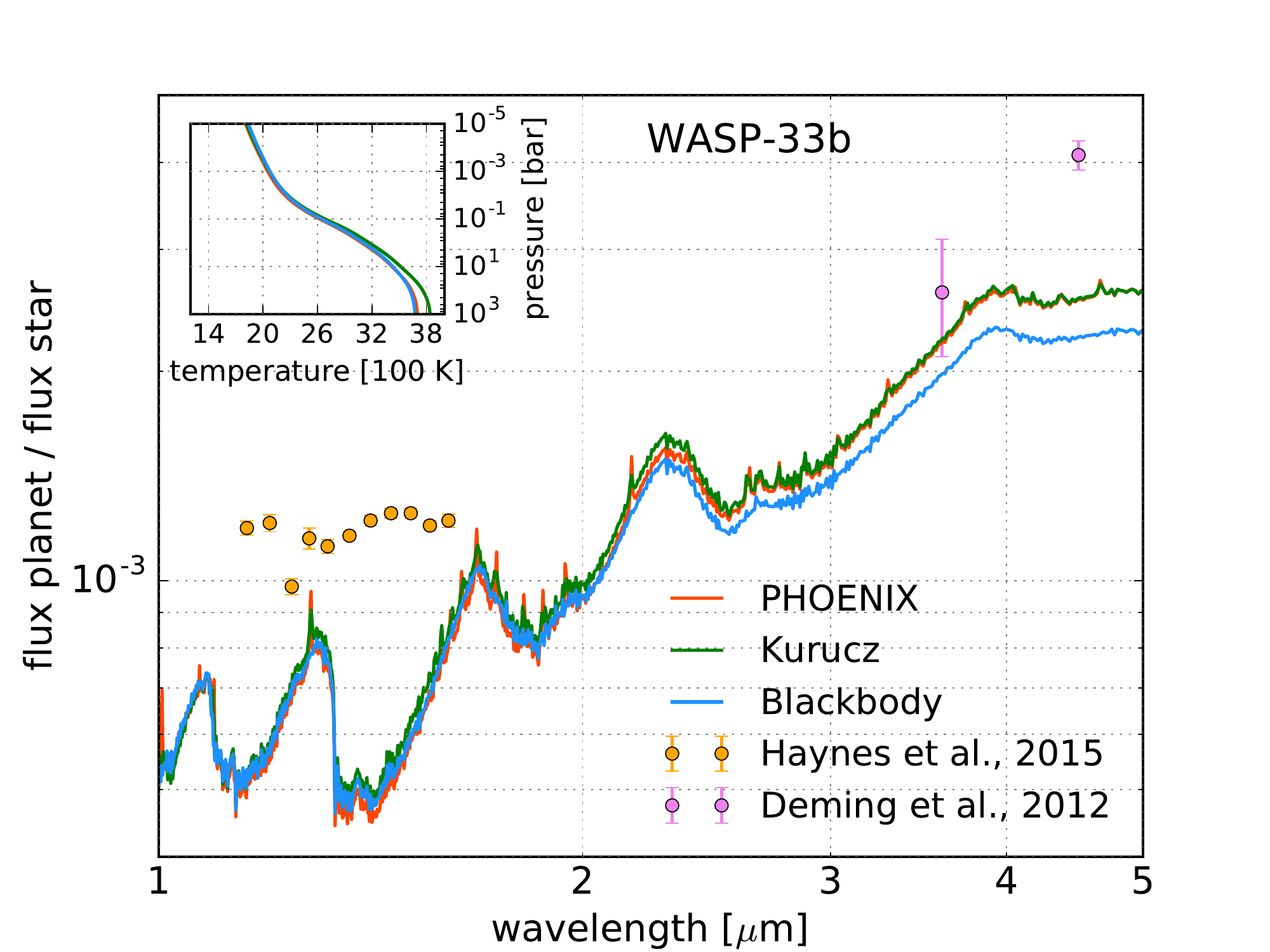}
\end{minipage}
\hfill
\begin{minipage}[t]{0.48\textwidth}
\includegraphics[width=\textwidth]{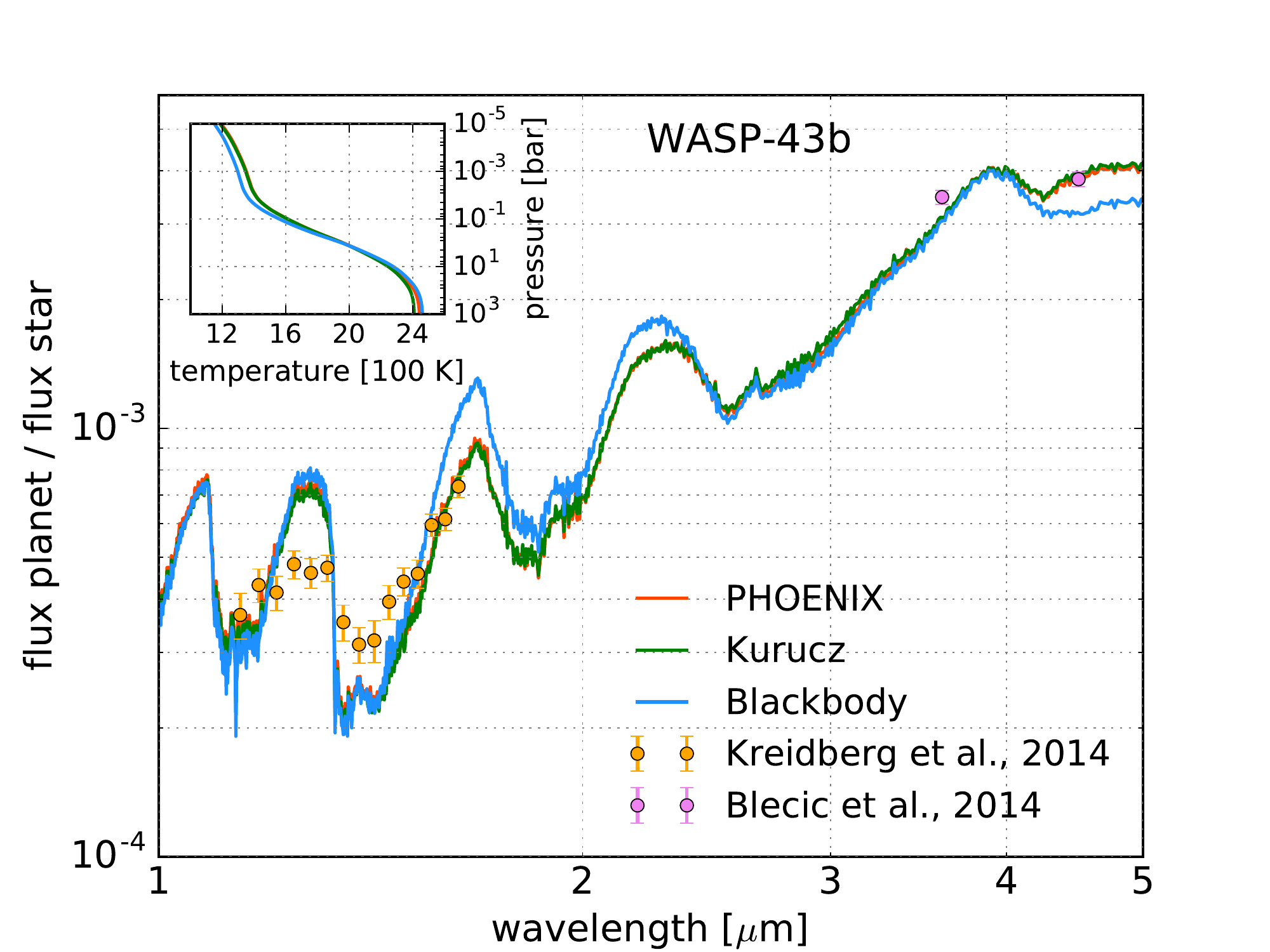}
\end{minipage}
\vspace{-0.1cm}
\caption{Null-hypothesis models for the 6 hot Jupiters in our current study: 1D, plane-parallel model atmospheres in chemical and radiative equilibrium, with solar metallicity/abundances.  The predicted dayside emission spectra were compared to published data (see text for details).  For each case study, we computed three models using the \texttt{PHOENIX} and Kurucz stellar models as well as a stellar blackbody.  For each assumption of the stellar irradiation flux, we iterated the model atmosphere to attain radiative equilibrium (see text for details).}
\label{fig:bench_star}
\end{center}
\end{figure*}

Figure \ref{fig:bench_star} shows the null-hypothesis models for all 6 studies.  We have computed synthetic spectra and temperature-pressure profiles using a stellar blackbody, a Kurucz stellar model and a \texttt{PHOENIX} stellar model.  All of the stellar models were customized for each case study by specifying, as input parameters, the stellar effective temperature, surface gravity and metallicity.  The synthetic spectra in all three cases are qualitatively similar.  The largest difference occurs between 3 and 10 $\mu$m.  These differences appear to be more pronounced for the hottest hot Jupiters (i.e., WASP-12b and WASP-33b).  Interestingly, the choice of stellar model affects the strength of the water-band features between 1.5 and 2.5 $\mu$m, which are partially probed by the WFC3 instrument on the Hubble Space Telescope.  This discrepancy between the models is somewhat apparent for HD 189733b and WASP-43b.  The shapes of the temperature-pressure profiles, in all 6 cases, are very similar with the largest discrepancies in either the very high optically thin or deep optically thick layers, which are less important for the planetary emission.

Overall, HD 189733b appears to be consistent with a null hypothesis and its dayside emission spectrum is reasonably described by a 1D, plane-parallel model in chemical and radiative equilibrium with solar metallicity.  WASP-43b is fairly well described by the null hypothesis.  However, our models for WASP-8b, WASP-12b, WASP-14b and WASP-33b consistently under-predict the infrared fluxes.  These discrepancies could be either due to an insufficient opacity implementation (lacking partial molecular absorption, aerosol extinction or inaccurate line profiles) or due to a limited methodological framework, lacking chemical disequilibrium (which requires a self-consistent calculation coupled to a chemical kinetics solver), radiative disequilibrium (which requires another self-consistent calculation coupled to atmospheric dynamics) or non-1D effects (which a 1D model prescription with $f$ cannot characterize and which would ideally require coupling to a 3D spatially resolved general circulation model).  We will defer this investigation to future work.

\begin{figure*}
\begin{center}
\begin{minipage}[t]{0.48\textwidth}
\includegraphics[width=\textwidth]{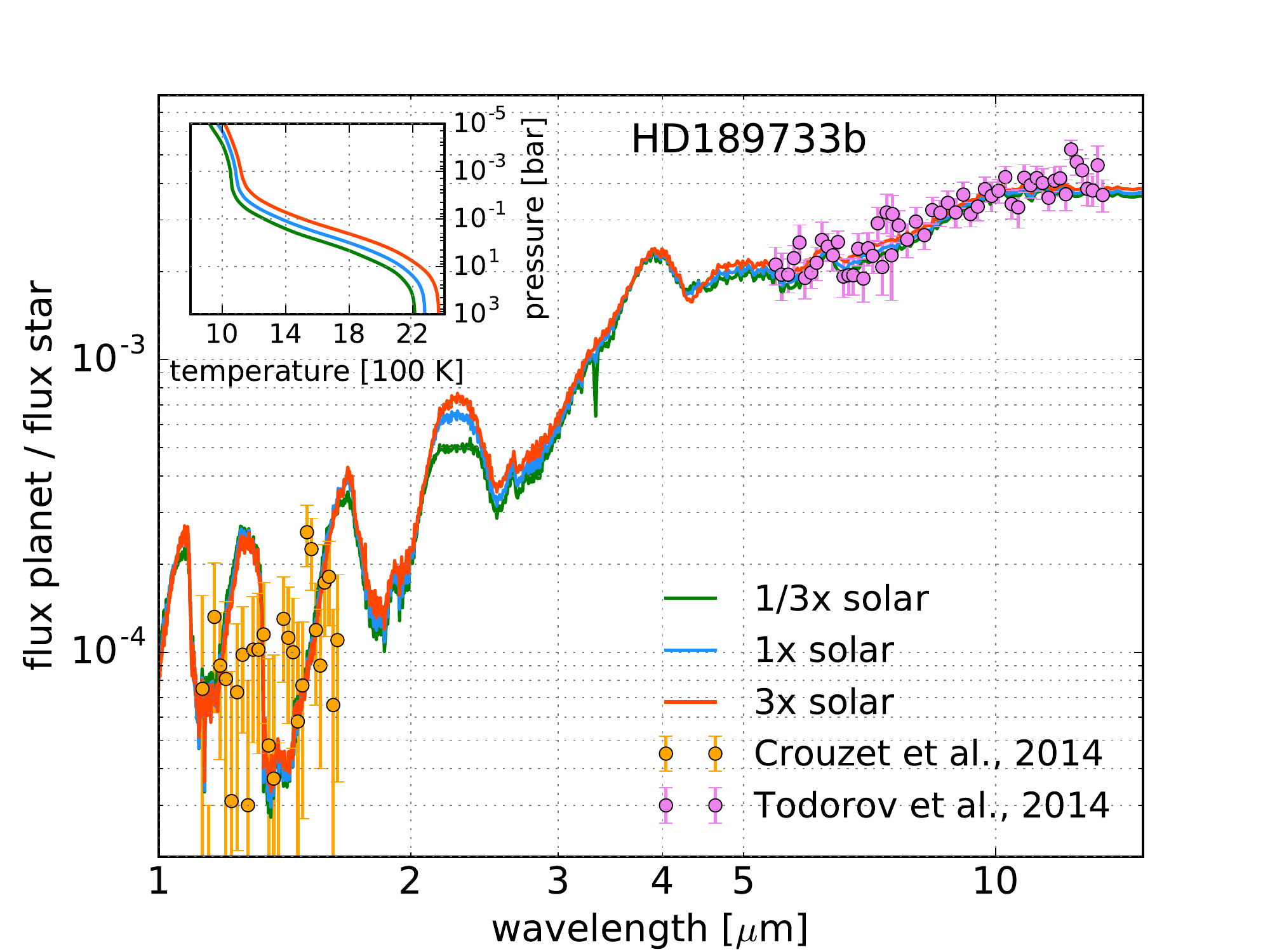}
\end{minipage}
\hfill
\begin{minipage}[t]{0.48\textwidth}
\includegraphics[width=\textwidth]{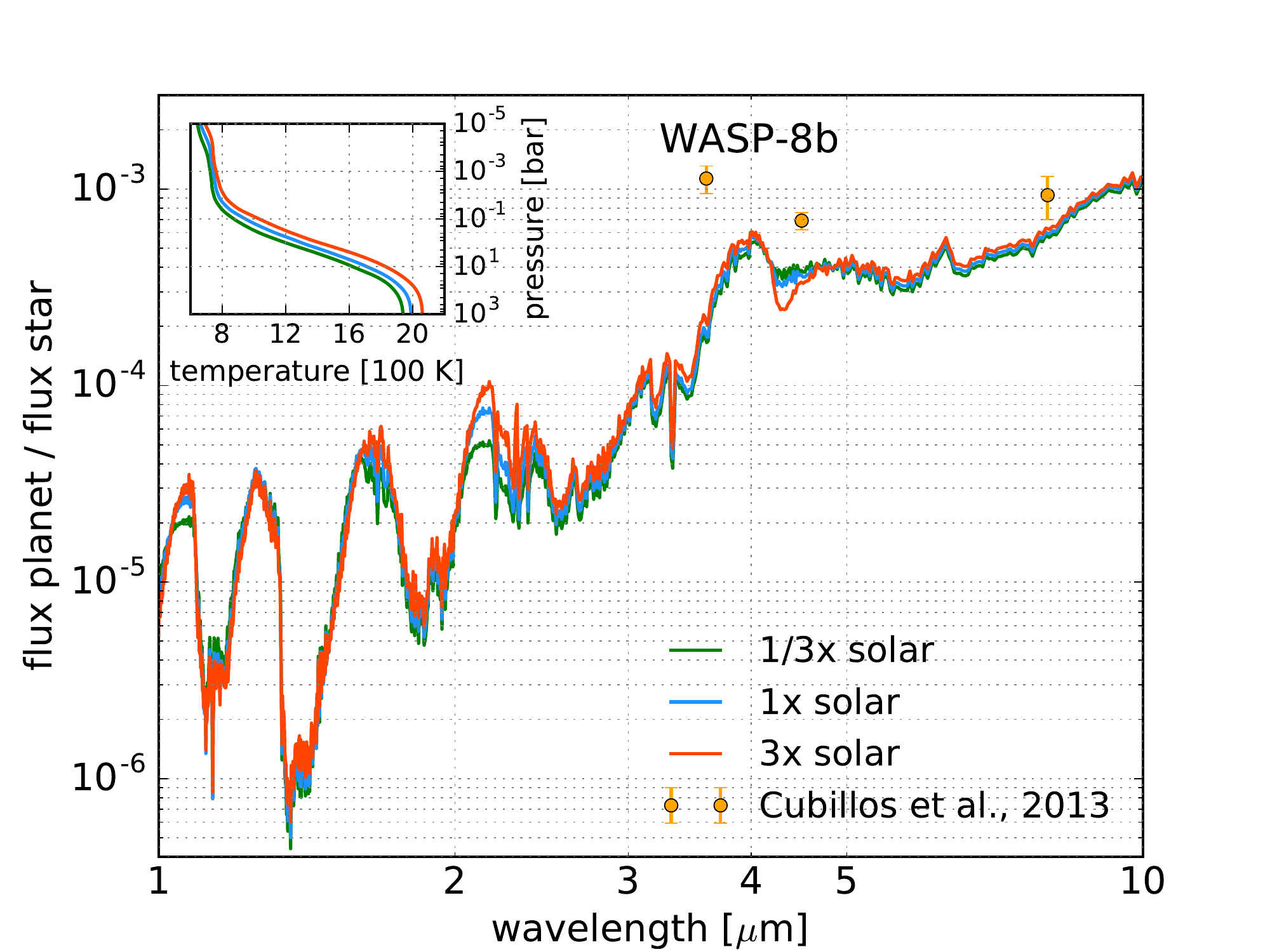}
\end{minipage}
\begin{minipage}[t]{0.48\textwidth}
\includegraphics[width=\textwidth]{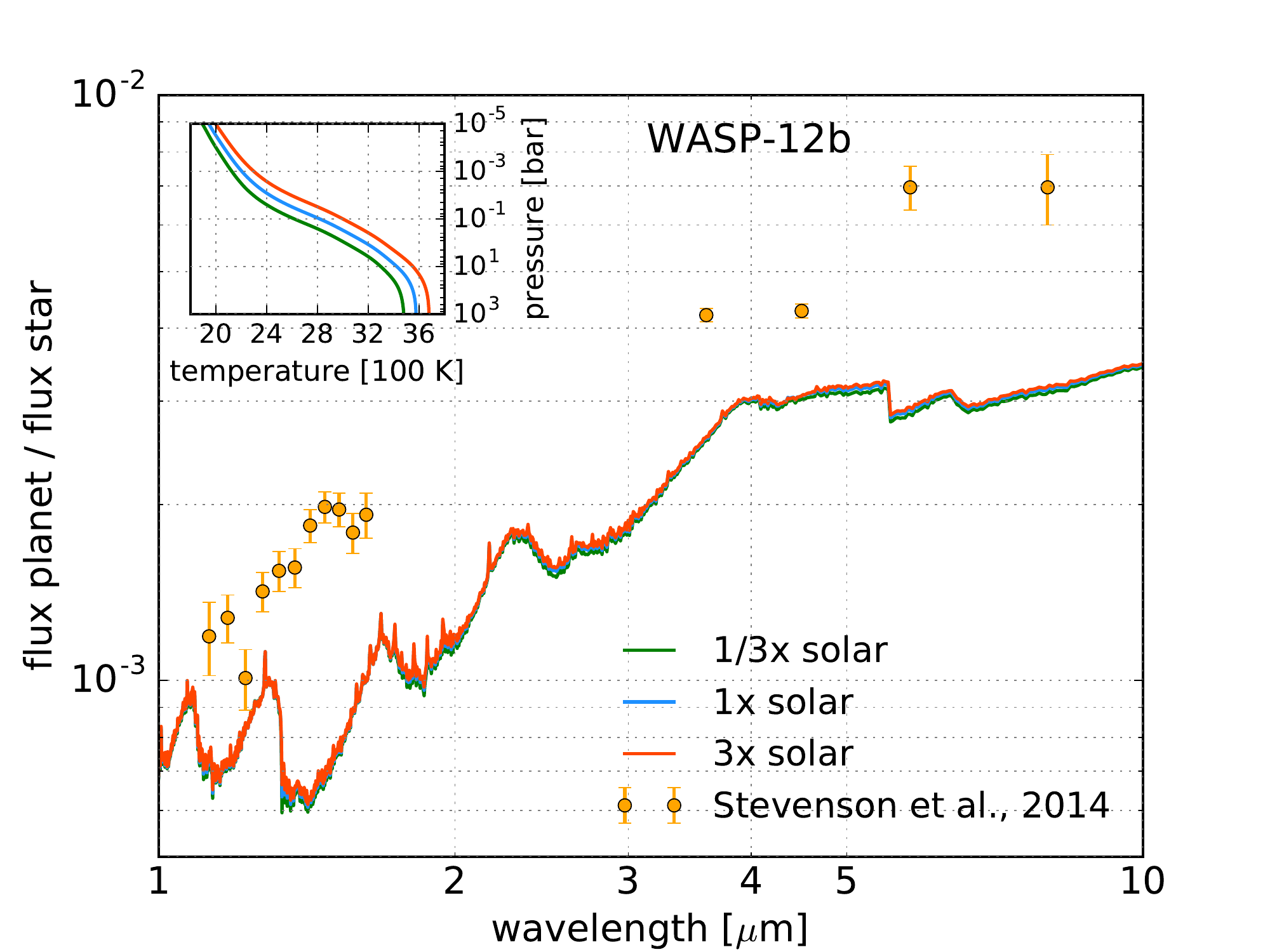}
\end{minipage}
\hfill
\begin{minipage}[t]{0.48\textwidth}
\includegraphics[width=\textwidth]{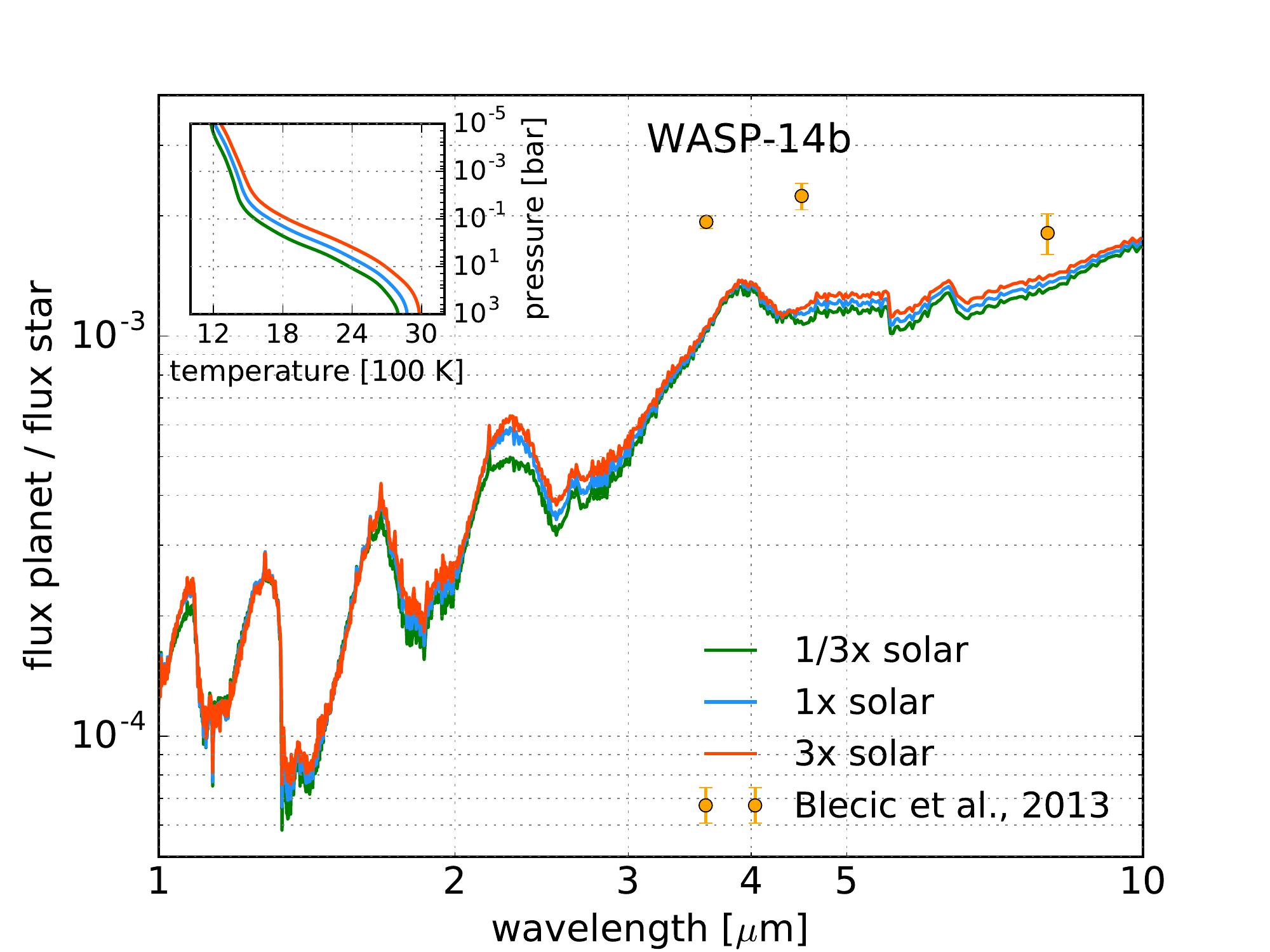}
\end{minipage}
\begin{minipage}[t]{0.48\textwidth}
\includegraphics[width=\textwidth]{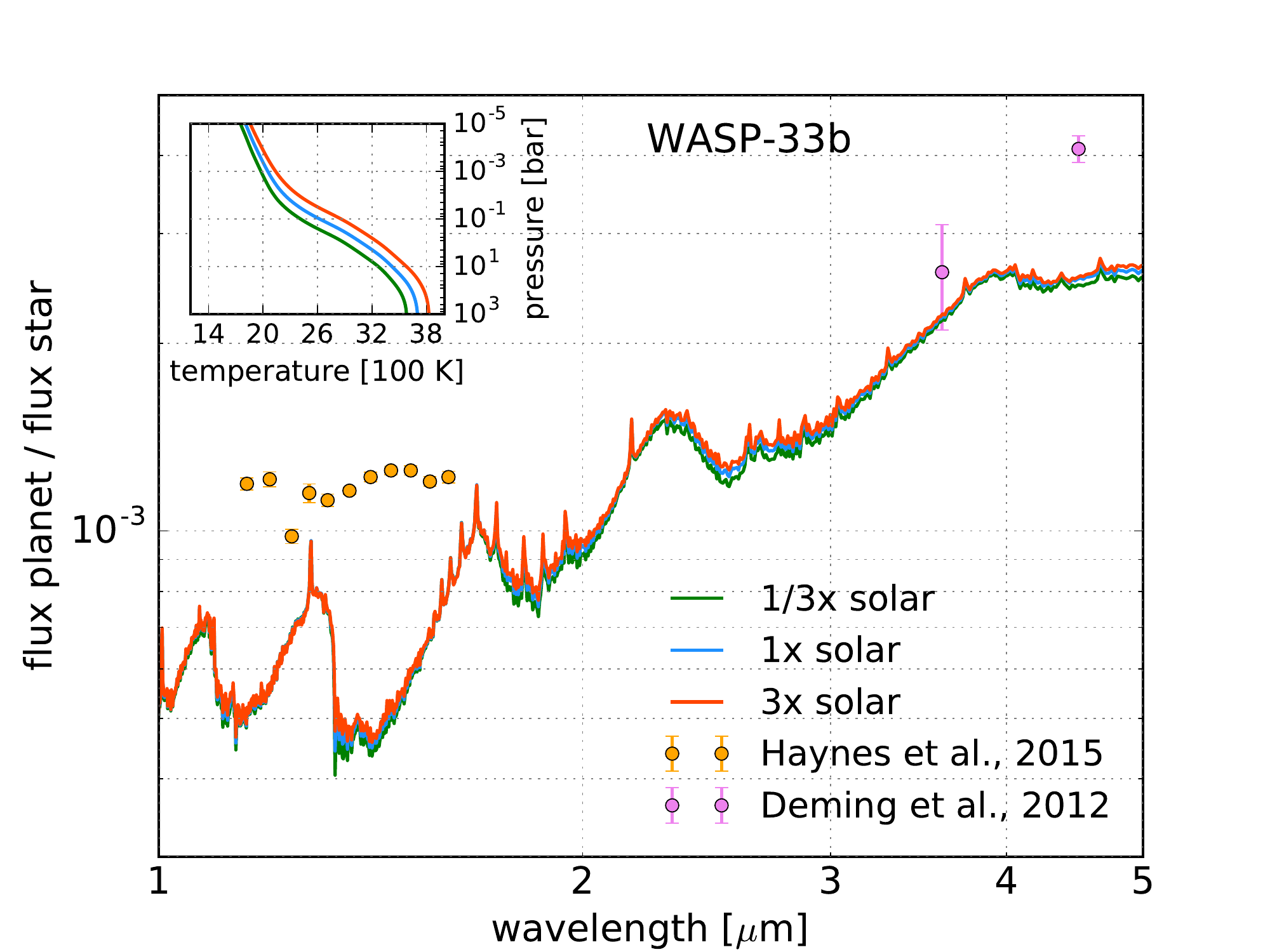}
\end{minipage}
\hfill
\begin{minipage}[t]{0.48\textwidth}
\includegraphics[width=\textwidth]{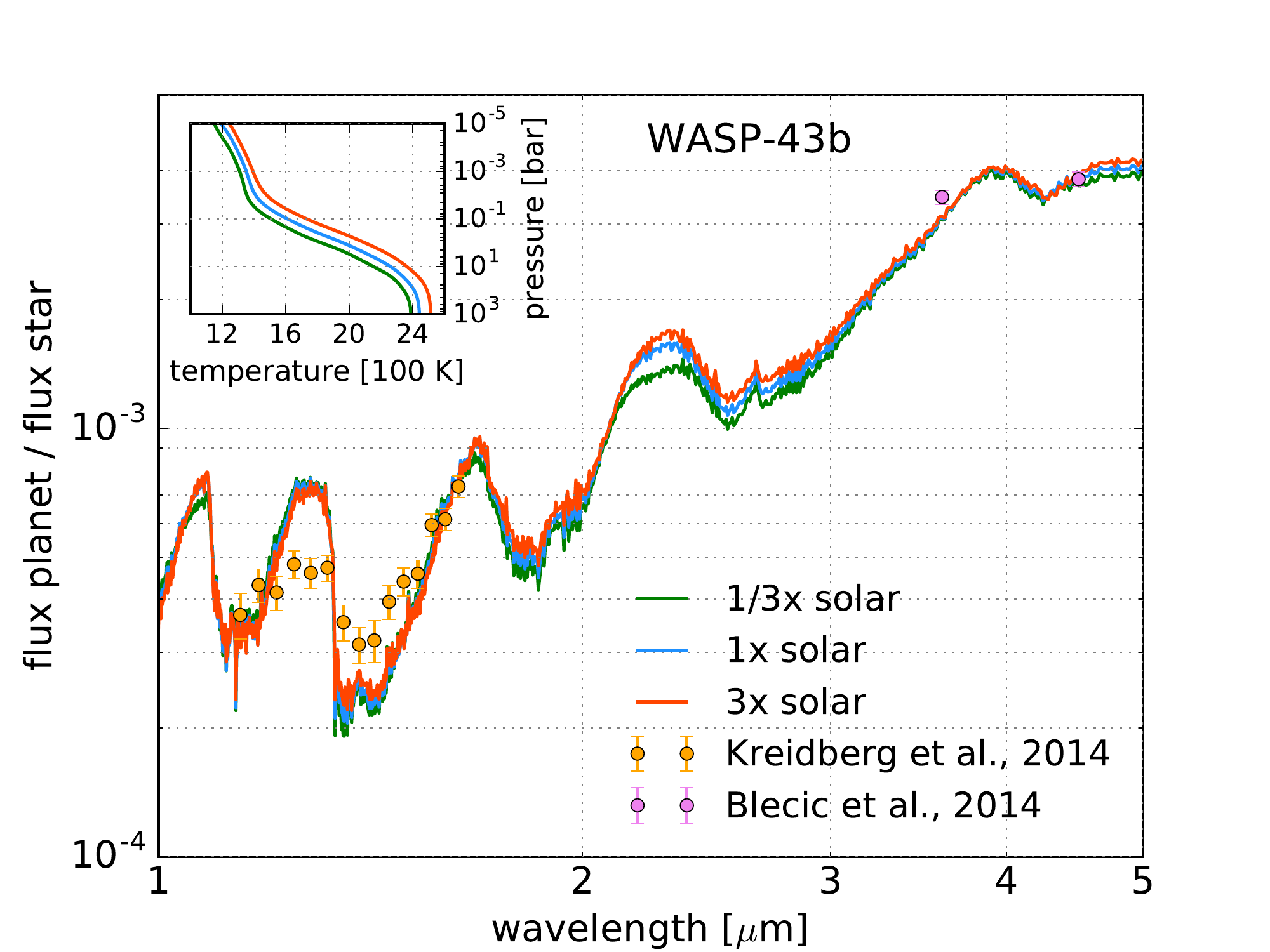}
\end{minipage}
\vspace{-0.1cm}
\caption{Same as Figure \ref{fig:bench_star}, but using only the \texttt{PHOENIX} stellar model and examining the effects of varying the metallicity of the model atmospheres.}
\label{fig:bench_metal}
\end{center}
\end{figure*}

For further variations on the theme, we retain the \texttt{PHOENIX} stellar models as they offer higher spectral resolution and more updated atomic/molecular line lists than the Kurucz stellar models.  In Figure \ref{fig:bench_metal}, we repeat our calculations with 1/3$\times$, 1$\times$ and 3$\times$ solar metallicity.  We find the expected trend that a higher metallicity leads to generally hotter model atmospheres, which has the effect of strengthening the near-infrared water-band features.  However, compared to the null hypothesis, decreasing or increasing the metallicity by a factor of 3 appears to have a minimal effect on the synthetic spectra, which is consistent with the retrieval analysis conducted for WASP-43b in \cite{kr14}, where they obtain similar metallicity uncertainties based on data constraints.  Our conclusions are qualitatively identical to those visible in Figure \ref{fig:bench_star}.

\begin{figure*}
\begin{center}
\begin{minipage}[t]{0.48\textwidth}
\includegraphics[width=\textwidth]{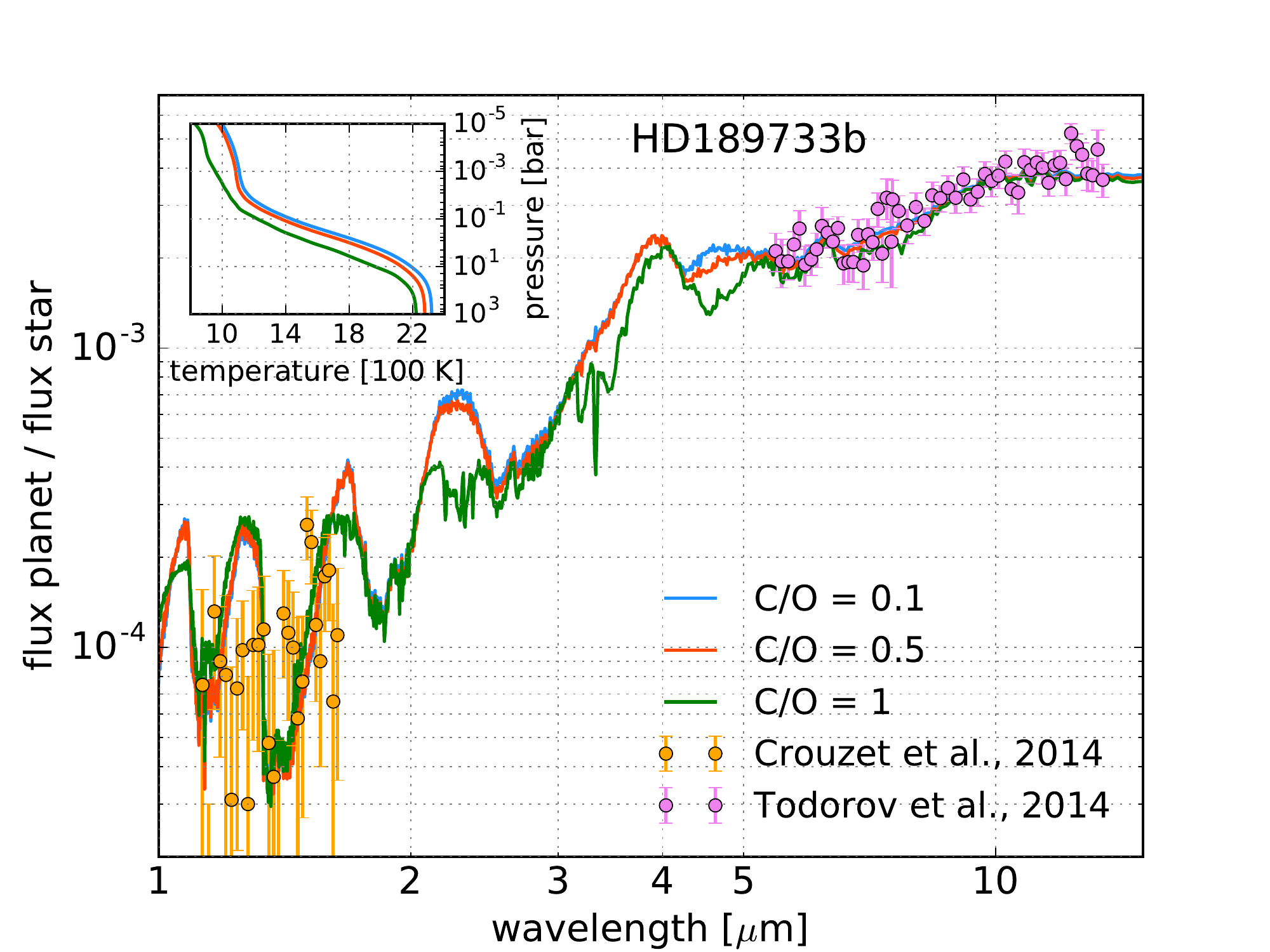}
\end{minipage}
\hfill
\begin{minipage}[t]{0.48\textwidth}
\includegraphics[width=\textwidth]{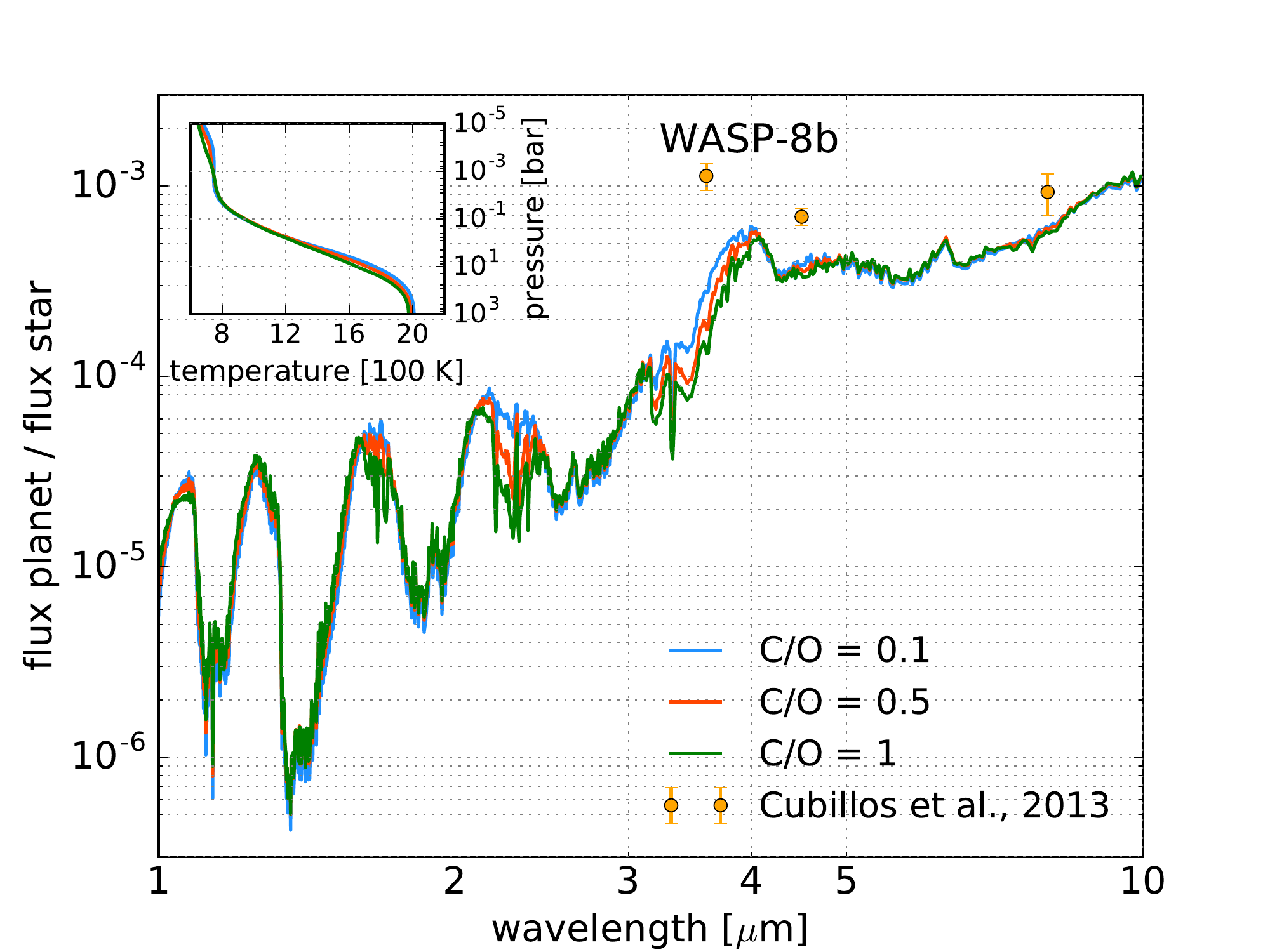}
\end{minipage}
\begin{minipage}[t]{0.48\textwidth}
\includegraphics[width=\textwidth]{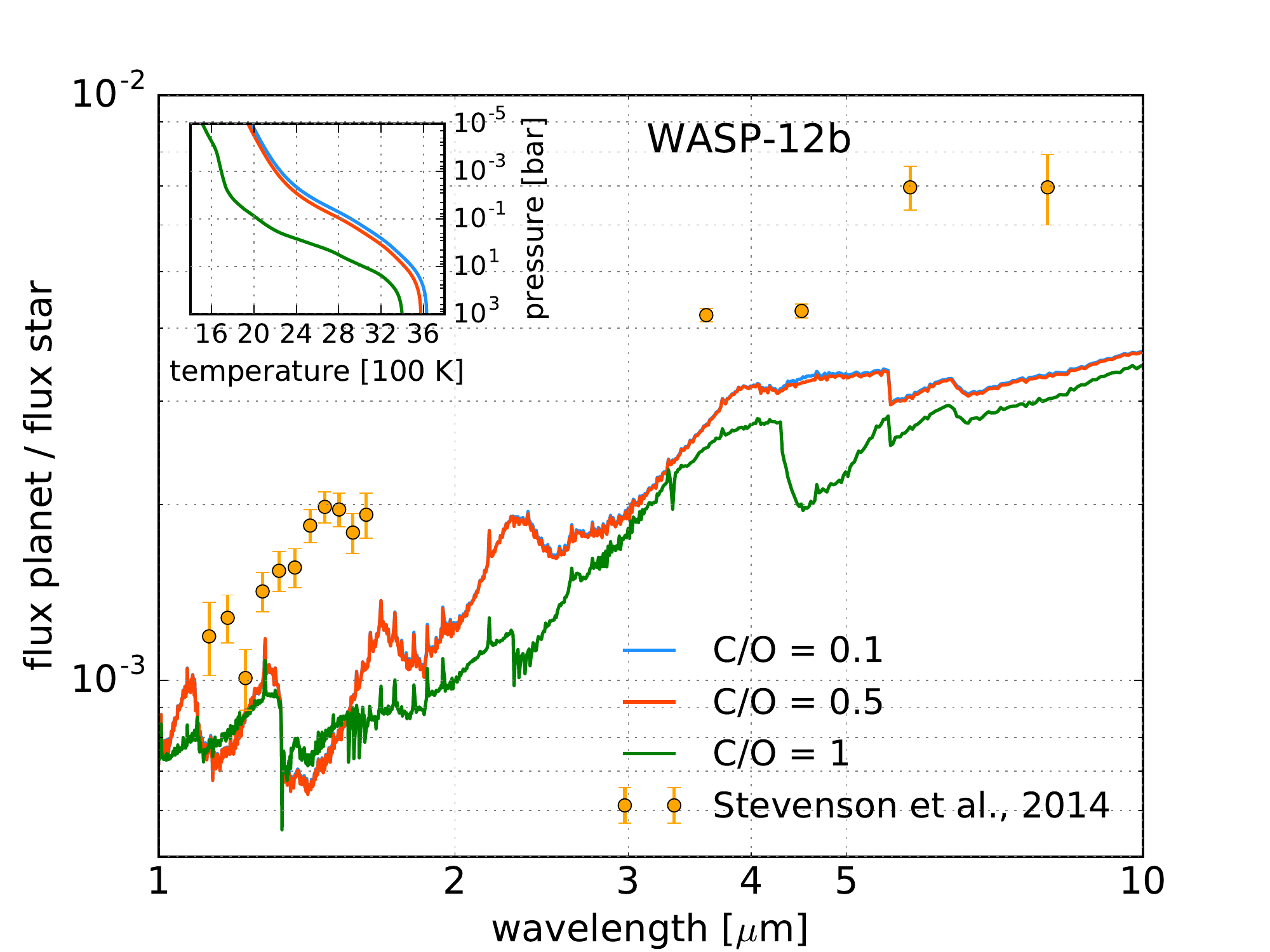}
\end{minipage}
\hfill
\begin{minipage}[t]{0.48\textwidth}
\includegraphics[width=\textwidth]{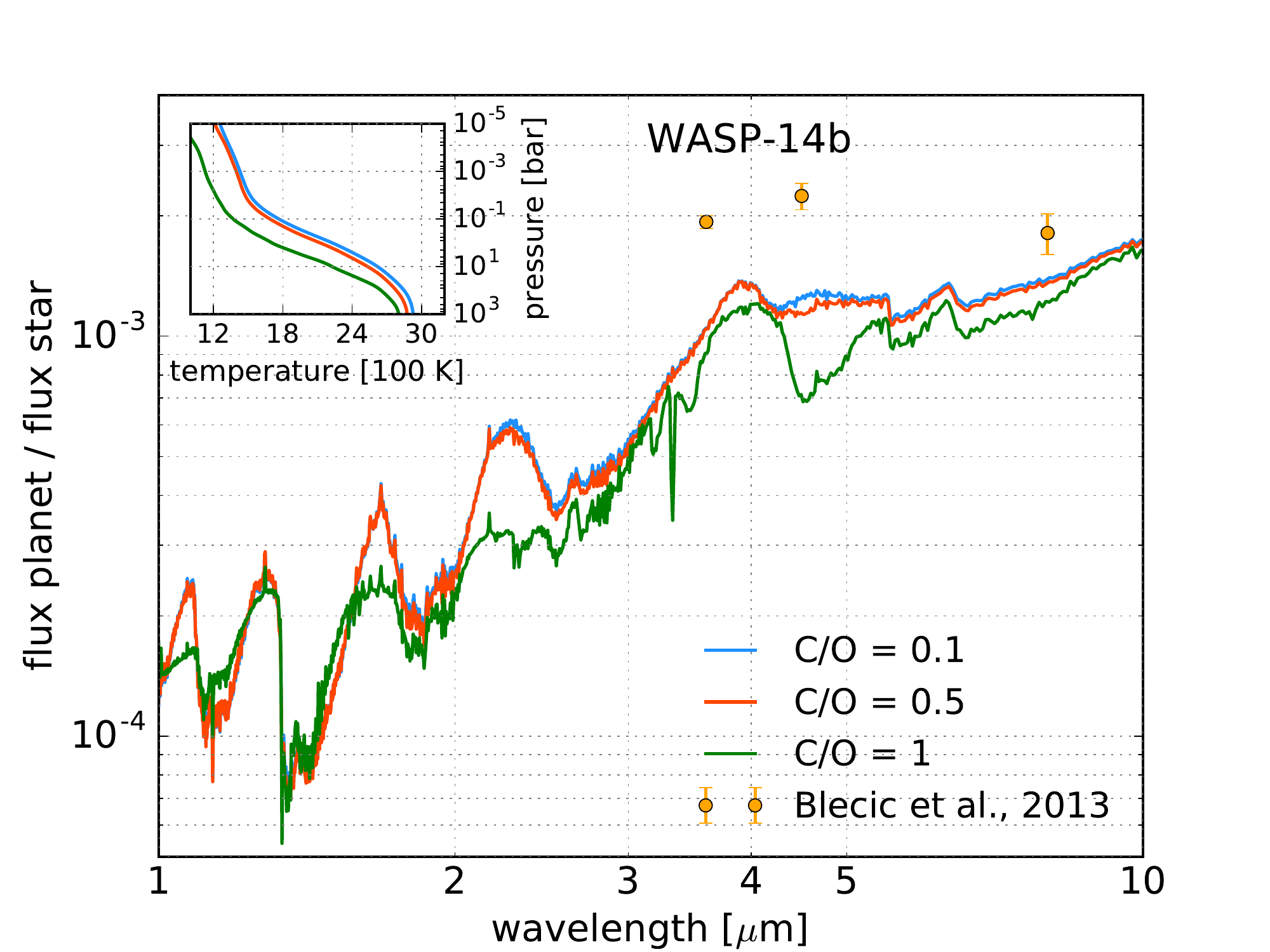}
\end{minipage}
\begin{minipage}[t]{0.48\textwidth}
\includegraphics[width=\textwidth]{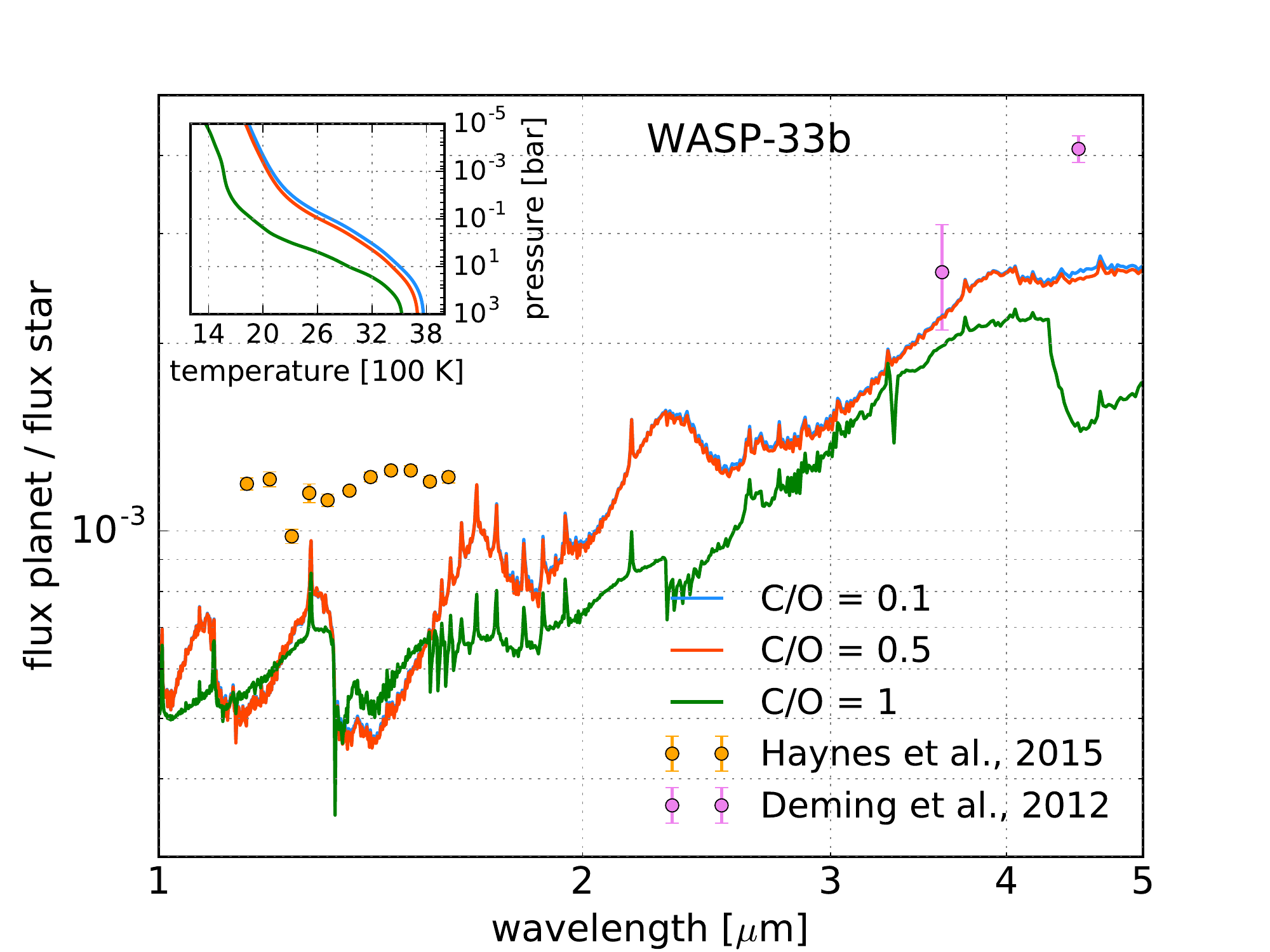}
\end{minipage}
\hfill
\begin{minipage}[t]{0.48\textwidth}
\includegraphics[width=\textwidth]{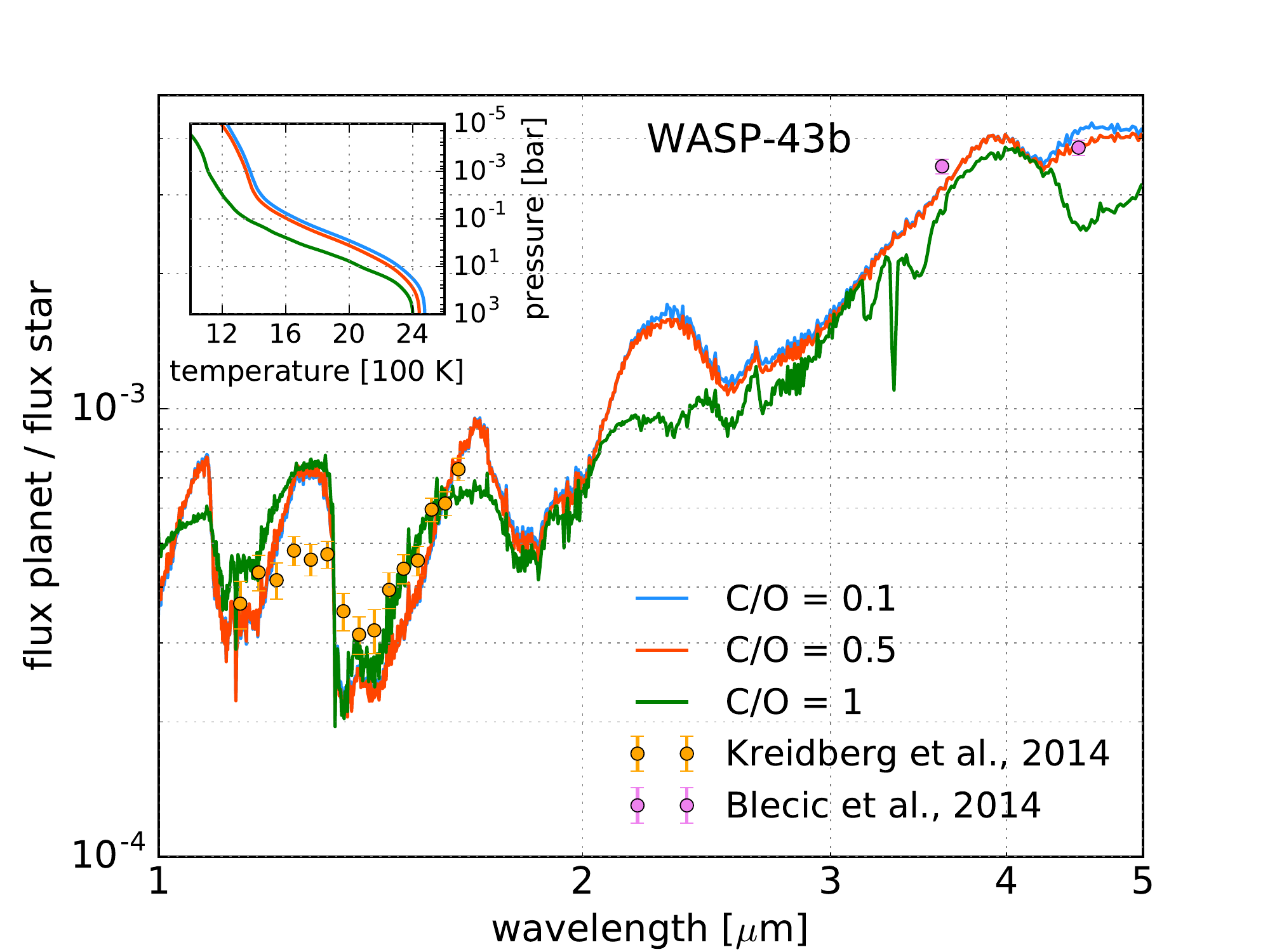}
\end{minipage}
\vspace{-0.1cm}
\caption{Same as Figure \ref{fig:bench_star}, but using only the \texttt{PHOENIX} stellar model and examining the effects of varying the C/O (0.1, 0.5 and 1) of the model atmospheres.}
\label{fig:bench_coratio}
\end{center}
\end{figure*}

Varying the C/O has a more marked effect, as we show in Figure \ref{fig:bench_coratio}.  Specifically, we examine water-rich (C/O$=0.1$), solar-abundance (C/O$=0.5$) and C/O$=1$ scenarios.  Generally, we find that the C/O$=1$ models have consistently colder temperature-pressure profiles, due to the lower abundance of H$_2$O as the oxygen atom is preferentially sequestered by CO, at high temperatures, compared to the water-rich and solar-abundance models.  The increasing abundance of CO also leads to stronger absorption features at 2.3, 4.5 and 4.8 $\mu$m, which render the model atmospheres darker (i.e., they have less flux in these bands).  This transition to the stronger CO features is more pronounced in the hotter objects (WASP-12b and WASP-33b).  Our qualitative conclusions appear to be unchanged: our models for WASP-8b, WASP-12b, WASP-14b and WASP-33b still under-predict the infrared fluxes.  It is somewhat difficult to judge if the data favours the water-rich or C/O$=1$ models, for HD 189733b and WASP-43b, without running a detailed atmospheric retrieval model, which we again defer to future work.

\section{Summary, Discussion \& Conclusions}
\label{sec:sum}

\subsection{Summary}

We have presented the new, extensible code, \texttt{HELIOS}, which solves the equation of radiative transfer for a 1D, plane-parallel atmosphere that allows for non-isotropic scattering via the specification of the functional forms of the single-scattering albedo and the scattering asymmetry factor.  It uses a staggered spatial grid with the options of specifying isothermal or non-isothermal layers. We have used \texttt{HELIOS-K} \citep{gr15} to compute the opacities of the four molecules, which are active in the infrared, and combined those by weighing them with the validated analytical formulae of \cite{he16b} and \cite{he16c}  for equilibrium chemistry. In order to combine the various gaseous absorbers we have employed a correlated-$k$ approximation, which assumes perfect correlation between the molecular bands. The boundary conditions are the stellar irradiation flux at the top of the model atmosphere and the internal heat flux at the bottom.  \texttt{HELIOS} further allows for the stellar irradiation flux to be specified as a simple Planck function or from a stellar model (e.g. Kurucz, \texttt{PHOENIX}).  We have constructed and optimized \texttt{HELIOS} to run on GPUs, which allows for fast computation on a single machine.  We have exploited this efficiency to explore the parameter space of stellar type, metallicity and C/O ratio.

\subsection{Comparison to previous work}

Several groups have made contributions to a rich body of literature on self-consistent radiative transfer models in exoplanetary atmospheres.  The work of \cite{bu06,bu07,bu08} uses the accelerated lambda iteration method, originally developed for stellar atmospheres \citep{hubeny95}.  The work of \cite{fo05,fo06,fo08,fo10} and \cite{mo13,mo15} use an atmosphere modeling code and radiative transfer methods with a heritage from brown dwarf and Solar System models \citep{mc89,toon89,burrows97,ma99}.  \cite{amundsen14} recently implemented a radiative transfer code using the two-stream approximation in the limit of pure absorption.  \cite{molli15} constructed a pure-absorption code using the ``variable Eddington factor" method, which has a heritage from the study of stellar atmospheres (e.g. \citealt{auer70}) and protoplanetary disks \citep{dullemond02}. Our approach and assembly of the various components (see above) and their collective implementation is a novel endeavor and we hope it will contribute to the advancement of this field.

\subsection{Discussion and Opportunities for Future Work}

In the current work, we have considered a small set of the four main infrared absorbers (H$_2$O, CO$_2$, CO, CH$_4$), and included the opacity associated with CIA from H$_2$-H$_2$ and H$_2$-He pairs.  Future work should include more opacity sources, especially that associated with C$_2$H$_2$ and HCN, if one is interested in C/O$>1$ models, and Na and K as these are major absorbers in the visible for very hot planets. Also, important at the higher-temperature end of exoplanets is continuum absorption by electrons moving freely in the field or being decoupled from the shell of neutral atoms (e.g. H, He), molecules (e.g. H$_2$) or ions (e.g. H$^-$) \citep{sh07}. Furthermore, it is important to conduct a study examining the accuracy of the employed correlated-$k$ approximation for different combinations of molecular absorbers, since this could be a potential source of error---a study similar to \cite{amundsen16} for the random-overlap scheme. Another opportunity for future work is the inclusion of aerosols and clouds, whose proper implementation remains a subject of debate. Additionally, we will implement convective adjustment as the next step in sophistication and we plan to investigate the effect of disequilibrium chemistry (induced by both atmospheric motion and photochemistry) and radiative disequilibrium by coupling \texttt{HELIOS} to a chemical kinetics code and a general circulation model.  Hot Jupiters are complex, three-dimensional entities (e.g., \citealt{burrows10}) and interpreting them, on a detailed case-by-case basis, requires a three-dimensional model (e.g., \citealt{kataria15}).  The exact interpretation of the molecular abundances associated with the 6 hot Jupiters may be performed using an atmospheric retrieval code.  \texttt{HELIOS} is a key component of the open-source Exoclimes Simulation Platform (ESP; \url{exoclime.net}), which includes a chemical kinetics code \citep{tsai16}, retrieval code \citep{lavie16} and general circulation models (\citealt{mendonca16}; Grosheintz et al., in preparation). The up-to-date version of \texttt{HELIOS} may be downloaded from its main repository \url{github.com/exoclime/HELIOS} and the version used to produce the results in this work is archived under the DOI: 10.5281/zenodo.164176.

\section*{Acknowledgements}

M.M., L.G., S.G., J.M., B.L., D.K., S.T. and K.H. thank the Swiss National Science Foundation (SNF), the Center for Space and Habitability (CSH), the PlanetS National Center of Competence in Research (NCCR) and the MERAC Foundation for partial financial support. 

\software{\\
\texttt{HELIOS-K} (\citealt{gr15}; \url{github.com/exoclime/HELIOS-K}), \\
\texttt{CUDA} \citep{cuda}, \\
\texttt{PyCUDA} \citep{pycuda}, \\
\texttt{python} \citep{python}, \\ 
\texttt{scipy} \citep{scipy}, \\ 
\texttt{numpy} \citep{numpy}, \\
\texttt{matplotlib} \citep{matplotlib}.
}

\appendix

\section{Licensing and Permission to use the \texttt{TEA} Code}

We thank the developers of the Thermochemical Equilibrium Abundances (\texttt{TEA}) code \citep{bl15}, initially developed at the University of Central Florida, Orlando, Florida, USA.  The Reproducible-Research Compendium (RRC) is available at \texttt{github.com/exoclime/HELIOS.publications}.

\bibliographystyle{apj}

\begin{thebibliography}{}
\expandafter\ifx\csname natexlab\endcsname\relax\def\natexlab#1{#1}\fi

\bibitem[{{Abramowitz} \& {Stegun}(1972)}]{abramowitz72}
{Abramowitz}, M., \& {Stegun}, I.~A. 1972, {Handbook of Mathematical Functions}

\bibitem[{{Allard} \& {Hauschildt}(1995)}]{al95}
{Allard}, F., \& {Hauschildt}, P.~H. 1995, \apj, 445, 433

\bibitem[{{Amundsen} {et~al.}(2014){Amundsen}, {Baraffe}, {Tremblin},
  {Manners}, {Hayek}, {Mayne}, \& {Acreman}}]{amundsen14}
{Amundsen}, D.~S., {Baraffe}, I., {Tremblin}, P., {et~al.} 2014, \aap, 564, A59

\bibitem[{{Amundsen} {et~al.}(2016){Amundsen}, {Tremblin}, {Manners},
  {Baraffe}, \& {Mayne}}]{amundsen16}
{Amundsen}, D.~S., {Tremblin}, P., {Manners}, J., {Baraffe}, I., \& {Mayne},
  N.~J. 2016, ArXiv e-prints, arXiv:1610.01389

\bibitem[{{Anglada-Escud{\'e}} {et~al.}(2013){Anglada-Escud{\'e}},
  {Rojas-Ayala}, {Boss}, {Weinberger}, \& {Lloyd}}]{an13}
{Anglada-Escud{\'e}}, G., {Rojas-Ayala}, B., {Boss}, A.~P., {Weinberger},
  A.~J., \& {Lloyd}, J.~P. 2013, \aap, 551, A48

\bibitem[{Armstrong(1969)}]{armstrong69}
Armstrong, B.~H. 1969, Journal of the Atmospheric Sciences, 26, 741

\bibitem[{{Auer} \& {Mihalas}(1970)}]{auer70}
{Auer}, L.~H., \& {Mihalas}, D. 1970, \mnras, 149, 65

\bibitem[{{Barman} {et~al.}(2005){Barman}, {Hauschildt}, \& {Allard}}]{ba05}
{Barman}, T.~S., {Hauschildt}, P.~H., \& {Allard}, F. 2005, \apj, 632, 1132

\bibitem[{{Blecic} {et~al.}(2015){Blecic}, {Harrington}, \& {Bowman}}]{bl15}
{Blecic}, J., {Harrington}, J., \& {Bowman}, M.~O. 2015, ArXiv e-prints,
  arXiv:1505.06392

\bibitem[{{Blecic} {et~al.}(2013){Blecic}, {Harrington}, {Madhusudhan},
  {Stevenson}, {Hardy}, {Cubillos}, {Hardin}, {Campo}, {Bowman}, {Nymeyer},
  {Loredo}, {Anderson}, \& {Maxted}}]{bl13}
{Blecic}, J., {Harrington}, J., {Madhusudhan}, N., {et~al.} 2013, \apj, 779, 5

\bibitem[{{Blecic} {et~al.}(2014){Blecic}, {Harrington}, {Madhusudhan},
  {Stevenson}, {Hardy}, {Cubillos}, {Hardin}, {Bowman}, {Nymeyer}, {Anderson},
  {Hellier}, {Smith}, \& {Collier Cameron}}]{bl14}
---. 2014, \apj, 781, 116

\bibitem[{{Bouchy} {et~al.}(2005){Bouchy}, {Udry}, {Mayor}, {Moutou}, {Pont},
  {Iribarne}, {da Silva}, {Ilovaisky}, {Queloz}, {Santos}, {S{\'e}gransan}, \&
  {Zucker}}]{bouchy05}
{Bouchy}, F., {Udry}, S., {Mayor}, M., {et~al.} 2005, \aap, 444, L15

\bibitem[{{Boyajian} {et~al.}(2015){Boyajian}, {von Braun}, {Feiden}, {Huber},
  {Basu}, {Demarque}, {Fischer}, {Schaefer}, {Mann}, {White}, {Maestro},
  {Brewer}, {Lamell}, {Spada}, {L{\'o}pez-Morales}, {Ireland}, {Farrington},
  {van Belle}, {Kane}, {Jones}, {ten Brummelaar}, {Ciardi}, {McAlister},
  {Ridgway}, {Goldfinger}, {Turner}, \& {Sturmann}}]{bo15}
{Boyajian}, T., {von Braun}, K., {Feiden}, G.~A., {et~al.} 2015, \mnras, 447,
  846

\bibitem[{{Brown} {et~al.}(2001){Brown}, {Charbonneau}, {Gilliland}, {Noyes},
  \& {Burrows}}]{br01}
{Brown}, T.~M., {Charbonneau}, D., {Gilliland}, R.~L., {Noyes}, R.~W., \&
  {Burrows}, A. 2001, \apj, 552, 699

\bibitem[{{Burrows} {et~al.}(2008){Burrows}, {Budaj}, \& {Hubeny}}]{bu08}
{Burrows}, A., {Budaj}, J., \& {Hubeny}, I. 2008, \apj, 678, 1436

\bibitem[{{Burrows} {et~al.}(2007){Burrows}, {Hubeny}, {Budaj}, {Knutson}, \&
  {Charbonneau}}]{bu07}
{Burrows}, A., {Hubeny}, I., {Budaj}, J., {Knutson}, H.~A., \& {Charbonneau},
  D. 2007, \apjl, 668, L171

\bibitem[{{Burrows} {et~al.}(2010){Burrows}, {Rauscher}, {Spiegel}, \&
  {Menou}}]{burrows10}
{Burrows}, A., {Rauscher}, E., {Spiegel}, D.~S., \& {Menou}, K. 2010, \apj,
  719, 341

\bibitem[{{Burrows} {et~al.}(2006){Burrows}, {Sudarsky}, \& {Hubeny}}]{bu06}
{Burrows}, A., {Sudarsky}, D., \& {Hubeny}, I. 2006, \apj, 650, 1140

\bibitem[{{Burrows} {et~al.}(1997){Burrows}, {Marley}, {Hubbard}, {Lunine},
  {Guillot}, {Saumon}, {Freedman}, {Sudarsky}, \& {Sharp}}]{burrows97}
{Burrows}, A., {Marley}, M., {Hubbard}, W.~B., {et~al.} 1997, \apj, 491, 856

\bibitem[{{Cahoy} {et~al.}(2010){Cahoy}, {Marley}, \& {Fortney}}]{cahoy10}
{Cahoy}, K.~L., {Marley}, M.~S., \& {Fortney}, J.~J. 2010, \apj, 724, 189

\bibitem[{{Chan} {et~al.}(2011){Chan}, {Ingemyr}, {Winn}, {Holman},
  {Sanchis-Ojeda}, {Esquerdo}, \& {Everett}}]{ch11}
{Chan}, T., {Ingemyr}, M., {Winn}, J.~N., {et~al.} 2011, \aj, 141, 179

\bibitem[{{Chandrasekhar}(1960)}]{ch60}
{Chandrasekhar}, S. 1960, {Radiative transfer}

\bibitem[{{Charbonneau} {et~al.}(2002){Charbonneau}, {Brown}, {Noyes}, \&
  {Gilliland}}]{ch02}
{Charbonneau}, D., {Brown}, T.~M., {Noyes}, R.~W., \& {Gilliland}, R.~L. 2002,
  \apj, 568, 377

\bibitem[{{Charbonneau} {et~al.}(2005){Charbonneau}, {Allen}, {Megeath},
  {Torres}, {Alonso}, {Brown}, {Gilliland}, {Latham}, {Mandushev}, {O'Donovan},
  \& {Sozzetti}}]{ch05}
{Charbonneau}, D., {Allen}, L.~E., {Megeath}, S.~T., {et~al.} 2005, \apj, 626,
  523

\bibitem[{{Collier Cameron} {et~al.}(2010){Collier Cameron}, {Guenther},
  {Smalley}, {McDonald}, {Hebb}, {Andersen}, {Augusteijn}, {Barros}, {Brown},
  {Cochran}, {Endl}, {Fossey}, {Hartmann}, {Maxted}, {Pollacco}, {Skillen},
  {Telting}, {Waldmann}, \& {West}}]{co10}
{Collier Cameron}, A., {Guenther}, E., {Smalley}, B., {et~al.} 2010, \mnras,
  407, 507

\bibitem[{{Crouzet} {et~al.}(2014){Crouzet}, {McCullough}, {Deming}, \&
  {Madhusudhan}}]{cr14}
{Crouzet}, N., {McCullough}, P.~R., {Deming}, D., \& {Madhusudhan}, N. 2014,
  \apj, 795, 166

\bibitem[{{Cubillos} {et~al.}(2013){Cubillos}, {Harrington}, {Madhusudhan},
  {Stevenson}, {Hardy}, {Blecic}, {Anderson}, {Hardin}, \& {Campo}}]{cu13}
{Cubillos}, P., {Harrington}, J., {Madhusudhan}, N., {et~al.} 2013, \apj, 768,
  42

\bibitem[{{de Kok} {et~al.}(2013){de Kok}, {Brogi}, {Snellen}, {Birkby},
  {Albrecht}, \& {de Mooij}}]{de13}
{de Kok}, R.~J., {Brogi}, M., {Snellen}, I.~A.~G., {et~al.} 2013, \aap, 554,
  A82

\bibitem[{{Deming} {et~al.}(2005){Deming}, {Seager}, {Richardson}, \&
  {Harrington}}]{deming05}
{Deming}, D., {Seager}, S., {Richardson}, L.~J., \& {Harrington}, J. 2005,
  \nat, 434, 740

\bibitem[{{Deming} {et~al.}(2012){Deming}, {Fraine}, {Sada}, {Madhusudhan},
  {Knutson}, {Harrington}, {Blecic}, {Nymeyer}, {Smith}, \& {Jackson}}]{de12}
{Deming}, D., {Fraine}, J.~D., {Sada}, P.~V., {et~al.} 2012, \apj, 754, 106

\bibitem[{{Deming} {et~al.}(2015){Deming}, {Knutson}, {Kammer}, {Fulton},
  {Ingalls}, {Carey}, {Burrows}, {Fortney}, {Todorov}, {Agol}, {Cowan},
  {Desert}, {Fraine}, {Langton}, {Morley}, \& {Showman}}]{deming15}
{Deming}, D., {Knutson}, H., {Kammer}, J., {et~al.} 2015, \apj, 805, 132

\bibitem[{{Dullemond}(2002)}]{dullemond02}
{Dullemond}, C.~P. 2002, \aap, 395, 853

\bibitem[{{Fortney} {et~al.}(2006){Fortney}, {Cooper}, {Showman}, {Marley}, \&
  {Freedman}}]{fo06}
{Fortney}, J.~J., {Cooper}, C.~S., {Showman}, A.~P., {Marley}, M.~S., \&
  {Freedman}, R.~S. 2006, \apj, 652, 746

\bibitem[{{Fortney} {et~al.}(2008){Fortney}, {Lodders}, {Marley}, \&
  {Freedman}}]{fo08}
{Fortney}, J.~J., {Lodders}, K., {Marley}, M.~S., \& {Freedman}, R.~S. 2008,
  \apj, 678, 1419

\bibitem[{{Fortney} {et~al.}(2005){Fortney}, {Marley}, {Lodders}, {Saumon}, \&
  {Freedman}}]{fo05}
{Fortney}, J.~J., {Marley}, M.~S., {Lodders}, K., {Saumon}, D., \& {Freedman},
  R. 2005, \apjl, 627, L69

\bibitem[{{Fortney} {et~al.}(2010){Fortney}, {Shabram}, {Showman}, {Lian},
  {Freedman}, {Marley}, \& {Lewis}}]{fo10}
{Fortney}, J.~J., {Shabram}, M., {Showman}, A.~P., {et~al.} 2010, \apj, 709,
  1396

\bibitem[{{Freedman} {et~al.}(2008){Freedman}, {Marley}, \& {Lodders}}]{fr08}
{Freedman}, R.~S., {Marley}, M.~S., \& {Lodders}, K. 2008, \apjs, 174, 504

\bibitem[{{Gillon} {et~al.}(2012){Gillon}, {Triaud}, {Fortney}, {Demory},
  {Jehin}, {Lendl}, {Magain}, {Kabath}, {Queloz}, {Alonso}, {Anderson},
  {Collier Cameron}, {Fumel}, {Hebb}, {Hellier}, {Lanotte}, {Maxted},
  {Mowlavi}, \& {Smalley}}]{gi12}
{Gillon}, M., {Triaud}, A.~H.~M.~J., {Fortney}, J.~J., {et~al.} 2012, \aap,
  542, A4

\bibitem[{{Goody} \& {Yung}(1989)}]{go89}
{Goody}, R.~M., \& {Yung}, Y.~L. 1989, {Atmospheric radiation : theoretical
  basis}

\bibitem[{{Grimm} \& {Heng}(2015)}]{gr15}
{Grimm}, S.~L., \& {Heng}, K. 2015, \apj, 808, 182

\bibitem[{{Hansen} {et~al.}(2014){Hansen}, {Schwartz}, \& {Cowan}}]{hansen14}
{Hansen}, C.~J., {Schwartz}, J.~C., \& {Cowan}, N.~B. 2014, \mnras, 444, 3632

\bibitem[{{Harps{\o}e} {et~al.}(2013){Harps{\o}e}, {Hardis}, {Hinse},
  {J{\o}rgensen}, {Mancini}, {Southworth}, {Alsubai}, {Bozza}, {Browne},
  {Burgdorf}, {Calchi Novati}, {Dodds}, {Dominik}, {Fang}, {Finet}, {Gerner},
  {Gu}, {Hundertmark}, {Jessen-Hansen}, {Kains}, {Kerins}, {Kjeldsen},
  {Liebig}, {Lund}, {Lundkvist}, {Mathiasen}, {Nesvorn{\'y}}, {Nikolov},
  {Penny}, {Proft}, {Rahvar}, {Ricci}, {Sahu}, {Scarpetta}, {Sch{\"a}fer},
  {Sch{\"o}nebeck}, {Snodgrass}, {Skottfelt}, {Surdej}, {Tregloan-Reed}, \&
  {Wertz}}]{ha13}
{Harps{\o}e}, K.~B.~W., {Hardis}, S., {Hinse}, T.~C., {et~al.} 2013, \aap, 549,
  A10

\bibitem[{{Hauschildt} {et~al.}(1999){Hauschildt}, {Allard}, \&
  {Baron}}]{hauschildt99}
{Hauschildt}, P.~H., {Allard}, F., \& {Baron}, E. 1999, \apj, 512, 377

\bibitem[{{Haynes} {et~al.}(2015){Haynes}, {Mandell}, {Madhusudhan}, {Deming},
  \& {Knutson}}]{ha15}
{Haynes}, K., {Mandell}, A.~M., {Madhusudhan}, N., {Deming}, D., \& {Knutson},
  H. 2015, \apj, 806, 146

\bibitem[{{Hebb} {et~al.}(2009){Hebb}, {Collier-Cameron}, {Loeillet},
  {Pollacco}, {H{\'e}brard}, {Street}, {Bouchy}, {Stempels}, {Moutou},
  {Simpson}, {Udry}, {Joshi}, {West}, {Skillen}, {Wilson}, {McDonald},
  {Gibson}, {Aigrain}, {Anderson}, {Benn}, {Christian}, {Enoch}, {Haswell},
  {Hellier}, {Horne}, {Irwin}, {Lister}, {Maxted}, {Mayor}, {Norton}, {Parley},
  {Pont}, {Queloz}, {Smalley}, \& {Wheatley}}]{he09}
{Hebb}, L., {Collier-Cameron}, A., {Loeillet}, B., {et~al.} 2009, \apj, 693,
  1920

\bibitem[{{Heng} \& {Lyons}(2016)}]{he16b}
{Heng}, K., \& {Lyons}, J.~R. 2016, \apj, 817, 149

\bibitem[{{Heng} {et~al.}(2016){Heng}, {Lyons}, \& {Tsai}}]{he16a}
{Heng}, K., {Lyons}, J.~R., \& {Tsai}, S.-M. 2016, \apj, 816, 96

\bibitem[{{Heng} {et~al.}(2014){Heng}, {Mendon{\c c}a}, \& {Lee}}]{he14}
{Heng}, K., {Mendon{\c c}a}, J.~M., \& {Lee}, J.-M. 2014, \apjs, 215, 4

\bibitem[{{Heng} \& {Tsai}(2016)}]{he16c}
{Heng}, K., \& {Tsai}, S.-M. 2016, ArXiv e-prints, arXiv:1603.05418

\bibitem[{{Hubeny} \& {Lanz}(1995)}]{hubeny95}
{Hubeny}, I., \& {Lanz}, T. 1995, \apj, 439, 875

\bibitem[{Hunter(2007)}]{matplotlib}
Hunter, J.~D. 2007, Computing In Science \& Engineering, 9, 90

\bibitem[{{Husser} {et~al.}(2013){Husser}, {Wende-von Berg}, {Dreizler},
  {Homeier}, {Reiners}, {Barman}, \& {Hauschildt}}]{hu13}
{Husser}, T.-O., {Wende-von Berg}, S., {Dreizler}, S., {et~al.} 2013, \aap,
  553, A6

\bibitem[{{Ingalls} {et~al.}(2016){Ingalls}, {Krick}, {Carey}, {Stauffer},
  {Lawrence}, {Grillmair}, {Buzasi}, {Deming}, {Diamond-Lowe}, {Evans},
  {Morello}, {Stevenson}, {Wong}, {Capak}, {Glaccum}, {Laine}, {Surace}, \&
  {Storrie-Lombardi}}]{ingalls16}
{Ingalls}, J.~G., {Krick}, J.~E., {Carey}, S.~J., {et~al.} 2016, ArXiv
  e-prints, arXiv:1601.05101

\bibitem[{{Joshi} {et~al.}(2009){Joshi}, {Pollacco}, {Collier Cameron},
  {Skillen}, {Simpson}, {Steele}, {Street}, {Stempels}, {Christian}, {Hebb},
  {Bouchy}, {Gibson}, {H{\'e}brard}, {Keenan}, {Loeillet}, {Meaburn}, {Moutou},
  {Smalley}, {Todd}, {West}, {Anderson}, {Bentley}, {Enoch}, {Haswell},
  {Hellier}, {Horne}, {Irwin}, {Lister}, {McDonald}, {Maxted}, {Mayor},
  {Norton}, {Parley}, {Perrier}, {Pont}, {Queloz}, {Ryans}, {Smith}, {Udry},
  {Wheatley}, \& {Wilson}}]{jo09}
{Joshi}, Y.~C., {Pollacco}, D., {Collier Cameron}, A., {et~al.} 2009, \mnras,
  392, 1532

\bibitem[{{Kataria} {et~al.}(2015){Kataria}, {Showman}, {Fortney}, {Stevenson},
  {Line}, {Kreidberg}, {Bean}, \& {D{\'e}sert}}]{kataria15}
{Kataria}, T., {Showman}, A.~P., {Fortney}, J.~J., {et~al.} 2015, \apj, 801, 86

\bibitem[{{Kl{\"o}ckner} {et~al.}(2012){Kl{\"o}ckner}, {Pinto}, {Lee},
  {Catanzaro}, {Ivanov}, \& {Fasih}}]{pycuda}
{Kl{\"o}ckner}, A., {Pinto}, N., {Lee}, Y., {et~al.} 2012, Parallel Computing,
  38, 157

\bibitem[{{Kov{\'a}cs} {et~al.}(2013){Kov{\'a}cs}, {Kov{\'a}cs}, {Hartman},
  {Bakos}, {Bieryla}, {Latham}, {Noyes}, {Reg{\'a}ly}, \& {Esquerdo}}]{ko13}
{Kov{\'a}cs}, G., {Kov{\'a}cs}, T., {Hartman}, J.~D., {et~al.} 2013, \aap, 553,
  A44

\bibitem[{{Kreidberg} {et~al.}(2014){Kreidberg}, {Bean}, {D{\'e}sert}, {Line},
  {Fortney}, {Madhusudhan}, {Stevenson}, {Showman}, {Charbonneau},
  {McCullough}, {Seager}, {Burrows}, {Henry}, {Williamson}, {Kataria}, \&
  {Homeier}}]{kr14}
{Kreidberg}, L., {Bean}, J.~L., {D{\'e}sert}, J.-M., {et~al.} 2014, \apjl, 793,
  L27

\bibitem[{{Kurucz}(1979)}]{ku79}
{Kurucz}, R.~L. 1979, \apjs, 40, 1

\bibitem[{{Lacis} \& {Oinas}(1991)}]{lacis91}
{Lacis}, A.~A., \& {Oinas}, V. 1991, \jgr, 96, 9027

\bibitem[{{Lavie} {et~al.}(2016){Lavie}, {Mendon{\c c}a}, {Mordasini}, {Malik},
  {Bonnefoy}, {Demory}, {Oreshenko}, {Grimm}, {Ehrenreich}, \&
  {Heng}}]{lavie16}
{Lavie}, B., {Mendon{\c c}a}, J.~M., {Mordasini}, C., {et~al.} 2016, ArXiv
  e-prints, arXiv:1610.03216

\bibitem[{{Lehmann} {et~al.}(2015){Lehmann}, {Guenther}, {Sebastian},
  {D{\"o}llinger}, {Hartmann}, \& {Mkrtichian}}]{le15}
{Lehmann}, H., {Guenther}, E., {Sebastian}, D., {et~al.} 2015, \aap, 578, L4

\bibitem[{{Lodders} \& {Fegley}(2002)}]{lo02}
{Lodders}, K., \& {Fegley}, B. 2002, \icarus, 155, 393

\bibitem[{{Lodders} \& {Fegley}(2006)}]{lo06}
{Lodders}, K., \& {Fegley}, Jr., B. 2006, {Chemistry of Low Mass Substellar
  Objects}, ed. J.~W. {Mason}, 1

\bibitem[{{Marley} \& {McKay}(1999)}]{ma99}
{Marley}, M.~S., \& {McKay}, C.~P. 1999, \icarus, 138, 268

\bibitem[{{Marley} \& {Robinson}(2015)}]{marley15}
{Marley}, M.~S., \& {Robinson}, T.~D. 2015, \araa, 53, 279

\bibitem[{{McKay} {et~al.}(1989){McKay}, {Pollack}, \& {Courtin}}]{mc89}
{McKay}, C.~P., {Pollack}, J.~B., \& {Courtin}, R. 1989, \icarus, 80, 23

\bibitem[{{Meador} \& {Weaver}(1980)}]{me80}
{Meador}, W.~E., \& {Weaver}, W.~R. 1980, Journal of Atmospheric Sciences, 37,
  630

\bibitem[{{Mendon{\c c}a} {et~al.}(2016){Mendon{\c c}a}, {Grimm}, {Grosheintz},
  \& {Heng}}]{mendonca16}
{Mendon{\c c}a}, J.~M., {Grimm}, S.~L., {Grosheintz}, L., \& {Heng}, K. 2016,
  \apj, 829, 115

\bibitem[{{Mendon{\c c}a} {et~al.}(2015){Mendon{\c c}a}, {Read}, {Wilson}, \&
  {Lee}}]{mendonca2015}
{Mendon{\c c}a}, J.~M., {Read}, P.~L., {Wilson}, C.~F., \& {Lee}, C. 2015,
  \planss, 105, 80

\bibitem[{{Mihalas}(1970)}]{mi70}
{Mihalas}, D. 1970, {Stellar atmospheres}

\bibitem[{{Miller-Ricci} \& {Fortney}(2010)}]{mi10}
{Miller-Ricci}, E., \& {Fortney}, J.~J. 2010, \apjl, 716, L74

\bibitem[{{Molli{\`e}re} {et~al.}(2015){Molli{\`e}re}, {van Boekel},
  {Dullemond}, {Henning}, \& {Mordasini}}]{molli15}
{Molli{\`e}re}, P., {van Boekel}, R., {Dullemond}, C., {Henning}, T., \&
  {Mordasini}, C. 2015, \apj, 813, 47

\bibitem[{{Morley} {et~al.}(2013){Morley}, {Fortney}, {Kempton}, {Marley},
  {Visscher}, \& {Zahnle}}]{mo13}
{Morley}, C.~V., {Fortney}, J.~J., {Kempton}, E.~M.-R., {et~al.} 2013, \apj,
  775, 33

\bibitem[{{Morley} {et~al.}(2015){Morley}, {Fortney}, {Marley}, {Zahnle},
  {Line}, {Kempton}, {Lewis}, \& {Cahoy}}]{mo15}
{Morley}, C.~V., {Fortney}, J.~J., {Marley}, M.~S., {et~al.} 2015, \apj, 815,
  110

\bibitem[{{Munari} {et~al.}(2005){Munari}, {Sordo}, {Castelli}, \&
  {Zwitter}}]{mu05}
{Munari}, U., {Sordo}, R., {Castelli}, F., \& {Zwitter}, T. 2005, \aap, 442,
  1127

\bibitem[{{Murphy} \& {Meiksin}(2004)}]{mu04}
{Murphy}, T., \& {Meiksin}, A. 2004, \mnras, 351, 1430

\bibitem[{Nickolls {et~al.}(2008)Nickolls, Buck, Garland, \& Skadron}]{cuda}
Nickolls, J., Buck, I., Garland, M., \& Skadron, K. 2008, Queue, 6, 40

\bibitem[{{Oliphant}(2007)}]{scipy}
{Oliphant}, T.~E. 2007, {Computing in Science \& Engineering}, 9,
  doi:10.1109/MCSE.2007.58

\bibitem[{{Pierrehumbert}(2010)}]{pi10}
{Pierrehumbert}, R.~T. 2010, {Principles of Planetary Climate}

\bibitem[{{Queloz} {et~al.}(2010){Queloz}, {Anderson}, {Collier Cameron},
  {Gillon}, {Hebb}, {Hellier}, {Maxted}, {Pepe}, {Pollacco}, {S{\'e}gransan},
  {Smalley}, {Triaud}, {Udry}, \& {West}}]{qu10}
{Queloz}, D., {Anderson}, D.~R., {Collier Cameron}, A., {et~al.} 2010, \aap,
  517, L1

\bibitem[{{Richard} {et~al.}(2012){Richard}, {Gordon}, {Rothman}, {Abel},
  {Frommhold}, {Gustafsson}, {Hartmann}, {Hermans}, {Lafferty}, {Orton},
  {Smith}, \& {Tran}}]{ri12}
{Richard}, C., {Gordon}, I.~E., {Rothman}, L.~S., {et~al.} 2012, \jqsrt, 113,
  1276

\bibitem[{{Rothman} {et~al.}(2010){Rothman}, {Gordon}, {Barber}, {Dothe},
  {Gamache}, {Goldman}, {Perevalov}, {Tashkun}, \& {Tennyson}}]{ro10}
{Rothman}, L.~S., {Gordon}, I.~E., {Barber}, R.~J., {et~al.} 2010, \jqsrt, 111,
  2139

\bibitem[{{Rothman} {et~al.}(2013){Rothman}, {Gordon}, {Babikov}, {Barbe},
  {Chris Benner}, {Bernath}, {Birk}, {Bizzocchi}, {Boudon}, {Brown},
  {Campargue}, {Chance}, {Cohen}, {Coudert}, {Devi}, {Drouin}, {Fayt}, {Flaud},
  {Gamache}, {Harrison}, {Hartmann}, {Hill}, {Hodges}, {Jacquemart}, {Jolly},
  {Lamouroux}, {Le Roy}, {Li}, {Long}, {Lyulin}, {Mackie}, {Massie},
  {Mikhailenko}, {M{\"u}ller}, {Naumenko}, {Nikitin}, {Orphal}, {Perevalov},
  {Perrin}, {Polovtseva}, {Richard}, {Smith}, {Starikova}, {Sung}, {Tashkun},
  {Tennyson}, {Toon}, {Tyuterev}, \& {Wagner}}]{ro13}
{Rothman}, L.~S., {Gordon}, I.~E., {Babikov}, Y., {et~al.} 2013, \jqsrt, 130, 4

\bibitem[{{Seager} \& {Sasselov}(2000)}]{se00}
{Seager}, S., \& {Sasselov}, D.~D. 2000, \apj, 537, 916

\bibitem[{{Sharp} \& {Burrows}(2007)}]{sh07}
{Sharp}, C.~M., \& {Burrows}, A. 2007, \apjs, 168, 140

\bibitem[{{Sneep} \& {Ubachs}(2005)}]{sn05}
{Sneep}, M., \& {Ubachs}, W. 2005, \jqsrt, 92, 293

\bibitem[{{Southworth}(2010)}]{so10}
{Southworth}, J. 2010, \mnras, 408, 1689

\bibitem[{{Spiegel} \& {Burrows}(2010)}]{sb10}
{Spiegel}, D.~S., \& {Burrows}, A. 2010, \apj, 722, 871

\bibitem[{{Stevenson} {et~al.}(2014){Stevenson}, {Bean}, {Madhusudhan}, \&
  {Harrington}}]{st14}
{Stevenson}, K.~B., {Bean}, J.~L., {Madhusudhan}, N., \& {Harrington}, J. 2014,
  \apj, 791, 36

\bibitem[{{Stevenson} {et~al.}(2012){Stevenson}, {Harrington}, {Fortney},
  {Loredo}, {Hardy}, {Nymeyer}, {Bowman}, {Cubillos}, {Bowman}, \&
  {Hardin}}]{stevenson12}
{Stevenson}, K.~B., {Harrington}, J., {Fortney}, J.~J., {et~al.} 2012, \apj,
  754, 136

\bibitem[{{Sudarsky} {et~al.}(2003){Sudarsky}, {Burrows}, \& {Hubeny}}]{su03}
{Sudarsky}, D., {Burrows}, A., \& {Hubeny}, I. 2003, \apj, 588, 1121

\bibitem[{{Todorov} {et~al.}(2014){Todorov}, {Deming}, {Burrows}, \&
  {Grillmair}}]{to14}
{Todorov}, K.~O., {Deming}, D., {Burrows}, A., \& {Grillmair}, C.~J. 2014,
  \apj, 796, 100

\bibitem[{{Toon} {et~al.}(1989){Toon}, {McKay}, {Ackerman}, \&
  {Santhanam}}]{toon89}
{Toon}, O.~B., {McKay}, C.~P., {Ackerman}, T.~P., \& {Santhanam}, K. 1989,
  \jgr, 94, 16287

\bibitem[{{Tsai} {et~al.}(2016){Tsai}, {Lyons}, {Grosheintz}, {Rimmer},
  {Kitzmann}, \& {Heng}}]{tsai16}
{Tsai}, S.-M., {Lyons}, J.~R., {Grosheintz}, L., {et~al.} 2016, ArXiv e-prints,
  arXiv:1607.00409

\bibitem[{{van der Walt} {et~al.}(2011){van der Walt}, C., \&
  {Varoquaux}}]{numpy}
{van der Walt}, S., C., C., \& {Varoquaux}, G. 2011, Computing in Science \&
  Engineering, 13, 22

\bibitem[{{van Rossum}(1995)}]{python}
{van Rossum}, G. 1995, {Python} tutorial, Report CS-R9526

\end{thebibliography}

\end{document}